\documentclass{jfm}
\usepackage{graphicx}
\usepackage{comment}
\usepackage{newtxtext}
\usepackage{newtxmath}
\usepackage{natbib}

\usepackage{epstopdf, epsfig}
\usepackage{amsmath} 
\usepackage{amssymb}
\usepackage{xparse}
\usepackage{ifthen}	
\usepackage{xspace}
\usepackage{hyperref}
\hypersetup{
	colorlinks = true,
	urlcolor   = blue,
	citecolor  = black,
}
\usepackage{siunitx}
\usepackage{multirow}
\usepackage{array}
\usepackage{subfig}
\usepackage[usenames,dvipsnames,svgnames]{xcolor}
\usepackage{float}
\usepackage[shortlabels]{enumitem}

\renewcommand{\vec}[1]{{\bf #1}}
\renewcommand{\Re}{\mathrm{Re}}
\newcommand{\norm}[1]{\left\lVert #1 \right\rVert}

\newcommand\zeropad[2]{%
  \ifnum#2<0\relax%
    {\ensuremath-}\zeropadA{#1}{\the\numexpr#2*-1\relax}%
  \else%
    \zeropadA{#1}{#2}%
  \fi%
}
\def\zeropadA#1#2{%
  \ifnum1#2<1#1
    \zeropadA{#1}{0#2}%
  \else%
    #2%
  \fi%
}

\shorttitle{ECSs in turbulent Taylor-Couette flow}
\shortauthor{M. C. Krygier, J. L. Pughe-Sanford, and R. O. Grigoriev}

\title{Exact coherent structures and shadowing in turbulent Taylor-Couette flow}

\author{Michael C. Krygier\aff{1}
 \corresp{\email{mkrygier1@gatech.edu}},
      Joshua L. Pughe-Sanford\aff{1},
 \and Roman O. Grigoriev\aff{1}}

\affiliation{\aff{1}School of Physics, Georgia Institute of Technology, 
Atlanta, GA 30332, USA}

\newcommand{\rpo}[1]{{RPO$_{#1}$}}
\newcommand\lobe[1]{lobe #1}
\newcommand\lobes[2]{lobes #1 and #2}

\newcommand{\re}[1]{{TW$_{#1}$}}

\DeclareDocumentCommand \nnums {o o} {%
	\IfNoValueTF {#2} 
	 {{\sisetup{
	round-mode 			= places,
	round-precision 	= 4,
	scientific-notation = false
}  \num{#1}}}
	 {{\sisetup{
	round-mode 			= places,
	round-precision 	= #2,
	scientific-notation = false
}  \num{#1}}}%
	\xspace 
}

\DeclareDocumentCommand \enums {o o} {%
	\IfNoValueTF {#2} 
	 {{\sisetup{
	round-mode 			= places,
	round-precision 	= 4,
	scientific-notation = true,
	output-exponent-marker = \text{e}
}  \num{#1}}}
	 {{\sisetup{
	round-mode 			= places,
	round-precision 	= #2,
	scientific-notation = true,
	output-exponent-marker = \text{e}
}  \num{#1}}}%
	\xspace 
}

\DeclareDocumentCommand \Rei {o} {%
	\IfNoValueTF {#1} 
	 {$Re_{i}$}
	 {${Re_{i}={#1}}$}%
	\xspace 
}
\DeclareDocumentCommand \Reo {o} {%
	\IfNoValueTF {#1} 
	 {$Re_{o}$}
	 {${Re_{o}={#1}}$}%
	\xspace
}

\begin{document}

\maketitle

\begin{abstract}
We investigate a theoretical framework for modeling fluid turbulence based on the formalism of exact coherent structures (ECSs). Although highly promising, existing evidence for the role of ECSs in turbulent flows is largely circumstantial and comes primarily from idealized numerical simulations. In particular, it remains unclear whether three-dimensional turbulent flows in experiment shadow any ECSs. In order to conclusively answer this question, a hierarchy of ECSs should be computed on a domain and with boundary conditions exactly matching experiment. The present study makes the first step in this direction by investigating a small-aspect-ratio Taylor-Couette flow with naturally periodic boundary condition in the azimuthal direction. We describe the structure of the chaotic set underlying turbulent flow driven by counter-rotating cylinders and present direct numerical evidence for shadowing of a collection of unstable relative periodic orbits and a traveling wave, setting the stage for further experimental tests of the framework.
\end{abstract}

\begin{keywords}
turbulence theory, chaos, nonlinear dynamical systems
\end{keywords}


%
%

\section{Introduction}
\label{sec:1}

Fluid turbulence has a unique place in science and engineering, due to both its ubiquity and tremendous practical importance as well as its resistance to progress despite a long history of systematic investigation. A statistical description, which dominated early theoretical studies, brought some advances, such as the Kolmogorov's scaling law \citep{kolmogorov1941} and the law of the wall \citep{vonkarman1930,nikuradse1932}. These advances, however, are based on general concepts such as dimensional analysis, spatial uniformity, and/or isotropy, that are not directly related to the equations governing fluid flow and shed little light onto the nature of the turbulent cascades or momentum transport in wall-bounded flows. Moreover, statistical approaches provide minimal insight into the prediction and control of fluid turbulence.

Existing statistical models of fluid turbulence fail to predict even such simple quantities as the mean energy dissipation (e.g., in isotropic turbulence) or momentum flux (e.g., in a wall-bounded flow). The calculation of the friction coefficient for pipe flow turbulence is a good example, where one has to rely on empirically derived Moody charts \citep{moody1944}. Statistical description also fails to account for the presence of coherent structures, which are well-known to play an important role in turbulence \citep{hussain1983}. Indeed, coherent structures break both spatial uniformity and isotropy and introduce new spatial and temporal scales, invalidating the entire foundation of statistical description.

The most promising alternative approach is to build a dynamical description of turbulence firmly based on the equations governing fluid flow. Unlike other classical field theories, such as electromagnetism, whose governing equations are linear, fluid turbulence is governed by the Navier-Stokes equation which is strongly nonlinear, making analytical solutions intractable. Recent advances in numerical methods brought a realization that coherent structures represent unstable solutions of Navier-Stokes with simple temporal dependence, with the earliest example provided by \citet{nagata1990}. This started a revolution in our understanding of fluid turbulence \citep{kawahara2012}. 
Termed exact coherent structures (ECSs), such solutions have been shown to play a key role in the transition from laminar flow to turbulence \citep{kerswell2005,eckhardt2008} and self-sustaining processes in boundary layers \citep{waleffe2005}. In some cases, a single ECS was found to reproduce certain statistical properties of weakly turbulent flow. 
A time-periodic solution obtained by \citet{kawahara2001} for plane Couette flow 
was found to reproduce, with fairly high accuracy, both the mean flow profile and the fluctuations in all three components of the velocity.

It is therefore natural to ask whether these results are coincidental or some collection of ECSs can in fact provide a dynamical and statistical description of fluid turbulence. The idea that turbulence can be thought of as a deterministic walk through a repertoire of patterns (which we now associate with different ECSs) goes back to Eberhard Hopf \citep{hopf1942,hopf1948}. In Hopf's view, the snapshot of turbulent flow in the physical space corresponds to a point in the associated infinite-dimensional state (or phase) space. This point traces out a one-dimensional trajectory as the flow evolves in time. This trajectory is confined, due to dissipation, to a finite-dimensional set embedded within this state space. This set can be either an attractor (for sustained turbulence) or a repeller (for transient turbulence).

It took two more decades to flesh out the details of Hopf's picture, when mathematical foundations of deterministic chaos, geometry of chaotic sets, and ergodic theory were developed \citep{lorenz1963,mandelbrot1967,arnold1968}. In particular, for uniformly hyperbolic
chaotic systems without continuous symmetries, unstable periodic orbits (UPOs) are dense in the chaotic set \citep{gaspard2005}, which has two important implications. The first one is dynamical: chaotic trajectories shadow nearby UPOs. The second one is statistical: temporal averages over a chaotic trajectory can be computed as a sum over an infinite hierarchy of UPOs \citep{arnold1968}, with the weight of each term predicted by periodic orbit theory \citep{auerbach1987,cvitanovic1988,lan2010}.

Turbulent fluid flows are chaotic, at least in so far as the sensitive dependence on initial conditions is concerned, so it is natural to ask whether the properties of shadowing and ergodicity also apply to turbulence. Originally conjectured by Hopf, the shadowing property is widely assumed \citep{cvitanovic2013} in the studies exploring the ECS-based framework, although there is very little direct evidence in its favor. 

Most of the evidence for turbulent flows visiting neighborhoods of ECSs has been obtained using numerical simulations of minimal flow units and, therefore, has to be taken with a grain of salt. For instance, unstable equilibria and UPOs were found to be visited in forced two-dimensional flow \citep{kazantsev1998,chandler2013,lucas2015}, isotropic turbulence \citep{vanVeen2006}, and plane Couette flow \citep{cvitanovic2010,kreilos2012}. Traveling waves (TWs) were found to be visited in plane Poiseuille flow \citep{park2015} and pipe flow \citep{schneider2007,kerswell2007,dennis2014} and relative periodic orbits (RPOs) in pipe flow \citep{willis2013,budanur2017}. Experiments provide more direct and solid evidence. For instance, TWs were found to be visited in pipe flows \citep{hof2004,delozar2012} and unstable equilibria in forced quasi-two-dimensional flows \citep{suri2017,suri2018}.

Visits by turbulence to a neighborhood of an ECS, however, do not imply shadowing, since, in practice, no ECSs are visited particularly closely. Shadowing requires that turbulent flow remain in the neighborhood of an ECS for an extended interval of time comparable to the characteristic escape time. It also requires that turbulent flow evolve in the same manner as the ECS. Such evidence is currently limited to UPOs in flows without continuous symmetries, e.g., Kolmogorov flow in two dimensions \citep{suri2020} and three dimensions \citep{yalniz2020}. However, it is not completely understood how the presence of symmetries affects the shadowing property.

To sum up, while a dynamical description of fluid turbulence based on exact coherent structures is promising, it is yet to be properly validated in both numerical and experimental setting, especially for flows with continuous symmetries. A key challenge is the disconnect between experiments that are conducted on large spatial domains and typically for open flows such as pipe, channel, or plane Couette flow and numerics: ECSs are typically computed on minimal flow units with unphysical (e.g., spatially periodic) boundary conditions and/or under unrealistic constraints (e.g., in highly symmetric subspaces). To ensure an apples-to-apples comparison, the dynamical description should be validated in a geometry and under conditions where numerical simulations match experiment as closely as possible. Hence, this study focuses on a weakly turbulent Taylor-Couette flow (TCF) between concentric cylinders, which is easy to realize in practice. This is a closed flow whose boundary conditions in the azimuthal direction are naturally periodic, reflecting continuous rotational symmetry. Furthermore, we do not restrict ECSs to lie in any symmetry subspace.


\begin{figure}
\center
\includegraphics[width=0.5\textwidth]{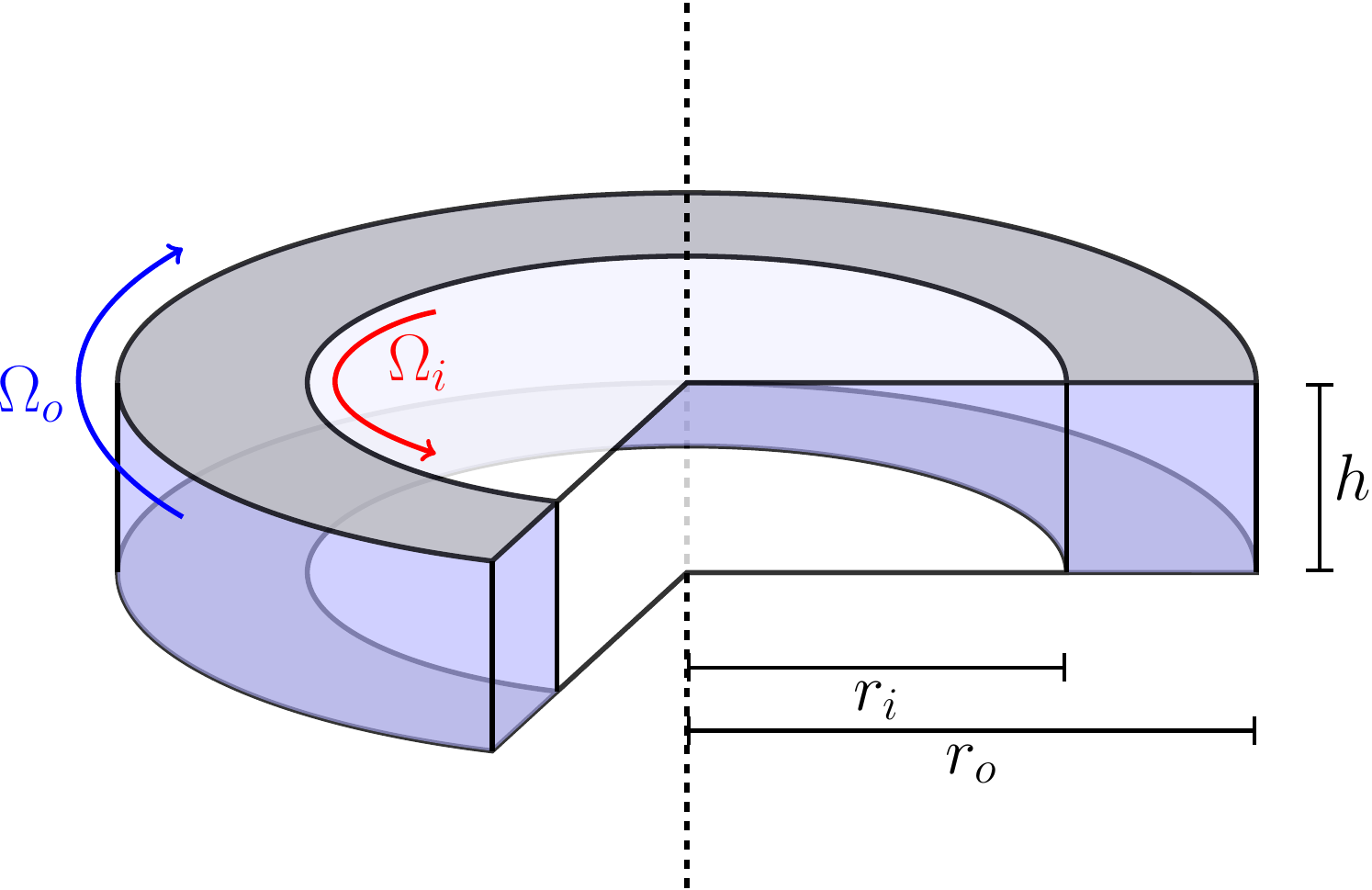}\caption{\label{fig:tc_setup} \small
Geometry of the Taylor-Couette flow driven by counter-rotating cylinders. The fluid domain is shaded purple.}
\end{figure}

The TCF considered in this study is characterized by four nondimensional parameters. Two of these are geometrical: $\Gamma=h/d$ and $\eta=r_i/r_o$, where $h$ is the height of the fluid layer and $r_i$ and $r_o$ are the radii of the inner and outer cylinder, respectively, as shown in \autoref{fig:tc_setup}. The other two are the Reynolds numbers $Re_i=\Omega_ir_id/\nu$ and $Re_o=\Omega_or_od/\nu$, where $d=r_o-r_i$ is the gap between the cylinders, $\nu$ is the kinematic viscosity of the fluid, and $\Omega_i$ and $\Omega_o$ are the angular velocities of the two cylinders. We focus on flows driven by counter-rotating cylinders ($Re_i=1200$, $Re_o=-1200$) in the wide-gap ($\eta=0.5$), small-aspect-ratio ($\Gamma=1$) geometry with stationary end-caps.

Previous studies of wide-gap, small-aspect-ratio TCF mainly focused on the flows driven by inner cylinder rotation with the outer cylinder and end-caps being stationary. The bifurcations between steady flows associated with the changes in $Re_i$, $\Gamma$, and $\eta$ have been studied by \citet{benjamin1981,mullin1982,cliffe1983,aitta1985,pfister1988} and \citet{mullin2002}. Time-dependent flows have been investigated by \citet{lorenzen1983,pfister1991,furukawa2002}. The onset of turbulent flows has been considered by \citet{streett1991,pfister1992} for $\Gamma>1$,
\citet{marques2006} for $\Gamma=1$, and 
\citet{buzug1992,buzug1993} for $\Gamma<1$.
The effect of the outer cylinder rotation ($Re_o\ne 0$) on the bifurcation sequence was considered by \citet{schulz2003} and \citet{altmeyer2012}, but neither study explored Reynolds numbers sufficiently high to observe turbulent flows.

Turbulence in counter-rotating TCF has been investigated mainly in small-gap ($0.9\lesssim\eta<1$), large-aspect-ratio ($\Gamma\gg1$) geometry \citep{coles1965,andereck1986}. A number of 
TWs, both stable and unstable, have been identified in this regime for $Re_o\lesssim -1200$. These TWs take the shape of rotating spiral waves \citep{meseguer2009}, ribbons \citep{deguchi2013}, or spatially localized vortex pairs \citep{deguchi2014}. 
However, no unstable \rpo{}s of TCF have been found so far.

The main objective of the present numerical study is to validate the key assumptions which underlie the dynamical description of fluid turbulence in a setting which is straightforward to replicate in experiment \citep{hochstrate2010,hoffmann2013,heise2013}.
Specifically, we are looking to compute a library of ECSs (both TWs and RPOs) that are dynamically prominent and  analyze turbulent flows for close passes to each ECS to identify shadowing events, if there are any. The paper is structured as follows. Numerical methods for solving the Navier-Stokes equation and computing ECSs are described in Section \ref{sec:math}. Our results are presented in Section \ref{sec:results} and analyzed in \ref{sec:disc}. Finally, Section \ref{sec:conc} presents our conclusions.

%
%

\section{Mathematical Formulation}
\label{sec:math}

\subsection{Direct numerical simulation} 
The flow is governed by the Navier-Stokes equation and the incompressibility condition which can be written in nondimensional form using the gap $d$ and the diffusive time scale $d^2/\nu$ as the length and time scale, respectively:
\begin{align}\label{eq:NSE}
\partial_{t} \vec{u} + ( \vec{u} \cdot \nabla)\vec{u} &= -\nabla p + \nabla^{2} \vec{u}, \nonumber\\
\nabla \cdot \vec{u} &= 0.
\end{align}
Here $\vec{u}=(u_{r},u_{\theta},u_{z})$ and $p$ are the nondimensional velocity and pressure. No-slip boundary conditions are imposed on all the walls:
\begin{align}\label{eq:bc}
\vec{u}(r_{i},\theta,z,t) &= (0,Re_{i},0), \nonumber\\
\vec{u}(r_{o},\theta,z,t) &= (0,Re_{o},0), \nonumber\\
\vec{u}(r,\theta,\pm h/2,t) &= (0,0,0).
\end{align}
The last relation describes the boundary conditions at the top and bottom of the fluid layer. 

Direct numerical simulations (DNS) of TCF were performed using a pseudospectral code \citep{avila2008, mercader2010, avila2012} which solves the governing equations in cylindrical coordinates $(r,\theta,z)$. 
The velocity field $\vec{u}$ at location $(r,\theta,z)$ and time $t$ is given by
\begin{equation}
\vec{u}(r,\theta,z,t)=\Re \sum_{k=0}^{N_r}\sum_{l=0}^{N_z}
\sum_{m=0}^{N_\theta/2}\vec{U}^{klm}(t)T_k(\rho)T_l(\zeta)e^{im\theta},
\end{equation}
where $\rho=(2r-r_i-r_o)/d$ and $\zeta=2z/h$. $N_r$, $N_z$, and $N_\theta$ are the number of spectral modes in the three coordinate directions,
$T_k(\cdot)$ is the Chebyshev polynomial of order $k$, and $\Re$ denotes the real part. 
The solution is advanced in time using a second order stiffly stable time-splitting scheme \citep{hugues1998}.
Advection terms are evaluated on the spatial grid $(r_k,z_l,\theta_m)$ in physical space, where
\begin{align}
r_k &= \frac{(r_{o}-r_{i})\cos(k\pi/N_{r})+r_{i}+r_{o}}{2}, \; k=0,\ldots,N_{r}, \nonumber\\
z_l &= \frac{\Gamma \cos(l \pi / N_{z})}{2}, \; l=0, \ldots, N_{z}
\end{align}
are Chebyshev collocation points and $\theta_m=2\pi m/N_\theta$ with $m=0,\cdots,N_\theta-1$. 
The Helmholtz and Poisson equations are solved efficiently using a complete diagonalization of the operators in both the radial and axial direction for each Fourier mode \citep{Orszag1983}. 

We set $N_{r}=32$, $N_{\theta}=128$, $N_{z}=48$ (which corresponds to $3 (N_r+1)(N_z+1)N_\theta=620928$ degrees of freedom) and used a time step $dt=O(10^{-6})$ to accurately resolve the spatial structure and temporal dependence of turbulent flow and all computed ECSs. The spatial resolution was chosen such that the magnitude of the spectral coefficients ${\bf U}^{klm}$ decreases by at least four orders of magnitude for ECSs (and at least three orders of magnitude for turbulent flows) as $k$, $l$, or $m$ increases from the smallest to the largest value.  

\subsection{Computation of Exact Coherent Structures}

Under the boundary conditions \eqref{eq:bc}, TCF is invariant under arbitrary rotations $R_\phi$ about the $z$-axis, and a reflection $K_z$ about the mid-plane $z=0$, where
\begin{align}
\label{eq:TCF_symmetries}
    R_\phi \vec{u}(r,\theta,z,t) &= \vec{u}(r,\theta+\phi,z,t), \\
    K_z \vec{u}(r,\theta,z,t) &= (u_{r},u_{\theta},-u_{z})(r,\theta,-z,t).
\end{align}
These transformations form a symmetry group $\mathcal{G}=\mathrm{SO}(2)\times \mathrm{Z}_{2}$.
The presence of continuous rotational symmetry and the lack of reflection symmetry in $\theta$ imply that the dynamically relevant ECSs in Taylor-Couette flow are relative, e.g., relative periodic orbits (\rpo{}s) and relative equilibria (\re{}s), which correspond to time-periodic and stationary states in a co-rotating reference frame. In particular, \rpo{}s satisfy
\begin{align}\label{eq:RPO}
     \vec{u}(T) - R_{\Phi}\vec{u}(0) = \vec{0},
\end{align}
where $\Phi$ and $T$ are the solution's rotational shift and period, respectively. The angular velocity of this co-rotating reference frame is $\Omega=\Phi/T$. \re{}s also satisfy \eqref{eq:RPO} for $T=\Phi/\Omega$ at arbitrary $\Phi$. For both \rpo{}s and \re{}s, some of which have an $N$-fold discrete rotational symmetry, we arbitrarily restrict $0\le\Phi<2\pi/N$ to make the definition of rotational shift unique.
To find an ECS, the nonlinear equation \eqref{eq:RPO} is solved for $\vec{u}(0)$, $T$ and/or $\Phi$ using a custom Newton-GMRES solver that takes advantage of a hookstep algorithm \citep{viswanath2007,viswanath2009critical}. A relative residual 
\begin{align}\label{eq:residual}
     \varepsilon=\frac{\|\vec{u}(T) - R_{\Phi}\vec{u}(0)\|}{\|\vec{u}(0)\|},
\end{align}
was used as the stopping condition for the solver: solutions are considered converged for $\varepsilon<10^{-11}$. A large set of initial conditions for the solver was generated using the natural measure of the flow, as described below. 

Turbulent flow was established using the following protocol. With fluid initially stationary in the entire flow domain, the outer cylinder angular velocity was set to a value $\Omega_o$ corresponding to the target $Re_o=-1200$ (with the inner cylinder stationary). The flow was then allowed to evolve for 8.8 time units until a (steady and azimuthally uniform) asymptotic state was established. All the Fourier modes except $m=0$ were then weakly perturbed in order to break the azimuthal symmetry of the flow, after which the inner cylinder angular velocity was set to a value $\Omega_i$ corresponding to the target $Re_i=1200$. The flow was then evolved for 1.6 time units until a statistically stationary state was reached.

Different turbulent field snapshots were then used as initial conditions to the Newton solver to generate a set of ECSs. To increase the likelihood that the computed ECSs are dynamically relevant, we initialized the Newton solver using deep minima of the recurrence function \citep{kawahara2001,viswanath2007,cvitanovic2010}
\begin{equation}
\label{eq:recurrence}
G(t,\tau) = \min_{\phi} \norm{\vec{u}(t+\tau) - R_{\phi}\vec{u}(t)},
\end{equation} 
where $\norm{\cdot}$ is the $L_2$-norm. 
Some of these minima correspond to close passes to \rpo{}s or \re{}s, and the corresponding flow fields $\vec{u}(t)$, time delays $\tau$, and rotation angles $\phi$ represent good initial conditions for the solver.
An alternative approach to identifying dynamically relevant initial conditions relies on dynamic mode decomposition \citep{page2020}.
Converged solutions were also numerically continued in $Re_i$ using pseudo-arclength continuation \citep{allgower2003}; some branches turned around, yielding several additional solutions. For this reason, all of the solutions used in this study were computed for $Re_i = 1200\pm 3$. Their properties are summarized in \autoref{tab:ECSinTCF}.
\begin{table}
\centering
\begin{tabular}{ c|c|c|c|c|
c|c|c|c|c
}
           $n$ & ${\bf u}_n$ & $T$ & $\Phi$ & $N_{\lambda}^{u}$  & $\gamma_n^{-1}$ & $\min_t D_n^1(t)$ & $\min_t D_n^2(t)$ & $\min_t D_n^3(t)$ & \multicolumn{1}{p{1.5cm}}{\centering Discrete \\ Symmetry} \\
\hline
1 & \re{01}       & \nnums[0.1304621044473397][4]  & $\pi$ & 10 & $0.0121$              	 & \nnums[0.1905267154312707][2] & \nnums[1.1695026723356305][2] & -- & $K_{z} R_{\pi/2}$ \\
2 & \re{02}       & \nnums[0.0127039504229837][4] & $\pi$ & 10 & $0.0026$  	 & \nnums[1.4348401188730640][2] & \nnums[0.6795786928412477][2] & \nnums[0.2715780129657656][2] & $R_{\pi}$   \\
3 & \re{03}       & \nnums[0.0458783639086701][4] & $2\pi$ & 43 & $0.0004$ & \nnums[1.3704080637019764][2] & \nnums[1.5758957216889453][2] & -- &  $K_{z} R_{\pi}$  \\
4 & \rpo{01}      & \nnums[0.0158624182706232][4] & \nnums[0.354129719612447][4]     & 4  & $0.0228$              & \nnums[0.1843551208706284][2] & \nnums[1.1648732957837167][2] & -- & $K_{z} R_{\pi/2}$ \\
5 & \rpo{02}      & \nnums[0.0501473983087829][4] & \nnums[0.82754100577114][4]      & 9  & $0.0063$             & \nnums[0.2204880921173422][2] & \nnums[1.1520643169672700][2] & -- & -- \\
6 & \rpo{03}      & \nnums[0.0506152236163852][4] & \nnums[0.881718176320125][4]     & 4  & $0.0095$              & \nnums[0.1791542762557419][2] & \nnums[1.1639677785117917][2] & -- & -- \\
7 & \rpo{04}      & \nnums[0.0512055162843001][4] & \nnums[0.967587754859682][4]     & 5  & $0.0205$              & \nnums[0.2247363639931177][2] & \nnums[1.1346878413365453][2] & -- & -- \\
8 & \rpo{05}      & \nnums[0.0261565883944357][4] & \nnums[3.41183263172747][4]      & 5  & $0.0227$              & \nnums[0.1370686486009707][2] & \nnums[1.1190250355695361][2] & -- & -- \\
9 & \rpo{06}      & \nnums[0.0062842437016105][4]   & \nnums[3.268769835409118141][4] & 6  & $0.0247$            & \nnums[0.2197363112599227][2] & \nnums[1.1729283912343604][2] & -- & -- \\
10 & \rpo{07}      & \nnums[0.0213448639393844][4]   & \nnums[0.158087406999197][4]     & 4  & $0.0210$           & \nnums[0.2995075500216911][2] & \nnums[1.1508282398088534][2] & -- & $R_{\pi}$   \\
11 & \rpo{08}      & \nnums[0.0492694942754892][4]   & \nnums[0.909119027478775][4]     & 8  & $0.0124$  & \nnums[0.2027755001294223][2] & \nnums[1.1641532320437311][2] & -- & -- \\
12 & \rpo{09}      & \nnums[0.0191413067242211][4]   & \nnums[3.49486789040229][4]      & 3  & $0.0214$  & \nnums[0.2297324110333579][2] & \nnums[1.1560326090103583][2] & -- & -- \\
13 & \rpo{10}     & \nnums[0.0194401408283861][4]   & \nnums[3.46150307760976][4]      & 4  & $0.0152$  & \nnums[0.2404452801822634][2] & \nnums[1.1485694534600233][2] & -- & -- \\
14 & \rpo{11}     & \nnums[0.0196435359606627][4]   & \nnums[3.42709286968503][4]      & 5  & $0.0159$  & \nnums[0.2315417768383805][2] & \nnums[1.1437983381970793][2] & -- & -- \\
15 & \rpo{12}     & \nnums[0.0217019478159313][4]   & \nnums[0.144164982451532][4]     & 5  & $0.0193$  & \nnums[0.2952389762892502][2] & \nnums[1.1596232149794203][2] & -- & $R_{\pi}$   \\
16 & \rpo{13}     & \nnums[0.00810388959264575][4]  & \nnums[1.16167275586289][4]      & 9  & $0.0051$  & \nnums[1.3549053109038098][2] & \nnums[0.7660823255603668][2] & \nnums[0.2690440475536456][2] & $R_{\pi}$   \\
17 & \rpo{14}    & \nnums[0.044638061905787][4]    & \nnums[5.00163546840678][4]      & 9  & $0.0077$  & \nnums[1.4498230049240082][2] & \nnums[0.8814047684112668][2] & \nnums[0.2758037692686085][2] & -- \\
18 & \rpo{15}     & \nnums[0.0073267969789072452][4] & \nnums[1.32982609982][4] & 10 & $0.0045$              & \nnums[1.4087444725530436][2] & \nnums[0.7026098547783486][2] & \nnums[0.1837980064332096][2] & $R_{\pi}$   \\
19 & \rpo{16}     & \nnums[0.00630959952508422][4]   & \nnums[1.64294470470632][4]     & 10 & $0.0039$  & \nnums[1.5180627193575083][2] & \nnums[1.0270991510450740][2] & \nnums[0.3902980982119430][2] & -- \\
20 & \rpo{17}     & \nnums[0.0121412284978695][4]    & \nnums[0.318603061149668][4]    & 15 & $0.0016$  & \nnums[1.5288516474139724][2] & \nnums[1.0065782327292341][2] & \nnums[0.5810798773445096][2] & $R_{\pi}$   \\
21 & \rpo{18}     & \nnums[0.055794884392551425][4]  & \nnums[2.2912426784348208][4]   & 7  & $0.0084$  & \nnums[1.4402377003664859][2] & \nnums[0.8154976237489269][2] & \nnums[0.2247593596973368][2] & -- \\
22 & \rpo{19}     & \nnums[0.045096829918751004][4]  & \nnums[0.74886278380068871][4]  & 12 & $0.0021$  & \nnums[0.4248532373602723][2] & \nnums[1.3148219463949371][2] & -- & -- \\
23 & \rpo{20}     & \nnums[0.036123079700446183][4]  & \nnums[3.939561286622311087][4] & 9  & $0.0084$  	        & \nnums[1.4224507284218044][2] & \nnums[0.7197785593484938][2] & \nnums[0.2438013401933967][2] & -- 
\end{tabular}
\caption{\label{tab:ECSinTCF} \small Properties of ECSs found in TCF for $\Gamma=1$, $\eta=0.5$, $Re_i\approx 1200$ and \Reo[-1200]: the temporal period $T$ and shift $\Phi$, the number of unstable directions $N_{\lambda}^{u}$, the inverse of the escape rate $\gamma_n^{-1}$, the minimal distance to the turbulent trajectory in lobes 1, 2, and 3, and the discrete symmetries, if any.
}
\end{table}

We used the shortest ECS (\rpo{06}) to verify that our numerical solutions are fully resolved. Specifically, after converging this solution using the standard spatial and temporal discretization, we recomputed it using a finer discretization and used the corresponding relative residual \eqref{eq:residual} to quantify the accuracy of the computed solution. Doubling the spatial resolution in all directions, while keeping the temporal resolution the same, yielded an acceptably small value $\varepsilon=6.1\times10^{-4}$. Doubling the temporal resolution, while keeping the spatial resolution the same, yielded an even smaller value $\varepsilon=1.2\times10^{-6}$.

\section{Results}
\label{sec:results}

\subsection{Exact coherent structures and state space geometry}
\label{sec:ECS}

\begin{figure}
	\center
	\includegraphics[width=\textwidth]{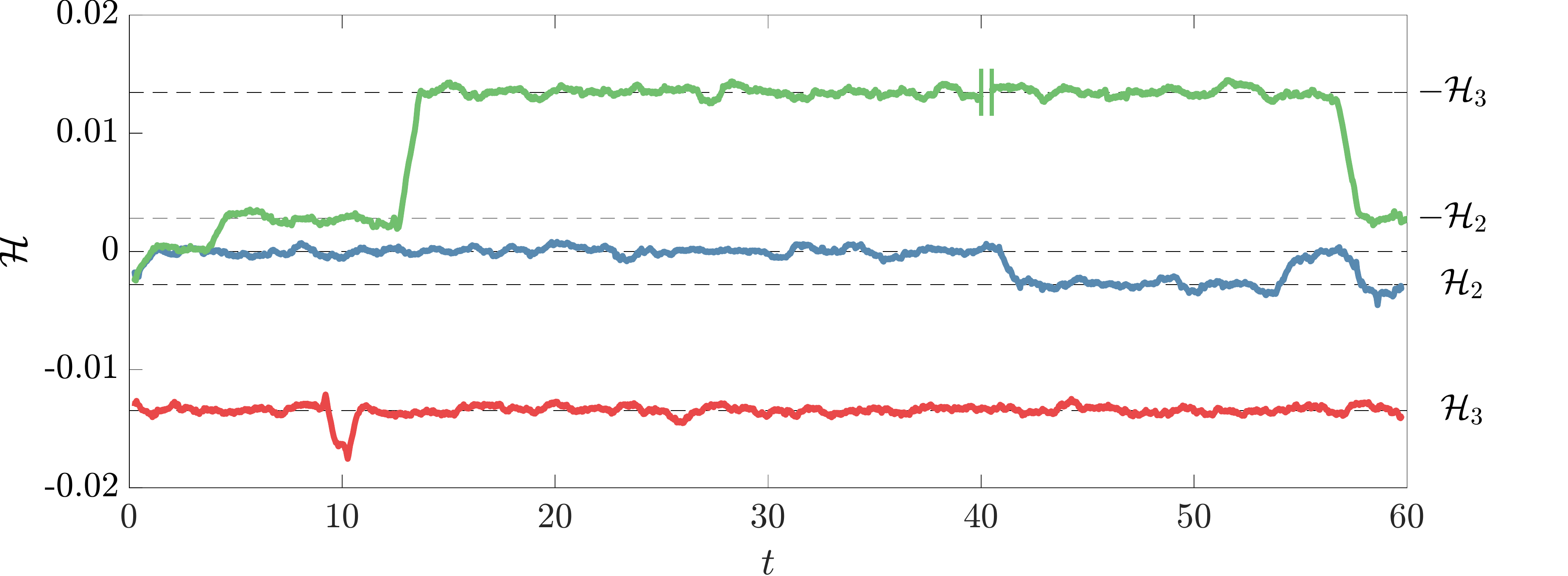} 
	\caption{\label{fig:connect} \small 
	Helicity for ${\bf u}^a(t)$ (blue), ${\bf u}^b(t)$ (red), ${\bf u}^c(t)$ (green). Shown is the running average taken over a window of $1$ time unit. Note that $10$ time units are removed from the signal of ${\bf u}^c(t)$ at $t=40$, during which the flow remains in \lobe{3}. The mean helicity of each lobe are illustrated with black dashed lines. }
\end{figure}

It is easier to understand the state space geometry for this problem and the structure of the chaotic set associated with turbulent flow using coordinates (observables) that are invariant under rotation around the axis. We found the following coordinates to provide a convenient projection: the normalized energy
\begin{equation}
    \mathcal{E} = \frac{1}{Re^2V}\int_{V} {\bf u}^2 \, dV,
\end{equation}
rate of energy dissipation
\begin{align}
    \mathcal{D} = \frac{1}{Re^2V}\int_{V} \boldsymbol\omega^2 \, dV, 
\end{align}
and helicity
\begin{equation}
    \label{eq:sym_z}
    \mathcal{H} = \frac{1}{Re^2V} \int_V {\bf u}\cdot\boldsymbol\omega \, dV.
\end{equation}
Here $Re=|Re_i-Re_o|/2$ is the characteristic scale for the velocity, $V$ is the volume of the flow domain, and $\boldsymbol\omega=\nabla\times\vec{u}$ is the vorticity. Note that $\mathcal{E}$ and $\mathcal{D}$ are invariant under both $R_\phi$ and $K_z$, while $\mathcal{H}$ is invariant under $R_\phi$ and changes sign under $K_z$. The running average of helicity for the turbulent flow initialized as described in the previous section is shown in \autoref{fig:connect} as a blue curve. It is clear that this turbulent trajectory (we will refer to it as ${\bf u}^a(t)$) explores distinct but connected regions of state space. For $1\lesssim t\lesssim 40$ and $53\lesssim t\lesssim 57$, ${\bf u}^a(t)$ is confined to the part of the chaotic set (which we will refer to as \lobe{1}) that is centered at $\mathcal{H}=0$. For $41\lesssim t\lesssim 53$, ${\bf u}^a(t)$ is confined to the part of the chaotic set (which we will refer to as \lobe{2}) that is centered at $\mathcal{H}=\mathcal{H}_2\approx-0.003$. The chaotic set has several other lobes, as will be discussed shortly.

\begin{figure}
\center
\includegraphics[width=\textwidth]{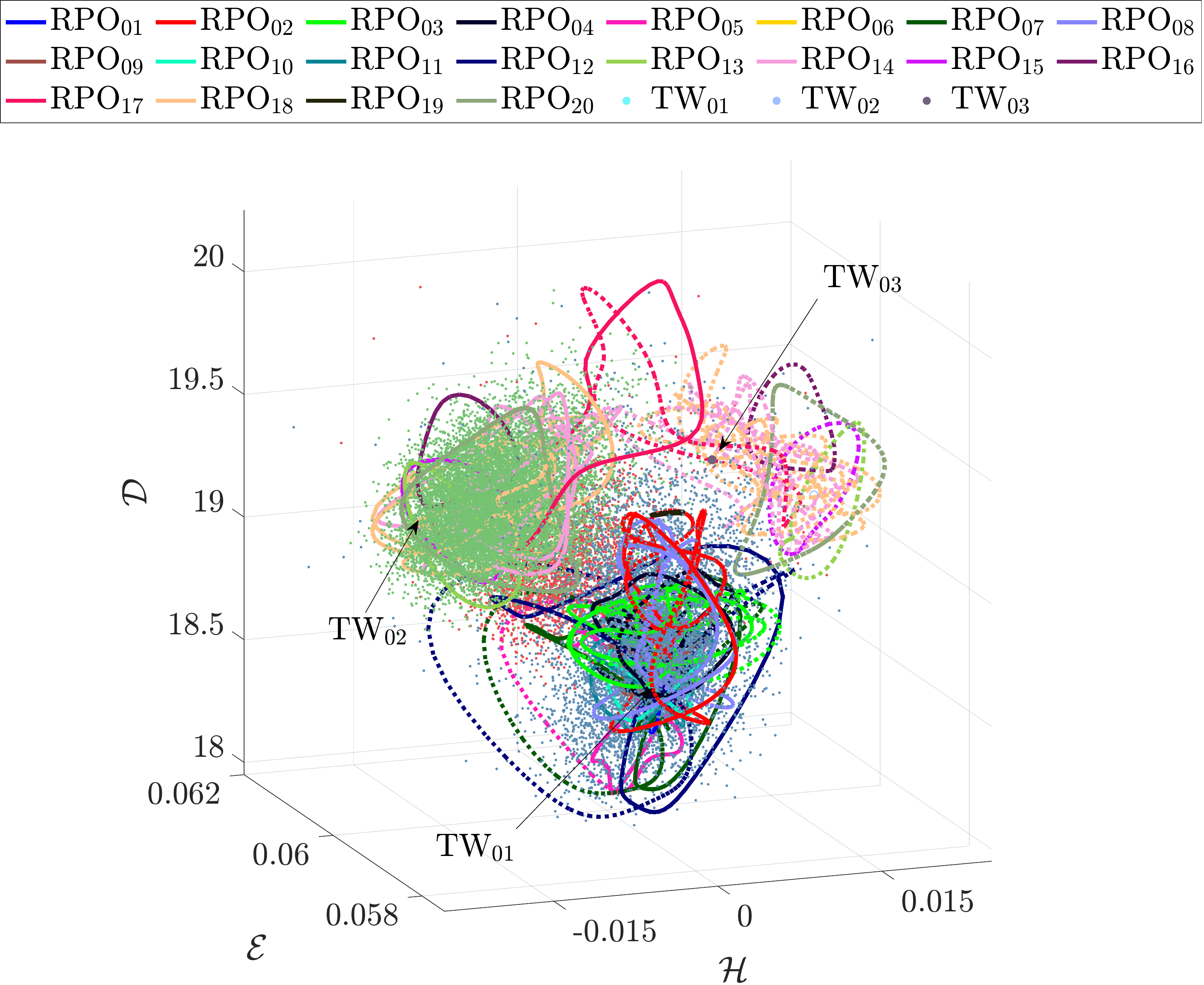} \\
\includegraphics[width=0.25\textwidth]{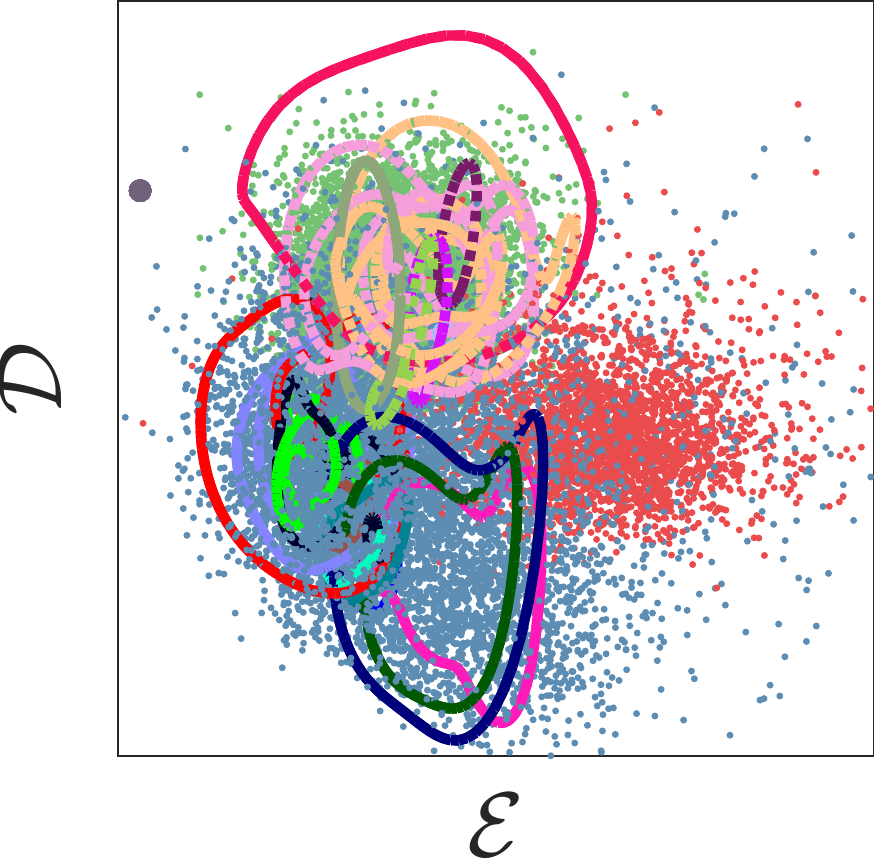} \hspace{3mm} 
\includegraphics[width=0.25\textwidth]{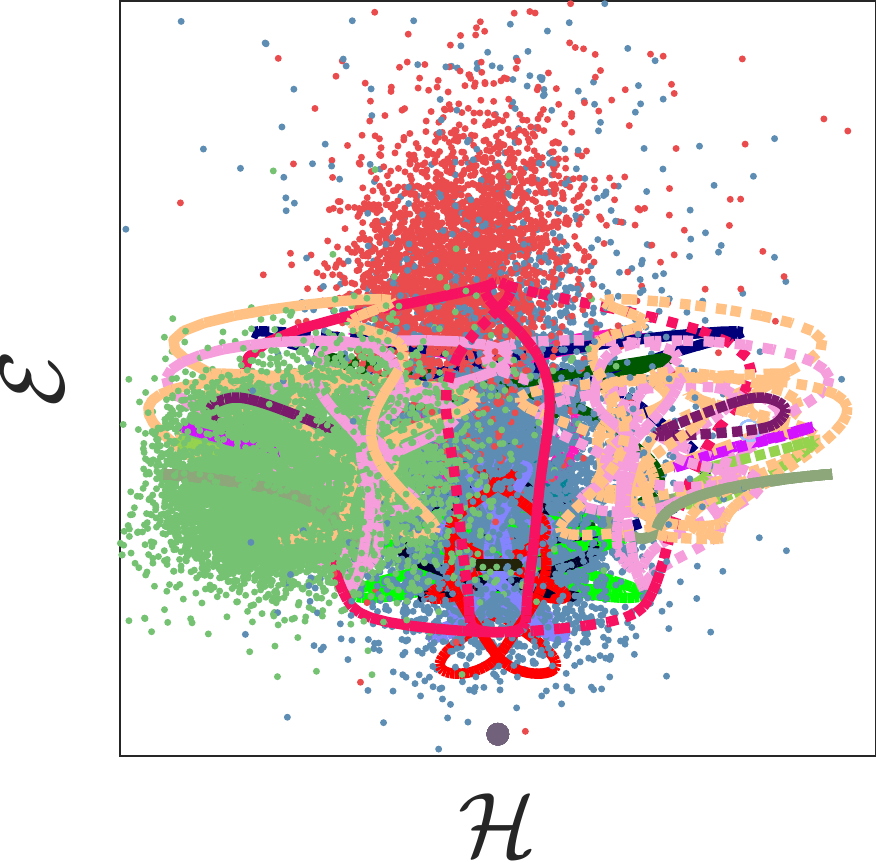} \hspace{3mm} 
\includegraphics[width=0.25\textwidth]{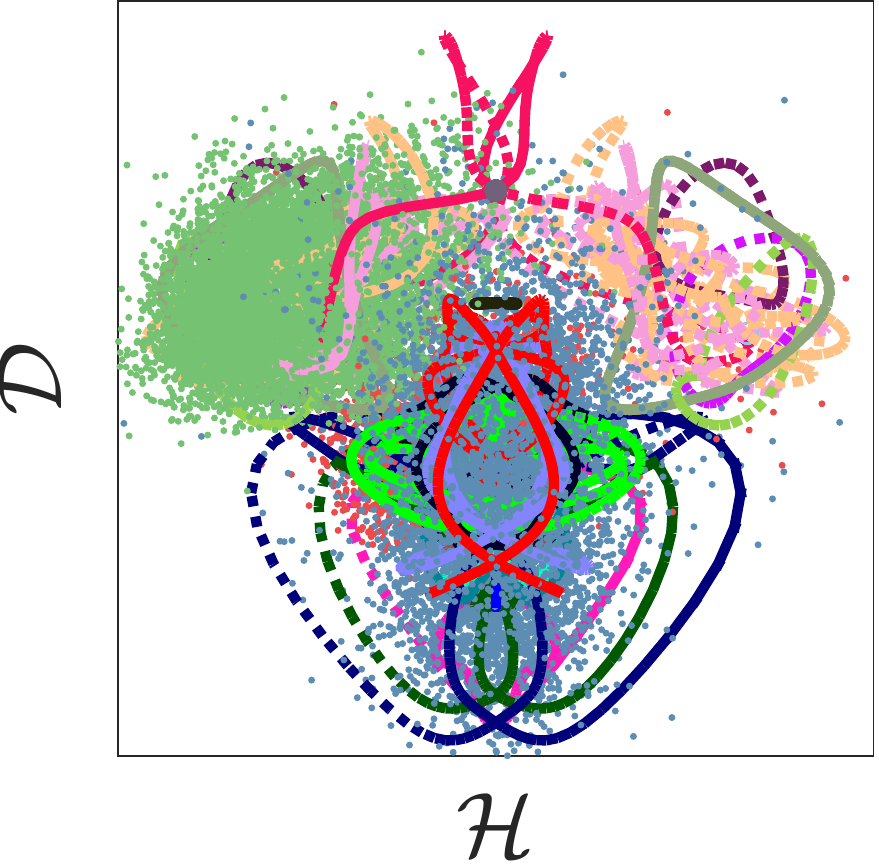}
\caption{\label{fig:proj1} \small 
Three-dimensional projection of the infinite-dimensional state space (top) and the corresponding two-dimensional projections (bottom). The coordinates used are the energy $\mathcal{E}$, the rate of energy dissipation $\mathcal{D}$, and the helicity $\mathcal{H}$. The individual lobes of the chaotic set are shown as clouds of points with different color; \lobe{1} shown in blue, \lobe{2} in red, and \lobe{3} in green. The computed \rpo{}s (\re{}s) are shown as solid lines (filled circles) and their symmetry-related copies as dotted lines (open circles) of the same color.}
\end{figure}

\begin{figure}
\center
\subfloat[]{\includegraphics[width=0.445\textwidth]{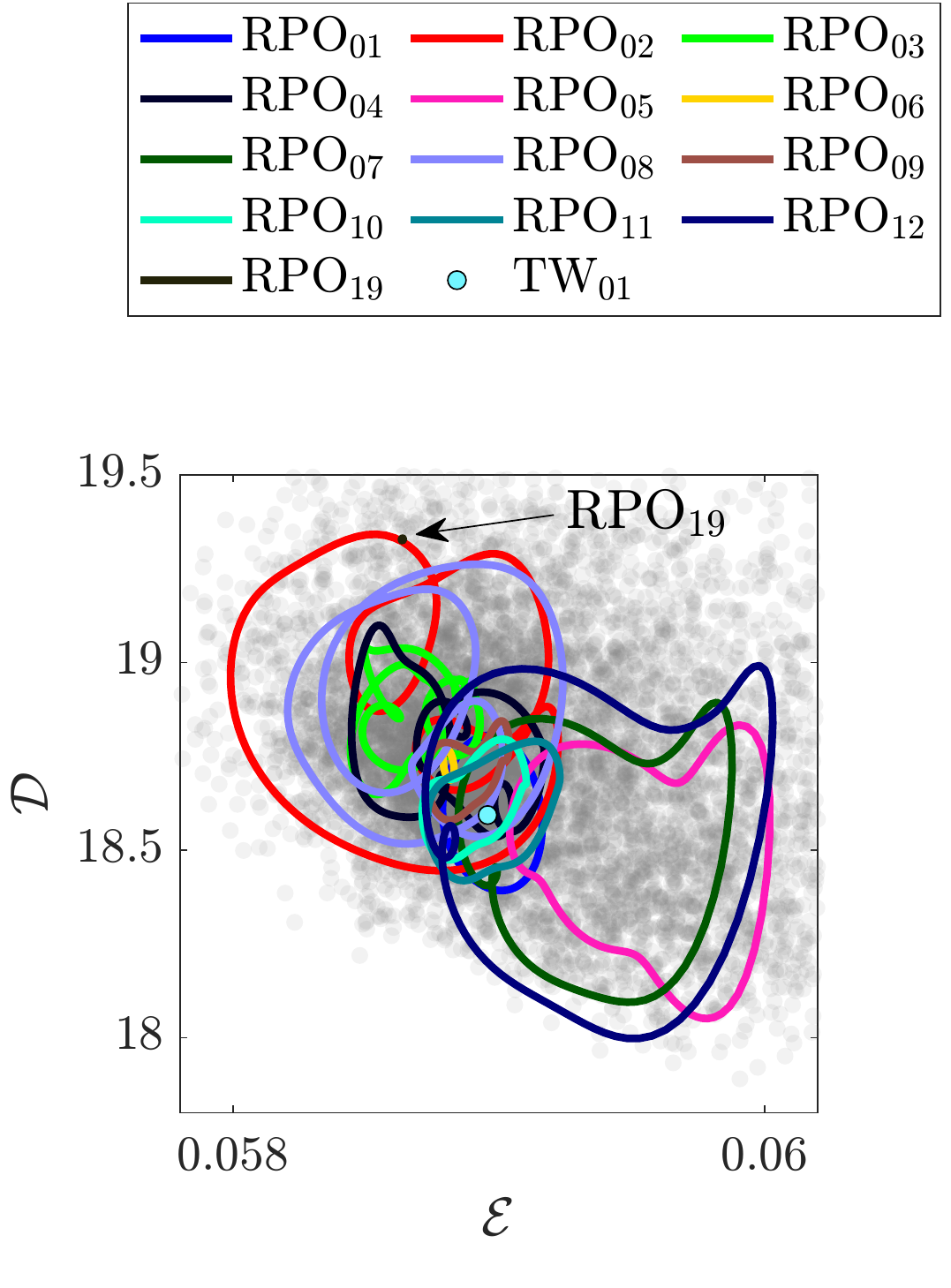}} \hspace{3mm}
\subfloat[]{\includegraphics[width=0.45\textwidth]{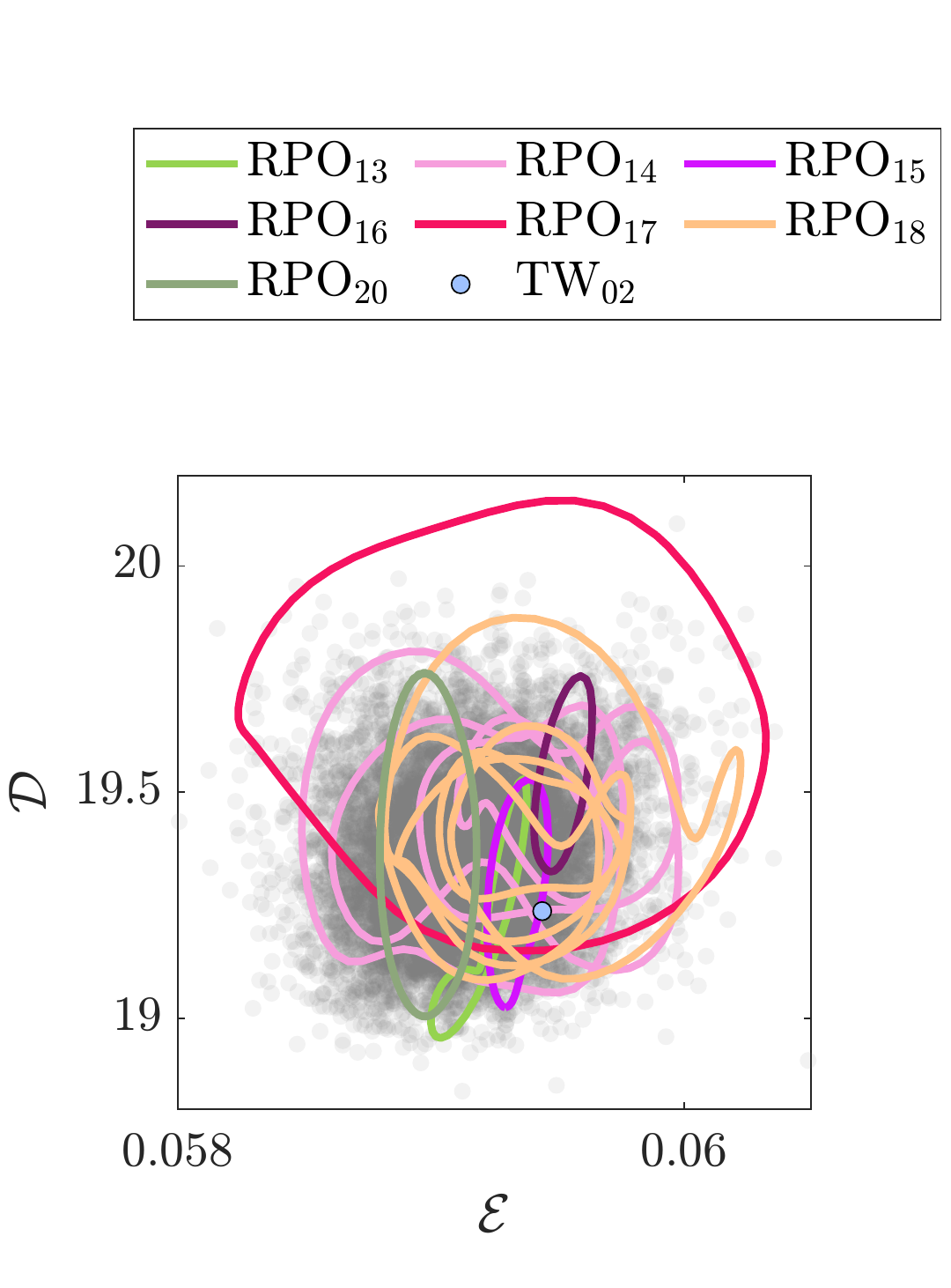}} 
\caption{\label{fig:proj2} \small 
Two-dimensional projection of (a) \lobe{1} and (b) \lobe{3}. In each lobe, the chaotic set (gray point cloud) is in the background. \rpo{}s and \re{}s collocated with each lobe in this projection are overlaid as solid curves and filled circles, respectively.}
\end{figure}

The relation between these lobes and ECSs can be seen in a low-dimensional projection of the state space onto $\mathcal{E}$, $\mathcal{D}$, and $\mathcal{H}$ shown in \autoref{fig:proj1}.
The portions of turbulent trajectory ${\bf u}^a(t)$ confined to \lobe{1} (\lobe{2}) are shown as a blue (red) cloud of points, representing different snapshots of the flow. For each ECS ${\bf u}_n$ that is not reflection-symmetric, both ${\bf u}_n$ and $K_z{\bf u}_n$ are shown in \autoref{fig:proj1}. Since all three coordinates are invariant with respect to rotations around the axis, a family of temporally periodic solutions corresponding to each \re{}, i.e., $R_\phi{\bf u}_n$ with $\phi\in[0,2\pi)$, is mapped to a single point. Similarly, a family of temporally quasi-periodic solutions corresponding to each \rpo{} is mapped to a single closed curve.
A large number of computed ECSs (\re{01}, \rpo{01}-\rpo{12}, and \rpo{19}) are collocated with \lobe{1}, as illustrated in \autoref{fig:proj2}(a), but we could not find any ECSs collocated with \lobe{2}. Instead, all initial conditions in \lobe{2} converged to ECSs that lie outside of both \lobes{1}{2}. Indeed, there is no guarantee that Newton iterations converge to a solution that is close to the near-recurrence used as the corresponding initial condition. At first glance, this appears to suggest that some of the ECSs we found are dynamically relevant and some are not.

Interestingly, most of the ECSs lying outside of \lobes{1}{2} were found to be grouped in the same region of the state space. This grouping suggests that these solutions may be a part of the skeleton underlying a different lobe of the same chaotic set or, perhaps, a different chaotic set. In fact, by using an \rpo{} lying in that region of the state space to initialize the flow, we computed a second turbulent trajectory, ${\bf u}^b(t)$, which remained confined to a region of the state space centered at $\mathcal{H}=\mathcal{H}_3\approx -0.013$ for 60 time units. The corresponding running average of helicity is shown in red in \autoref{fig:connect} and snapshots of ${\bf u}^{b}(t)$ are shown as a green point cloud in \autoref{fig:proj1}. \rpo{13}-\rpo{18} and \rpo{20} as well as \re{02} are collocated with this set, referred to as \lobe{3} below, as \autoref{fig:proj2}(b) illustrates.

To determine whether \lobe{3} is dynamically connected to \lobes{1}{2}, we continued the turbulent trajectory ${\bf u}^a(t)$ further in time. The running average of helicity for a portion of this trajectory (we will refer to it as ${\bf u}^c(t)$) is shown in green in \autoref{fig:connect}. It is clear that this trajectory visits \lobes{1}{2} as well as the symmetric (under reflection $K_z$) copies of \lobes{2}{3}. This suggests that \lobe{1} as well as \lobes{2}{3} and their symmetric copies are all dynamically connected. The turbulent trajectory ${\bf u}^c(t)$ will not be analyzed in the remainder of this paper. We will simply note that, after several hundred time units, turbulent flow eventually escapes the chaotic set composed of the five lobes mentioned previously and settles on what appears to be a stable quasiperiodic state. Hence, in our system, turbulence is a long-lived transient, similar to what was found in the same geometry at a larger aspect ratio \citep{hochstrate2010}.

\begin{figure}
\center
\subfloat[]{\includegraphics[width=0.28\columnwidth]{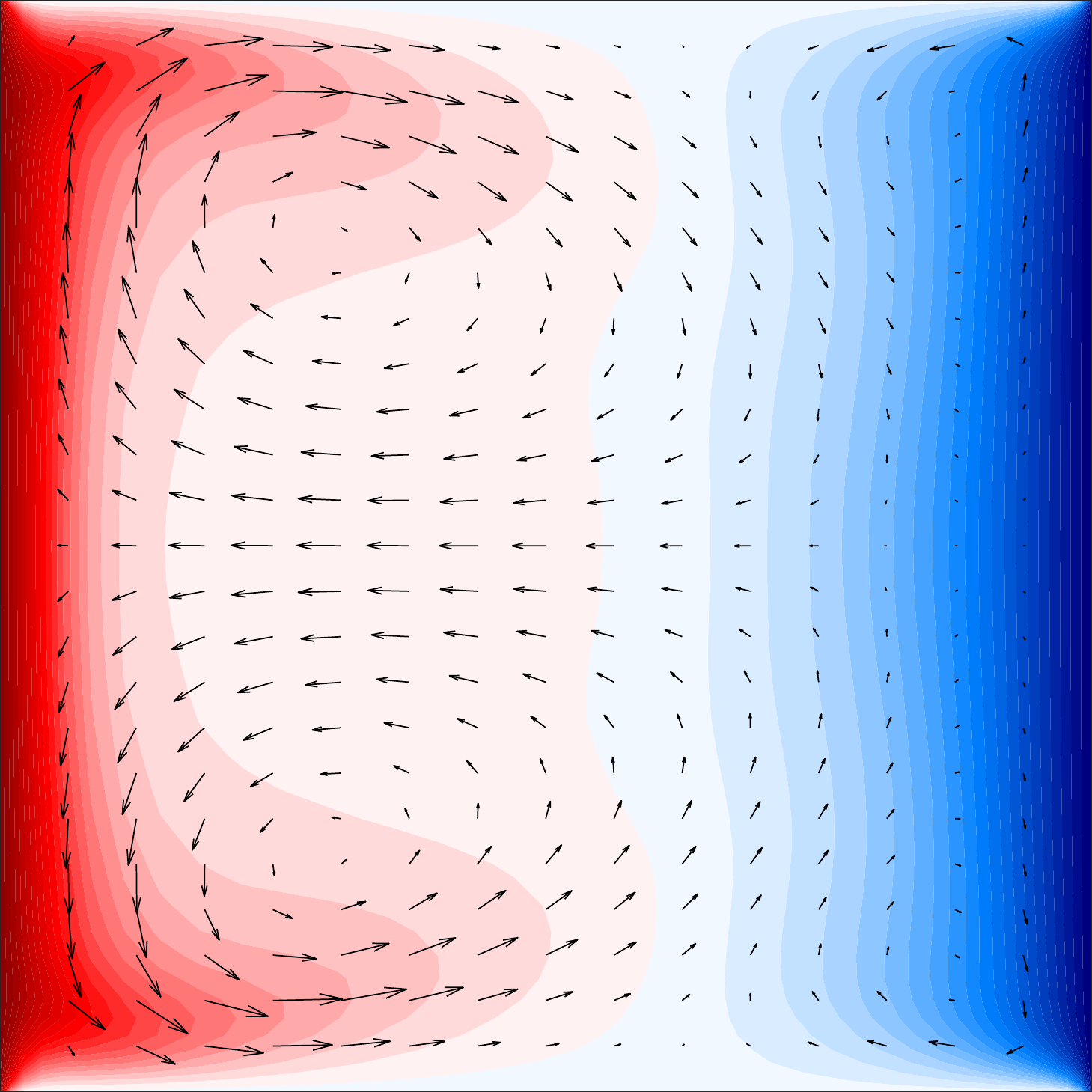}}\hspace{5mm}
\subfloat[]{\includegraphics[width=0.28\columnwidth]{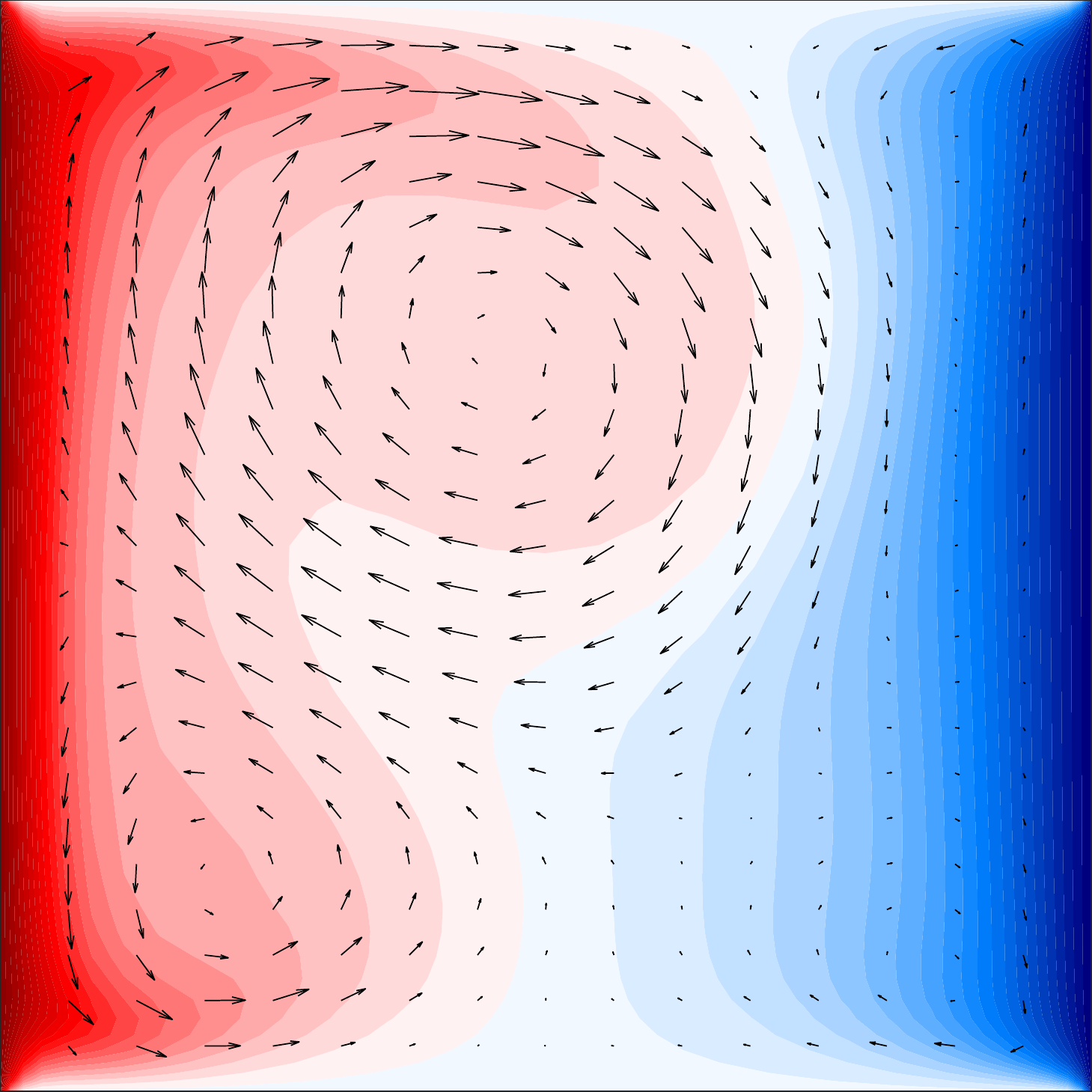}}\hspace{5mm}
\subfloat[]{\includegraphics[width=0.337\columnwidth]{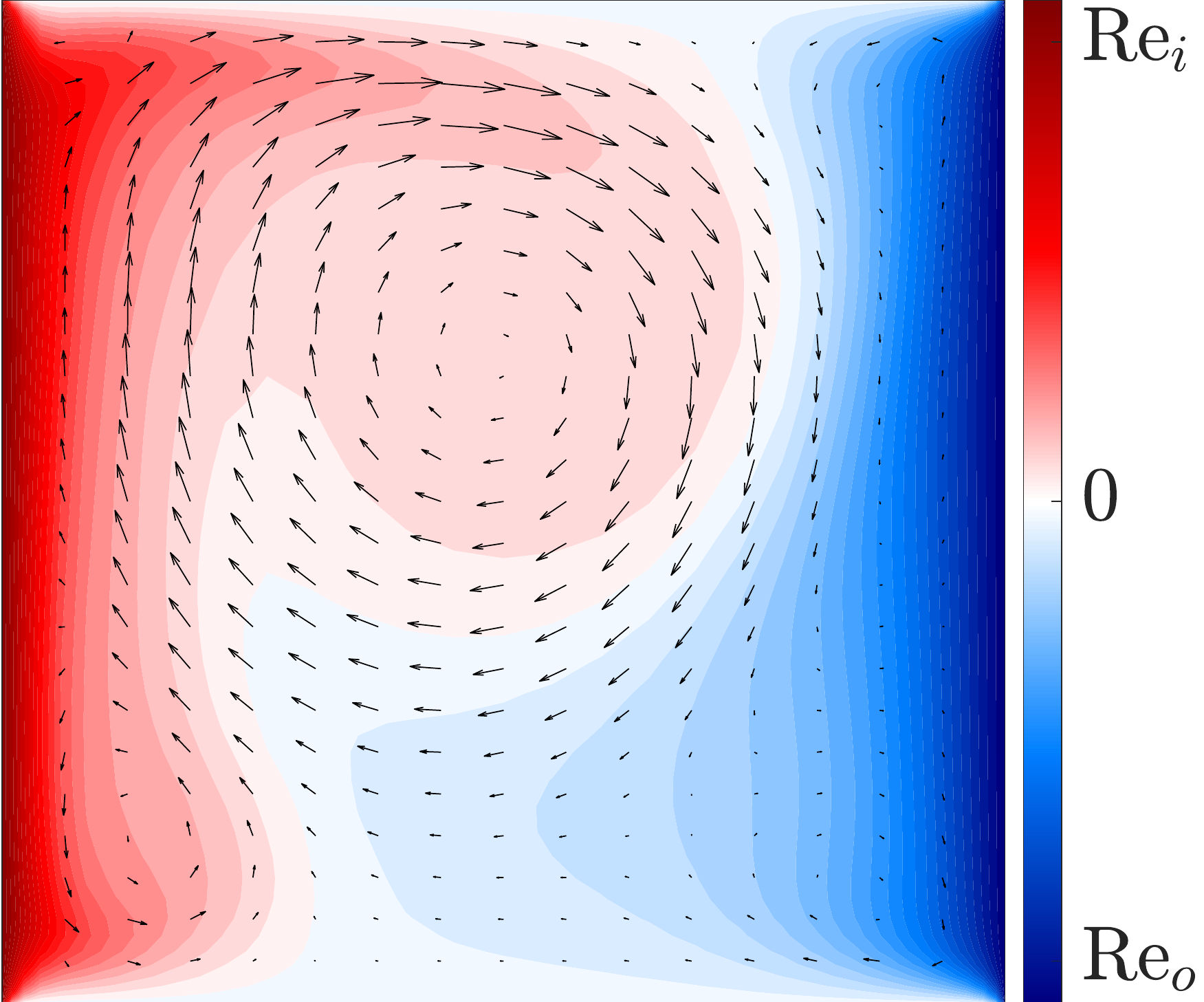}}
\caption{\label{fig:turb_lobes} \small
Mean flow field in (a) \lobe{1}, (b) \lobe{2}, and (c) \lobe{3}. In this figure and all following 2D slices, the arrows show the in-plane ($r$ and $z$) components of the flow and the color indicates the out-of-plane ($\theta$) component. 
}
\end{figure}

In the subsequent discussion, we will refer to the portion of ${\bf u}^a(t)$ restricted to \lobes{1}{2} of the chaotic set as ${\bf u}^1(t)$ and ${\bf u}^2(t)$, respectively. We will also identify ${\bf u}^3(t)={\bf u}^b(t)$, since ${\bf u}^b(t)$ remains in \lobe{3} for the entire interval over which it was computed. The characteristic flow fields in each lobe are compared in \autoref{fig:turb_lobes}, where the velocity field associated with ${\bf u}^1(t)$, ${\bf u}^2(t)$, or ${\bf u}^3(t)$ has been averaged in both time and $\theta$. Inside each lobe, we find two cellular vortical structures. In \lobe{1}, the two vortices are symmetric. In \lobes{2}{3}, one vortex is notably stronger than the other. This asymmetric flow structure illustrates the reflection symmetry breaking and explains the relative arrangement of the three lobes along the $\mathcal{H}$ axis in \autoref{fig:proj1}. 

At first glance, the spatial structure of the mean flow in different lobes appears to be similar to the structure of either single-cell (A1) or double-cell (N2) stationary cellular flows found for this (or similar) geometry in previous studies, which considered the case $Re_o=0$ \citep{cliffe1983,pfister1988,furukawa2002}. 
N2 was found to be stable for $Re_i\lesssim 133$ and is replaced by A1 (or its symmetric copy) at a higher $Re_i$.
The onset of time-dependence at $Re_i\approx 892$ does not appear to alter the structure of the flow dramatically \citep{marques2006}. Weak counter-rotation stabilizes symmetric flows (such as N2) at the expense of the asymmetric ones (such as A1), but does not notably modify the structure of the flows either \citep{schulz2003}. 

The setup considered here leads to a (nearly) symmetric flow in \lobe{1} that has one qualitative difference with N2: the sign of the azimuthal vorticity $\omega_\theta$ is opposite to that found in previous studies. In our case, the radial velocity is the largest and positive near the end-caps, as \autoref{fig:turb_lobes}(a) illustrates. For N2, the radial velocity is the largest and positive at the center plane $z=0$; it represents a strong central jet of angular momentum \citep{marques2006,altmeyer2012} which is not found in \lobe{1}. These differences are likely due to our choice of the outer cylinder rotating independently of the end-caps; in previous studies of TCF, the end-caps were assumed to rotate with the same angular velocity as the outer cylinder.

%
%

\subsection{A measure of closeness}
\label{sec:distance}

\autoref{fig:proj1} suggests that most of the ECSs we found are embedded into either \lobe{1} or \lobe{3} of the chaotic set and, therefore, are dynamically relevant. 
However, low-dimensional projections can be misleading. 
To determine that an ECS is truly dynamically relevant, we need to ensure
\begin{enumerate}[label=(\roman*),labelindent=1\parindent,leftmargin=4\parindent,labelsep=\parindent,itemindent=0em]
    \item that turbulent flow comes close to an ECS, at least occasionally, in the full state space,
    \item that turbulent flow remains in the neighborhood of the ECS over a characteristic time scale of the flow, and
    \item that turbulent flow shadows (i.e., evolves in a manner similar to) the ECS over time, while inside the neighborhood of the ECS.
\end{enumerate}

We will investigate whether these three conditions are satisfied in the remainder of the paper but, first, we need to define what ``close'' means in either the physical space or the corresponding high-dimensional state space. The term ``exact coherent structure'' reflects the visual similarity between (spatial) structures commonly observed in a turbulent flow and snapshots of (numerically) exact solutions of Navier-Stokes in the three-dimensional physical space. Visual similarity is, however, too qualitative to be of much use for the purposes of predicting dynamics, even on relatively short time scales; we need a more quantitative definition of similarity. 

Similarity is a relative term; before it is defined, we need to determine a yard stick for what is considered dissimilar. For parameters considered here, turbulent Taylor-Couette flow ${\bf u}(t)$ is characterized by a fluctuation $\tilde{\bf u}(t)={\bf u}(t)-\bar{\bf u}$ about an axially symmetric mean flow $\bar{\bf u}=\langle{\bf u}(t)\rangle_{t}$ that is small compared with this mean flow, i.e., $\|\tilde{\bf u}(t)\|\ll\|\bar{\bf u}\|$. Hence, the characteristic magnitude $\sigma=\langle\|\tilde{\bf u}(t)\|\rangle_{t}$ of the fluctuations about the mean flow --- effectively the ``radius'' of the chaotic set --- is a more appropriate scale for the dissimilarity in the physical space, or distance in the state space, between different flow states than the magnitude $\|\bar{\bf u}\|$ of the mean flow. 

\begin{figure}
\center
\subfloat[]{\includegraphics[width=0.9\columnwidth]{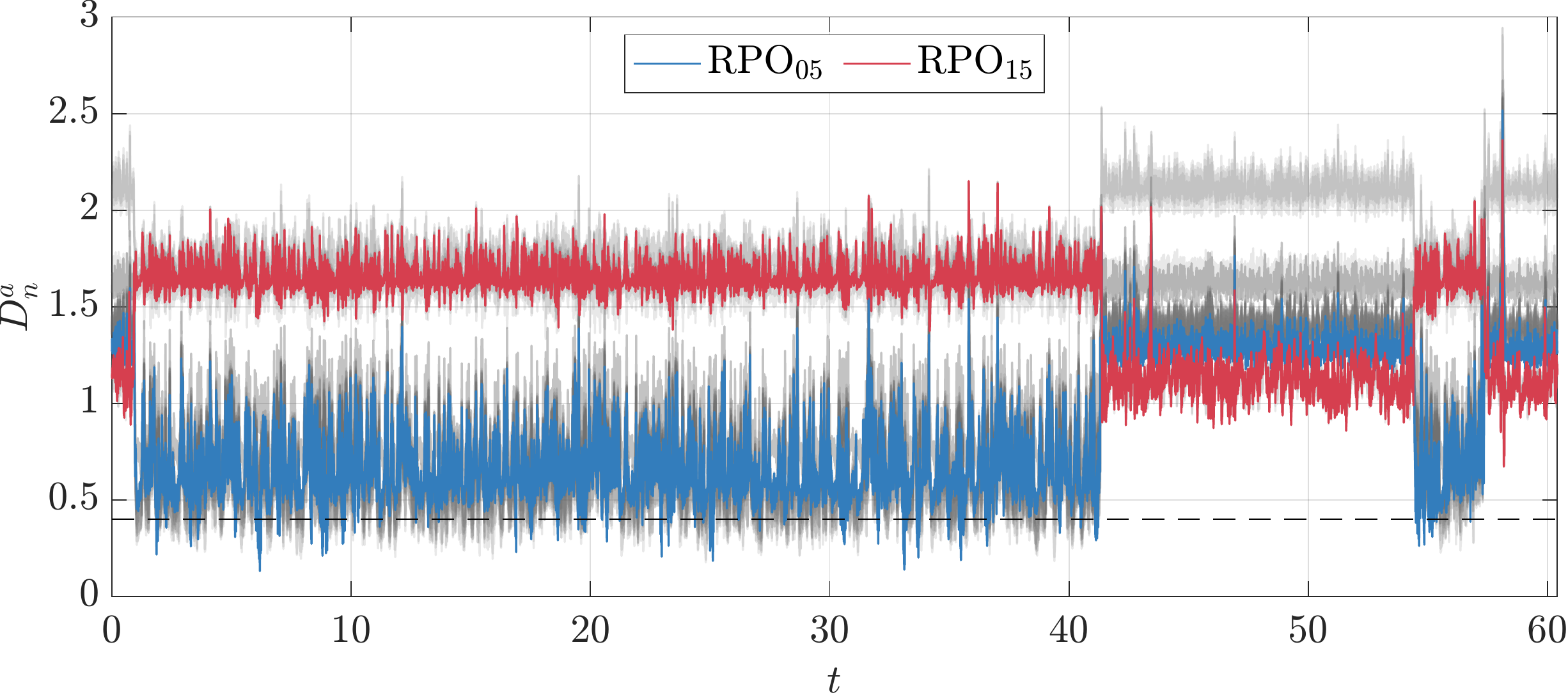}} \\
\subfloat[]{\includegraphics[width=0.9\columnwidth]{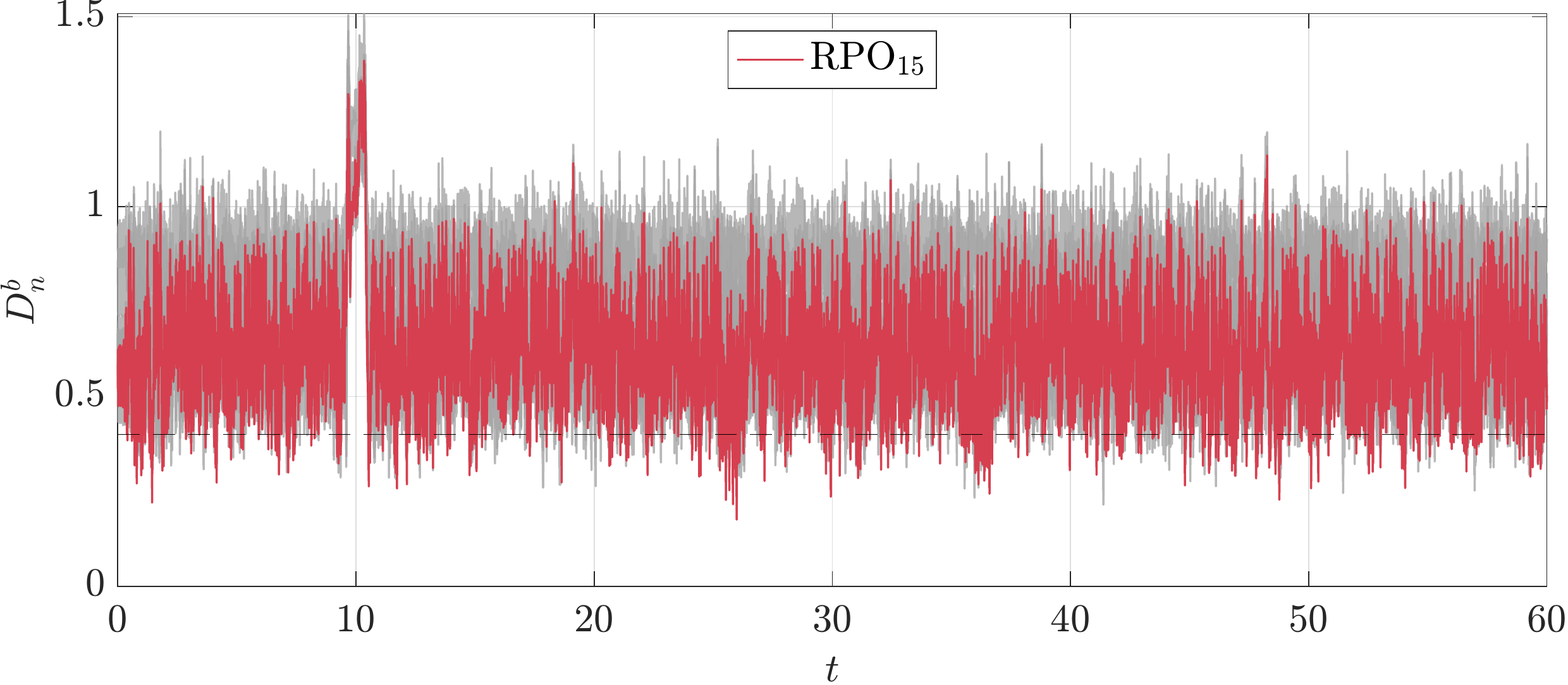}} 
\caption{\label{fig:distance} \small 
(a) The distance $D^a_n(t)$ between the turbulent trajectory ${\bf u}^a(t)$ confined to both \lobes{1}{2} and every ECS ${\bf u}_n$. All ECSs except for \rpo{05} and \rpo{15} are shown in gray. (b) The distance $D^b_n(t)$ between the turbulent trajectory ${\bf u}^b(t)$ inside \lobe{3} and ECSs collocated with this lobe. All ECSs except for \rpo{15} are shown in gray. The dashed line indicates our chosen threshold $\bar{D}$ below which two states are considered ``close."
}
\end{figure}

Since dynamics partition the chaotic set into three different lobes, there are many different ways to define $\sigma$. To avoid ambiguity, we define it using a turbulent trajectory that lies entirely inside \lobe{1}. Fortunately, the diameters of all three lobes are quite similar.
Next we define the (normalized) distance from an arbitrary point ${\bf u}$ in the state space to the family of solutions $R_\phi{\bf u}_n(\tau)$:
\begin{align}
D_n({\bf u}) = \min\limits_{\phi, \tau} \sigma^{-1}\| {\bf u} - R_\phi{\bf u}_n(\tau) \|,
\label{eq:distance}
\end{align}
where $\tau$ defines the temporal phase along the \rpo{} (this dependence is omitted for \re{}s for which time evolution is equivalent to rotation). 
The distance 
\begin{align}
D_n^j(t) = D_n({\bf u}^j(t))
\end{align}
measures how similar (close in the state space) the turbulent flow field snapshot ${\bf u}^j(t)$ is to the family of solutions $R_\phi{\bf u}_n(\tau)$. We will use $j=a,b$ when referring to the entire turbulent trajectories ${\bf u}^a(t)$ and ${\bf u}^b(t)$ and $j=1,2,3$ when referring to the portion of those trajectories confined to lobes 1, 2, or 3.
We will drop the indices when this does not cause confusion. 
The distances from both turbulent trajectories, ${\bf u}^a(t)$ and ${\bf u}^b(t)$, to various ECSs are shown in \autoref{fig:distance}, with the minimal values listed in \autoref{tab:ECSinTCF}.
The lengths of both turbulent trajectories correspond to $O(10^4)$ periods of the shortest \rpo{} we found, so they are likely long enough to sample a significant fraction of all three lobes. 

Let us first consider the turbulent trajectory ${\bf u}^a(t)$ confined to \lobes{1}{2}, which was used to generate the initial conditions for the Newton search. 
The distances to all the ECSs (as well as their symmetry-related copies) are shown in gray in \autoref{fig:distance}(a), except for two solutions: \rpo{05} (in blue) and \rpo{15} (in red). 
All of the distances jump sharply at $t\approx 1$, 42, 54, and 57, as can be clearly seen by focusing on the distance to \rpo{05} and \rpo{15}, characterized, respectively, by a low (high) average value of $|\mathcal{H}|$. 
These jumps correspond to the turbulent trajectory moving between \lobes{1}{2}. 
When the distance to \rpo{05} is low (high), the turbulent flow is inside \lobe{1} (\lobe{2}). Recall that \rpo{15} lies inside \lobe{3}, which is closer to \lobe{2}, so the distance to this \rpo{} shows the opposite trend.

Such behavior is an example of intermittency, which is a characteristic feature of all turbulent flows, but is particularly apparent in transitional flows \citep{chate1987} where it commonly manifests itself as spatiotemporal alternation between laminar and turbulent behavior \citep{eckhardt2007_pipe}. In the present case, intermittency represents a competition between different flow patterns shown in \autoref{fig:turb_lobes}(a) and (b) -- a phenomenon which has been observed in Taylor-Couette flow at large $\Gamma$ \citep{tsameret1994} as well as in many other non-equilibrium systems. 

\autoref{fig:distance}(a) shows that the turbulent trajectory ${\bf u}^a(t)$ spends most of the time in \lobe{1} (e.g., for $1\lesssim t\lesssim 42$). 
This is the likely reason most of the ECSs we found using Newton search also lie in that lobe. 
Indeed, each of the fourteen ECSs (\re{01}, \rpo{01}-\rpo{12}, and \rpo{19}) that is collocated with \lobe{1} in the projection shown in \autoref{fig:proj1} was also found to be collocated with this lobe in the full state space, with $D_n^1(t)\approx 0.6$ on average (and a minimum of around $0.2$) during the intervals when the turbulent flow is inside \lobe{1}.

The situation is similar for the turbulent trajectory ${\bf u}^b(t)$ confined to \lobe{3}. For the eight ECSs collocated with this lobe in the low-dimensional projection (\re{02}, \rpo{13}-\rpo{18}, and \rpo{20}), the distance is  $D_n^3(t)\approx 0.7$ on average and gets as low as $0.3$ or so, according to \autoref{fig:distance}(b). 
The spike around $t=10$ appears to represent an ``extreme'' event, where the turbulent trajectory explores the periphery of \lobe{3}.
The discussion in the remainder of the paper will focus mainly on \lobe{1}, which contains most of the ECSs found. 

\begin{figure}
\center
\subfloat[]{\includegraphics[width=0.45\textwidth]{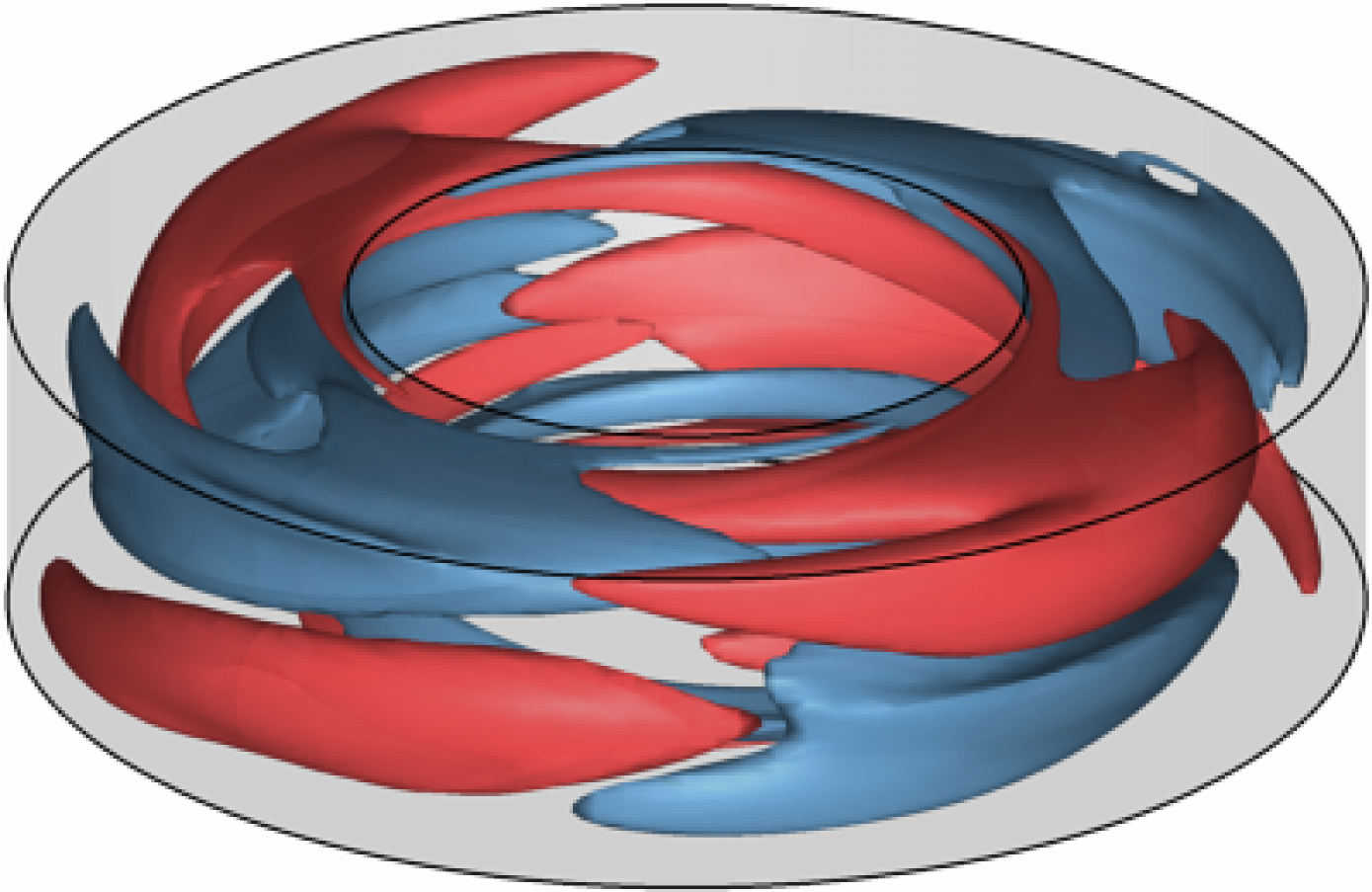}}\hspace{3mm}
\subfloat[]{\includegraphics[width=0.45\textwidth]{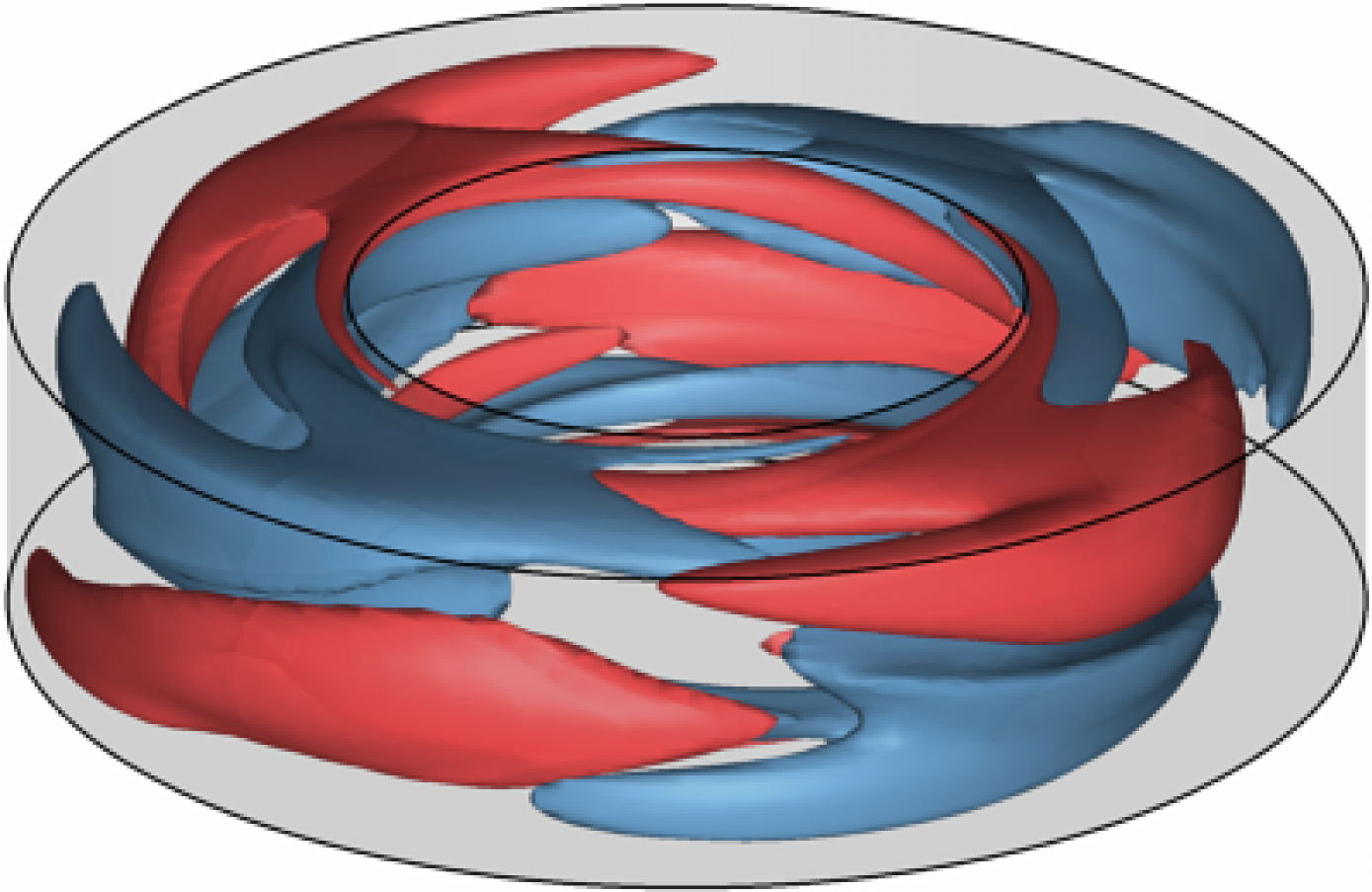}}  \\
\subfloat[]{\includegraphics[width=0.45\textwidth]{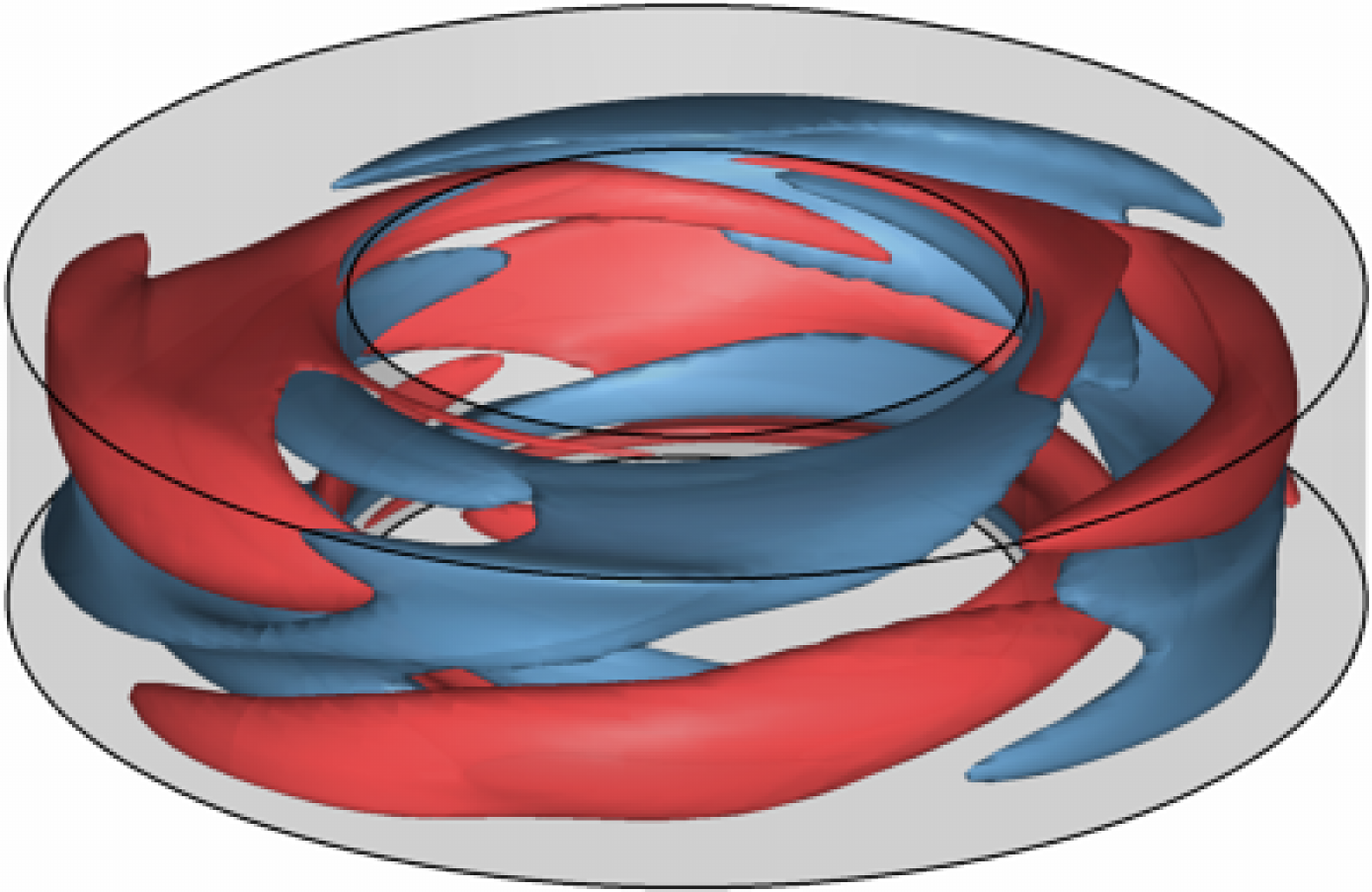}} \hspace{3mm}
\subfloat[]{\includegraphics[width=0.45\textwidth]{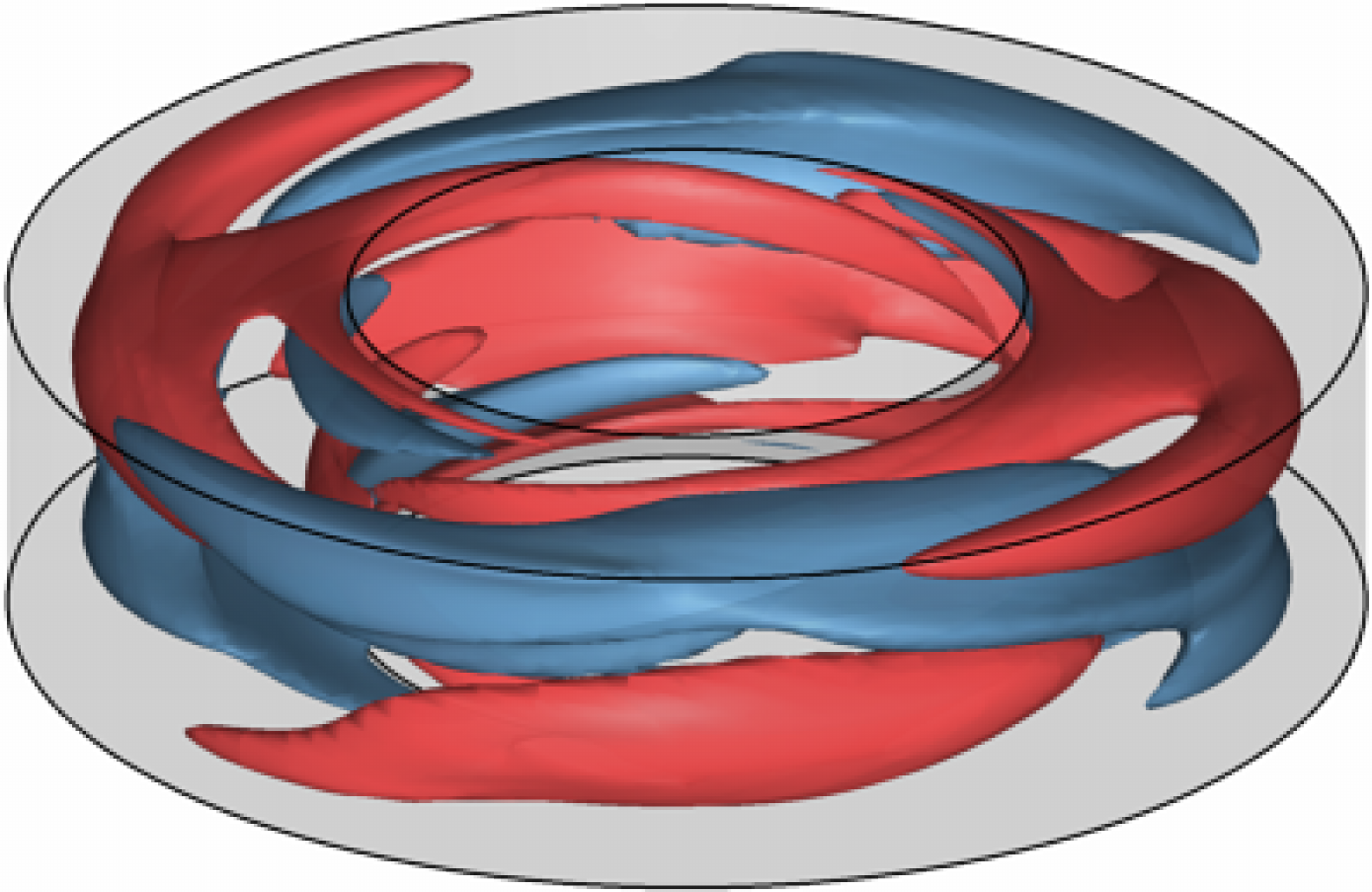}} 
\caption{\label{fig:distance_physicalSpace} \small 
Comparison of (a) the turbulent flow field ${\bf u}^a(t)$ and (b) \rpo{04} at an instant where $D^a_{7}(t) = 0.34$ is below threshold. Comparison of (c) the turbulent flow field ${\bf u}^a(t)$ and (d) \rpo{12} at an instant where $D^a_{15}(t) = 0.78$ is above threshold. 
To represent the flow structure in the entire flow domain, here and below, we show two level sets of $u_\theta$, one with a positive value (in red) and one with a negative value (in blue). In all plots, the mean flow has been subtracted off.
}
\end{figure}

To get a sense of the relationship between the magnitude of the distance in the state space and the similarity of the flow fields in the physical space, we compare snapshots of the turbulent flow field ${\bf u}^a(t)$ with a nearby ECS for various values of $D(t)$ in \autoref{fig:distance_physicalSpace}. The flow fields appear visually similar for $D(t)\lesssim \bar{D}$ with $\bar{D}=0.4$, while for $D(t)\gtrsim \bar{D}$ the differences become apparent. Based on this comparison, we define two flow states to be similar (dissimilar), or close (far) in the state space, for $D(t)<\bar{D}$ ($D(t)>\bar{D}$). The corresponding threshold $\bar{D}$ is shown as the dashed horizontal lines in \autoref{fig:distance}.
While using such an ``ocular norm'' to define closeness of different flow states may appear somewhat arbitrary, it is quite common in analyzing experimental data \citep{delozar2012}. 

\begin{figure}
\center
\subfloat[]{\includegraphics[height=0.4\columnwidth]{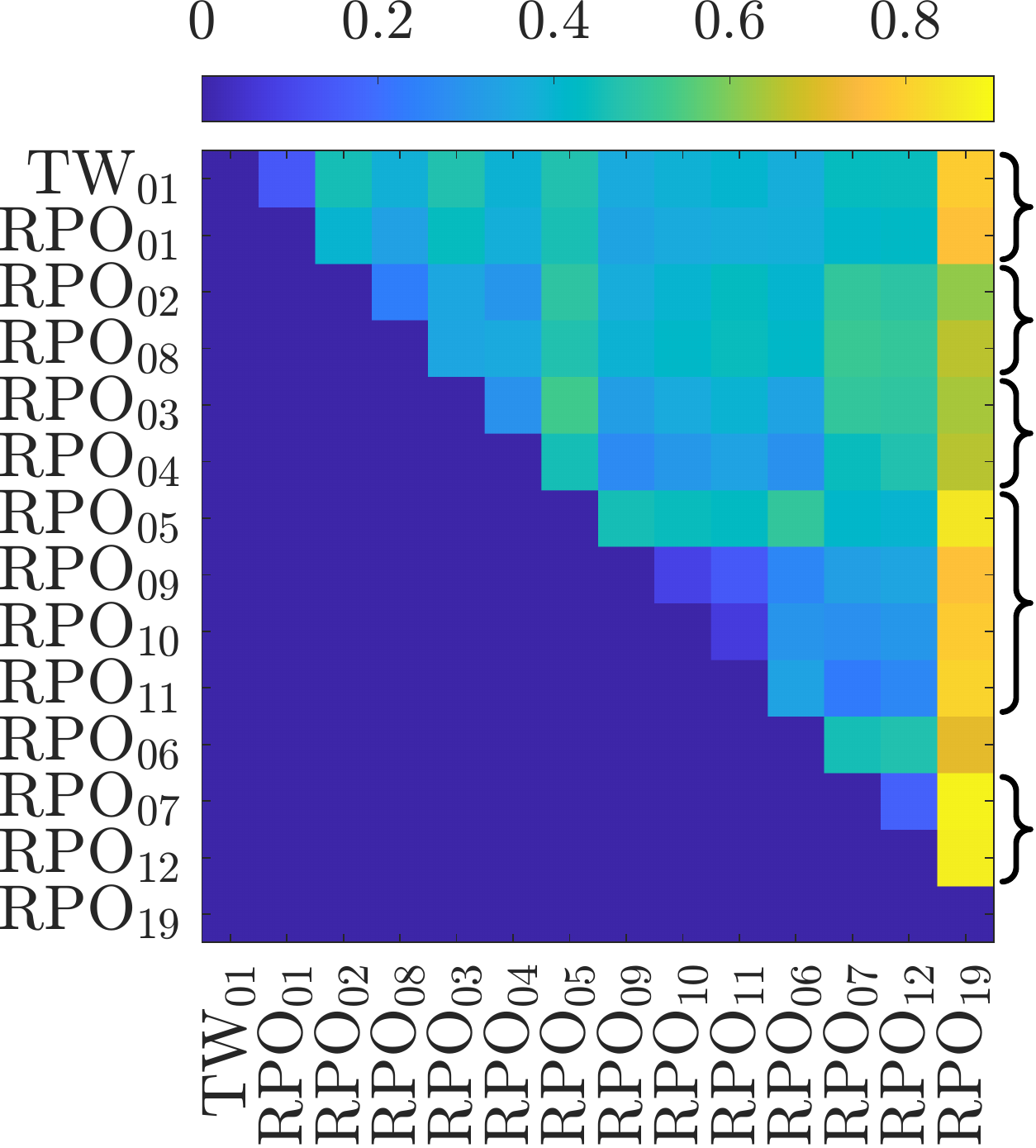}} \hspace{5mm}
\subfloat[]{\includegraphics[height=0.38\columnwidth]{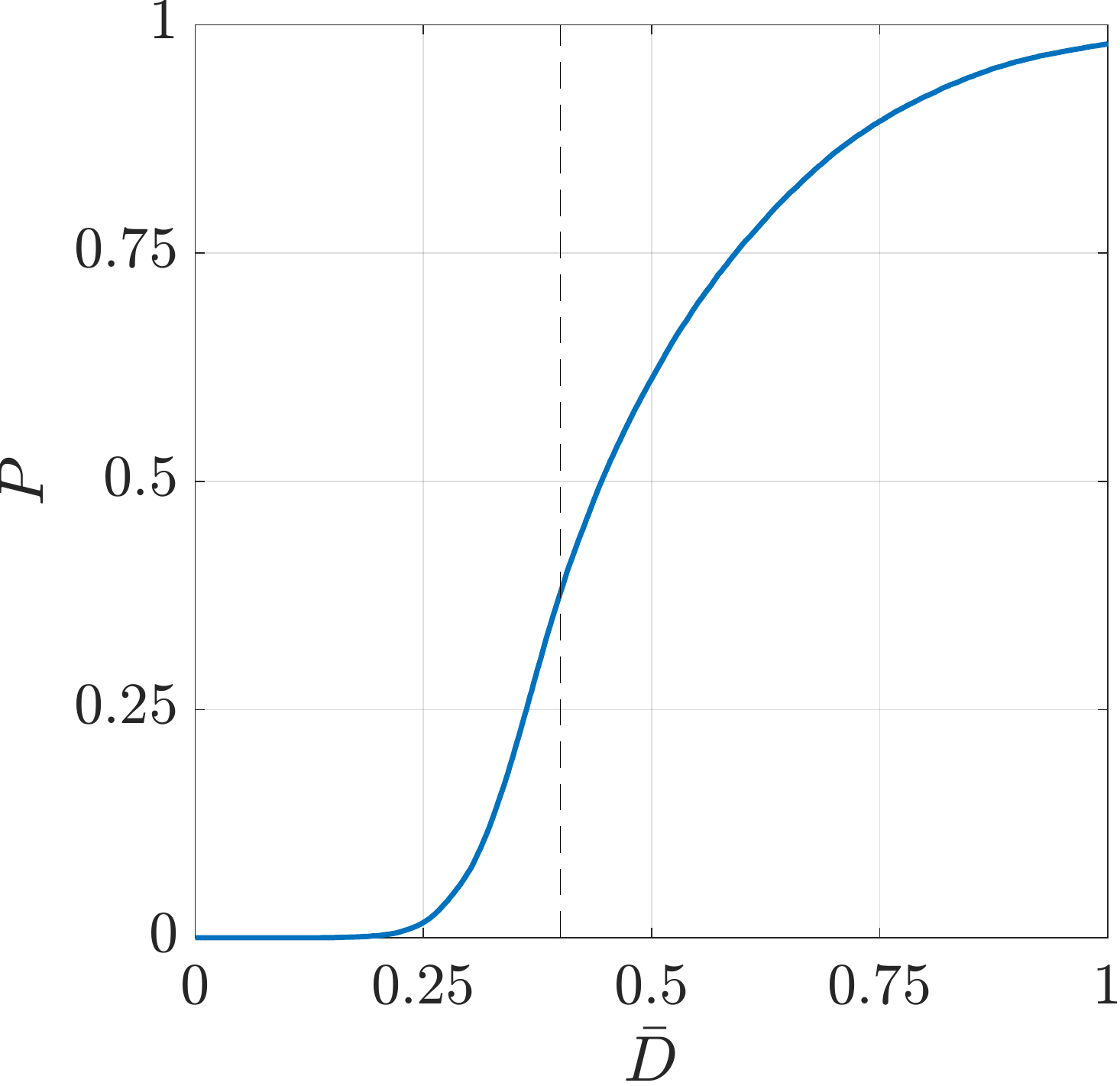}}
\caption{\label{fig:xcorr} \small 
(a) The minimal distance, $D^\text{min}_{nk}$, between \rpo{}s collocated with \lobe{1}. Braces on the right of the panel group together \rpo{}s that are connected by continuation in $Re_i$. (b) The probability that a snapshot of the turbulent trajectory will lie in a neighborhood of the ECS shown in (a), as a function of tubular neighborhood radius, $\bar{D}$. The dashed line indicates our chosen threshold below which two states are considered ``close." 
}
\end{figure}

Instead of relying on a somewhat subjective metric based on qualitative similarity of different flow fields, the distance threshold can be chosen based on a partition of the chaotic set induced by the set of computed ECSs. Let us partition \lobe{1} into tubular neighborhoods of different ECSs with sizes determined by $\bar{D}$. To get a sense of how large a fraction of \lobe{1} is covered by the union of these neighborhoods, we computed the probability $P$ that turbulent flow ${\bf u}^a(t)$, restricted to \lobe{1}, is in the neighborhood of any of the fourteen ECSs collocated with this lobe. As \autoref{fig:xcorr}(b) illustrates, these neighborhoods cover nearly 100\% of \lobe{1} for $\bar{D}\gtrsim 1$, about 40\% for $\bar{D}=0.4$, and less than 1\% for $\bar{D}\lesssim 0.2$.

Alternatively, $\bar{D}$ can be chosen based on the separation between different ECSs. At small $\bar{D}$, their neighborhoods will not overlap. As $\bar{D}$ is increased, different neighborhoods start to overlap more and more. As \autoref{fig:xcorr}(a) shows, the minimal distance
\begin{equation}
D^\text{min}_{nk} = \min\limits_{\phi,t,\tau} \sigma^{-1}\| {\bf u}_k(t) - R_\phi{\bf u}_n(\tau) \|,
\label{eq:xdistance}
\end{equation}
between pairs of ECSs is typically around 0.4 or larger, so our choice of $\bar{D}$ should be sufficient to distinguish different ECSs. The minimal distance between some pairs of ECSs (e.g., \rpo{09}-\rpo{11}) however is as low as 0.2, which means that individual snapshots of these solutions may not be visually distinct.

The results presented in this section show that a snapshot of turbulent flow ${\bf u}^j(t)$ is close to a family of ECSs $R_\phi{\bf u}_n$ provided $D_n^j(t)<\bar{D}$, where $\bar{D}=0.4$. As \autoref{fig:distance}(a) illustrates, the turbulent trajectory ${\bf u}^a(t)$ comes close to multiple \rpo{}s and one \re{} collocated with \lobe{1}; this happens quite frequently. Similarly, \autoref{fig:distance}(b) illustrates that turbulent trajectory ${\bf u}^b(t)$ comes close to multiple \rpo{}s and one \re{} collocated with \lobe{3}.
In particular we find that, condition (i) for dynamical relevance is satisfied for all ECSs collocated with \lobe{1}, with the possible exception of \rpo{19} (cf. \autoref{tab:ECSinTCF}). 
\autoref{fig:xcorr}(a) shows that \rpo{19} lies far from all other solutions collocated with \lobe{1}. This suggests that \re{01} and \rpo{01}-\rpo{12} are actually embedded in the primary lobe and may play a dynamically important role, while \rpo{19} does not.

%
%

\subsection{Shadowing of solutions}
\label{sec:shadowing}

Next we turn to checking conditions (ii) and (iii) of dynamical relevance using time intervals where $D^j_n(t)<\bar{D}$ for a given $n$. While condition (ii) is easy to verify by direct inspection, verifying shadowing condition (iii) requires more care. Formally, shadowing is defined for chaotic trajectories that come infinitesimally close to an unstable solution \citep{gaspard2005}. In our case, none of the computed ECS families are approached particularly closely by turbulent flow on time scales accessible to numerical simulations or experiments. One, therefore, has to define shadowing in practically meaningful terms. 

\citet{yalniz2020} proposed a topological approach to define shadowing of unstable periodic orbits in a turbulent flow confined to a subspace with only discrete symmetries. The approach is based on computing the number of connected components and holes for temporally discretized trajectories over one temporal period of the solution and comparing the corresponding persistence diagrams using the bottleneck distance. We define shadowing events differently for several reasons. First of all, the period of an ECS is not a proper time scale: in practice, ECSs with longer periods will never be shadowed fully. Instead, an ECS ${\bf u}_n$ will typically be shadowed for an interval of time comparable to the inverse of the escape rate 
\begin{align}
    \gamma_n = \sum_k\lambda_{n,k},
\end{align}
where the sum goes over the unstable directions associated with the ECS and $\lambda_{n,k}$ are the corresponding Floquet exponents. We only require an ECS to be shadowed for an interval equal to its expected escape time $\gamma_n^{-1}$. The mean escape times are $0.024$ for \lobe{1} and $0.006$ for \lobe{3}.

Furthermore, the approach used here is both less complicated and allows one to define shadowing for both \rpo{}s and \re{}s. It is essentially a generalization of the approach proposed by \citet{suri2020} for unstable periodic orbits and relies on the skew product decomposition of the dynamics \citep{FiSaScWu96,SaScWu99} in the vicinity of relative solutions induced by continuous symmetries. For each \rpo{}, the computed ECS ${\bf u}_n(\tau)$ represents a family of solutions that lie on a two-torus in the state space; all points of this two-torus can be parameterised as $R_\phi{\bf u}_n(\tau)$, where $0<\phi<2\pi/N$ and $0<\tau<T$. Let $R_\phi{\bf u}_n(\tau)$ be a point in the family of solutions that is closest to a point ${\bf u}^a(t)$ on the turbulent trajectory, such that
\begin{equation}
\{\tau(t), \phi(t)\} = \arg\min\limits_{\tau,\phi} \sigma^{-1}\|{\bf u}^a(t) - R_\phi{\bf u}_n(\tau)\|,
\label{eq:tprime}
\end{equation}
where $\phi$ is constrained to the interval $[0,2\pi/N)$ for solutions with an $N$-fold discrete rotational symmetry (e.g., $N=2$ for \rpo{01} and \rpo{15}).
In this decomposition, coordinates $\tau$ and $\phi$ describe the evolution along the group manifold (time translations and rotations), while $D_n^a$ describes the evolution transverse to the group manifold. Below, we will drop the subscript and superscript of $D$ since the context makes it clear which turbulent solution and ECS is discussed.

For $D=0$, a trajectory ${\bf u}(t)$ will have $d\tau/dt=1$ and $d\phi/dt=0$ for $0<\tau<T$. Once per period, both $\tau(t)$ and $\phi(t)$ will experience a jump by, respectively, $T$ and $\Phi$ (see \autoref{tab:ECSinTCF}). 
The discontinuities in $\tau(t)$ and $\phi(t)$ can be removed using coordinates $\tilde\tau$ and $\tilde\phi$ such that
\begin{align}
    \tau &= \tilde{\tau}\ \mathrm{mod}\ T,\nonumber \\
    \tilde\phi &={\phi}\ \mathrm{mod}\ \Phi.
\end{align}
Then, for $D=0$,  $d\tilde\tau/dt=1$ and $d\tilde\phi/dt=0$ for all $t$. For any trajectory that has a small but nonzero $D$, the dynamics of $\tau$ and $\phi$ should be qualitatively similar: $d\tilde\tau/dt\approx 1$ and $d\tilde\phi/dt\approx 0$. The evolution of $D$, $\tau$, and $\phi$ can therefore be used to define a set of natural criteria for shadowing of an \rpo{}.

Specifically, we will consider a turbulent trajectory to shadow an \rpo{} if, for a temporal interval $I(t)=[t-\gamma_n^{-1}/2,t+\gamma_n^{-1}/2]$, all three coordinates evolve similarly to the way they would for the \rpo{}. In practice, we define similarity in terms of normalized deviations
\begin{align}
    E_\tau &= \min_{t_0} \frac{1}{T\gamma_n^{-1}} \int_{I(t)} |t'+t_0-\tilde \tau(t')|^2 dt' \nonumber \\
    E_\phi &= \min_{\phi_0} \frac{N}{\pi\gamma_n^{-1}} \int_{I(t)} |\phi_0-\tilde \phi(t') |^2 dt'
\end{align}
of $\tilde{\tau}$ and $\tilde{\phi}$ from straight lines with slope one and zero, respectively. 
The turbulent trajectory is considered to be shadowing an \rpo{}, if the following three conditions are satisfied: 
\begin{enumerate}[label=(\alph*),labelindent=1\parindent,leftmargin=4\parindent,labelsep=\parindent,itemindent=0em]
    \item \label{itm:ShadowingA} $D(t')<\bar{D}$ for $t'\in I(t)$,
    \item \label{itm:ShadowingB} $E_\tau<\bar{E}_\tau$,
    \item \label{itm:ShadowingC} $E_\phi<\bar{E}_\phi$
\end{enumerate}
for an appropriate choice of the thresholds $\bar{E}_\tau$ and $\bar{E}_\phi$.
For \re{}s, the two coordinates describing the evolution along the group manifold are not independent, $d\tau/d\phi=T/\Phi$. In this case, shadowing is determined by two of the three conditions: (a) and (b).

\begin{figure}
\center
\subfloat[]{\includegraphics[width=\textwidth]{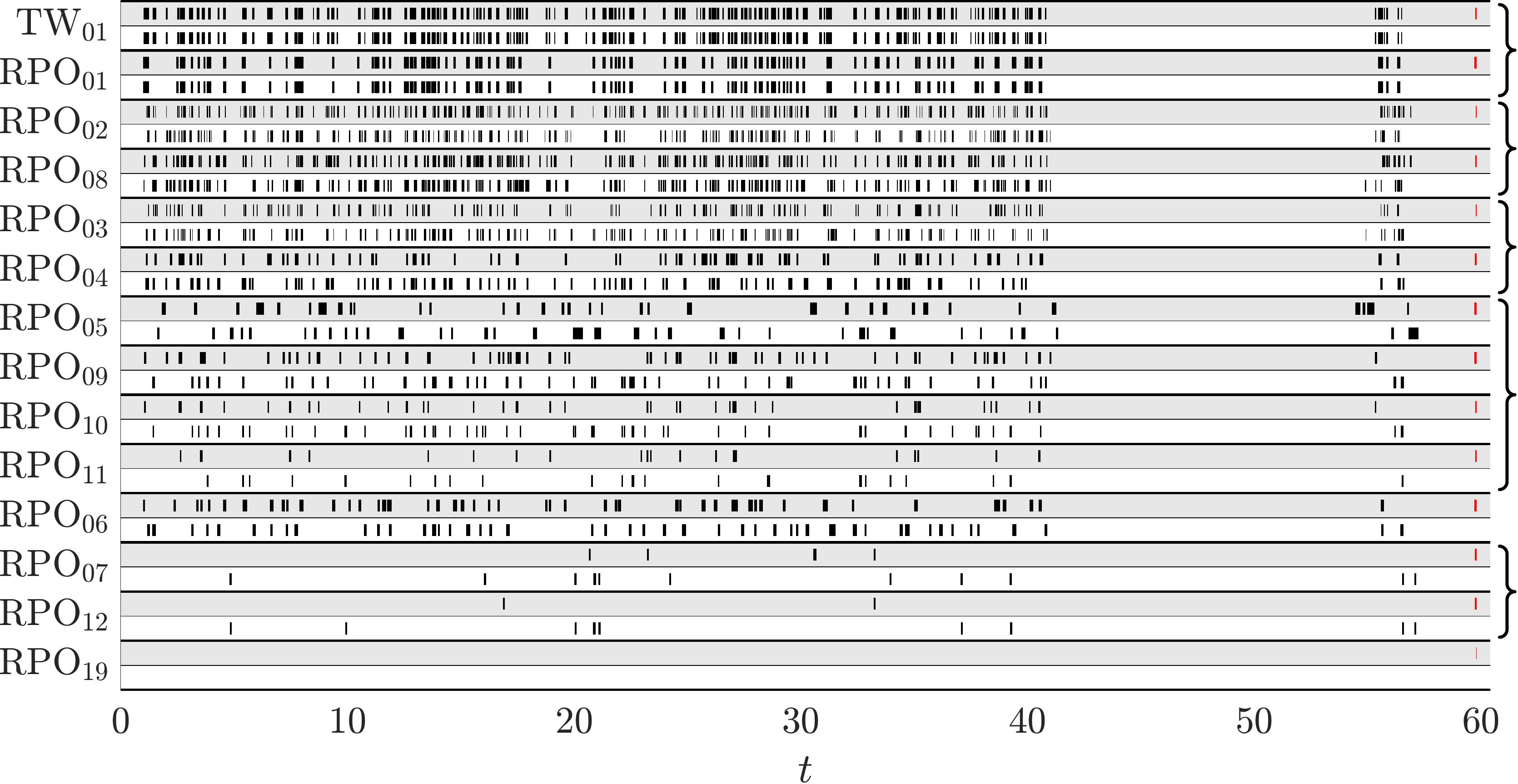}} \\
\subfloat[]{\includegraphics[width=\textwidth]{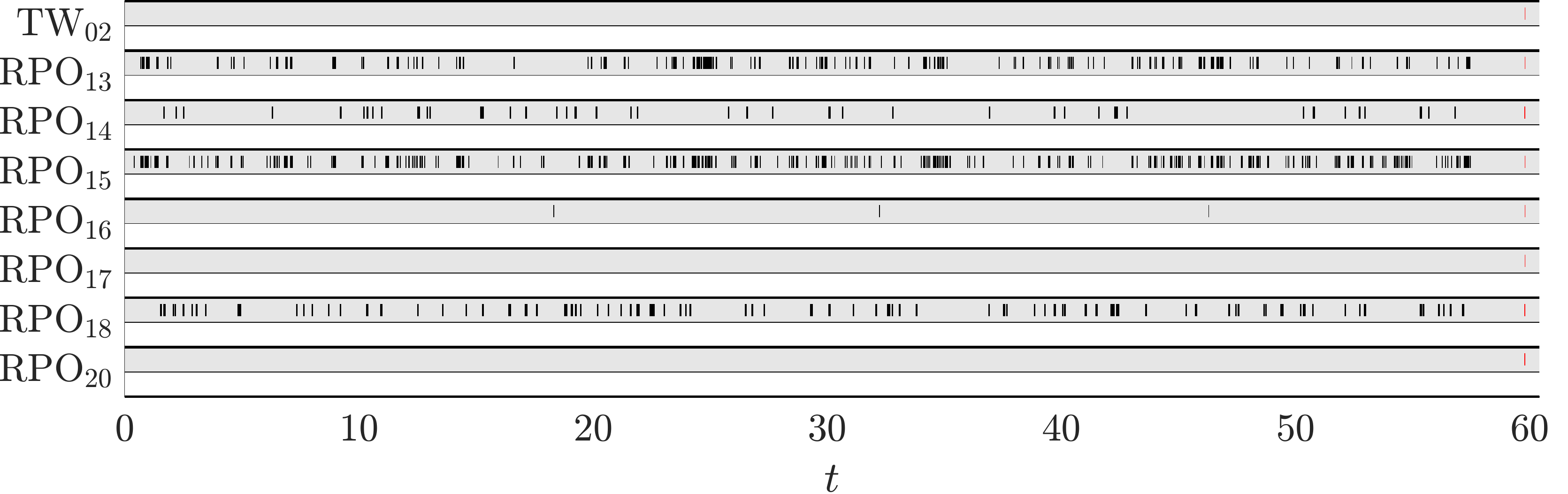}} 
\caption{\label{fig:shadow_summary} \small 
A summary of the shadowing events (marked with black bars) for each of the two turbulent trajectories, (a) ${\bf u}^a(t)$ and (b) ${\bf u}^b(t)$. Each ECS is represented by two rows, one for the numerically converged solution ${\bf u_{n}}$ (gray) and  one for its symmetric copy $K_{z} {\bf u_{n}}$ (white). Red bars on the right represent the escape times $\gamma_n^{-1}$ to scale. Curly braces on the far right group together \rpo{}s that are related via continuation in $Re_i$.
}
\end{figure}

We chose $\bar{E}_\tau=0.0001$ and $\bar{E}_\phi=0.0003$, so that conditions (a)-(c) are satisfied with roughly equal probability.
\autoref{fig:shadow_summary} summarizes the results for both the turbulent trajectory ${\bf u}^a(t)$ confined to \lobes{1}{2} and the turbulent trajectory ${\bf u}^b(t)$ confined to \lobe{3}. 
In the former case, we find that turbulence shadows both \rpo{}s and a \re. Moreover, if an ECS is shadowed, then so is its reflected copy. This suggests that turbulence does not break reflection symmetry in a statistical sense, i.e., the probability of shadowing any ECS is comparable to the probability of shadowing its reflected copy. As \autoref{fig:proj1} illustrates, this is not entirely unexpected, since \lobe{1} is fairly symmetric with respect to the reflection.

\begin{figure}
\vspace{5mm}
\center
\subfloat[]{\includegraphics[width=\textwidth]{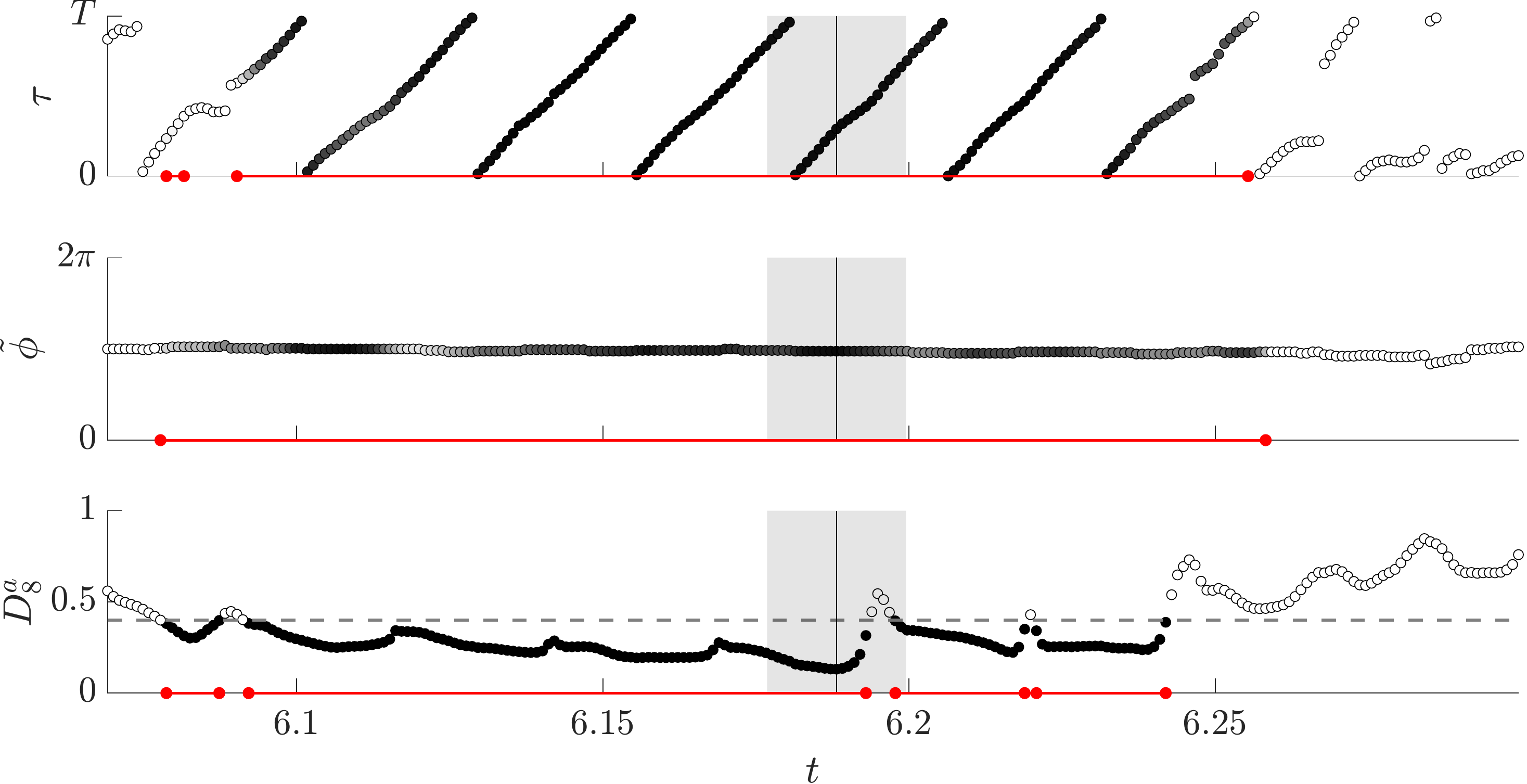}} \\
\subfloat[]{\includegraphics[width=0.4\textwidth]{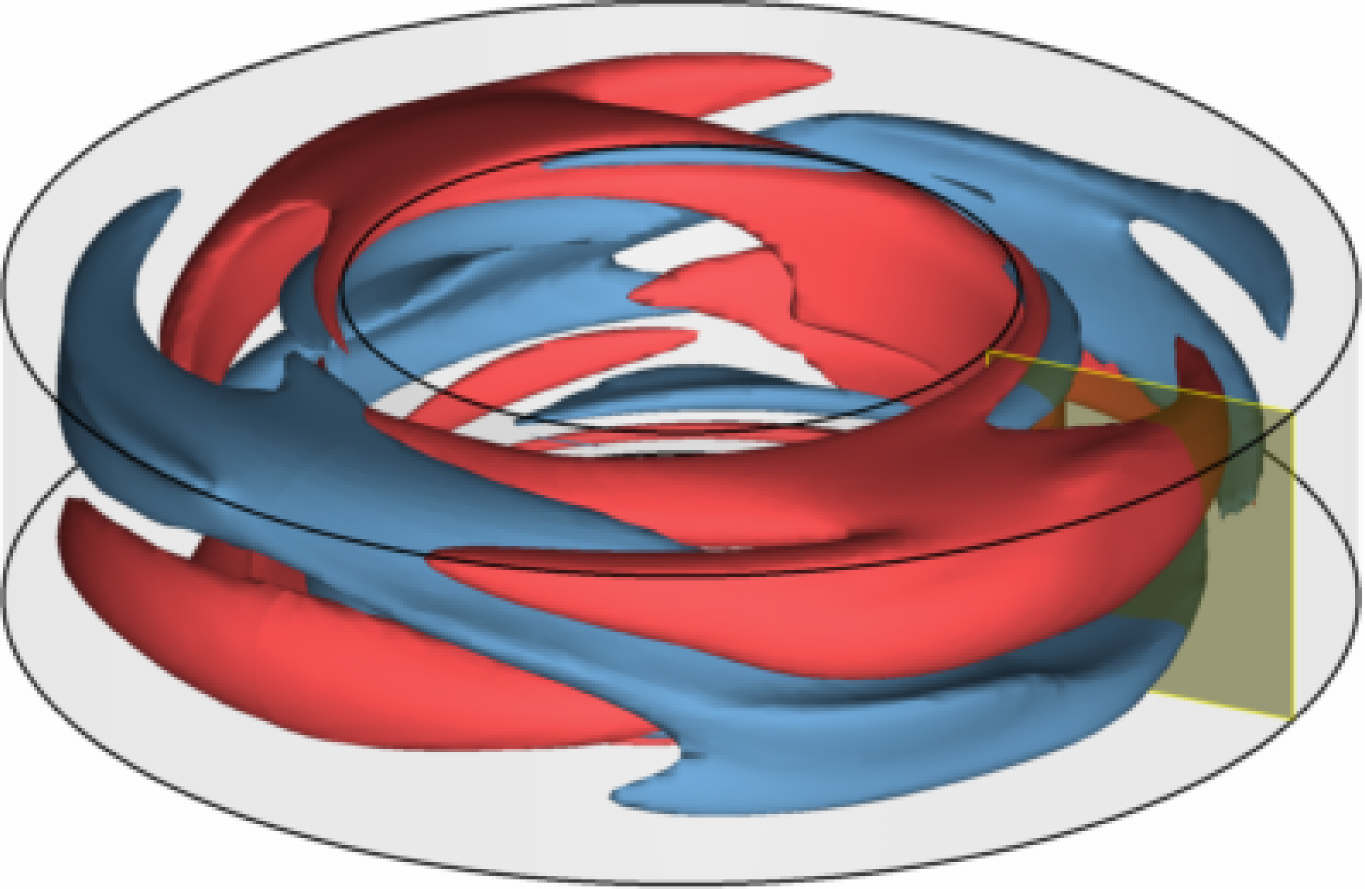}} \hspace{3mm}
\subfloat[]{\includegraphics[width=0.4\textwidth]{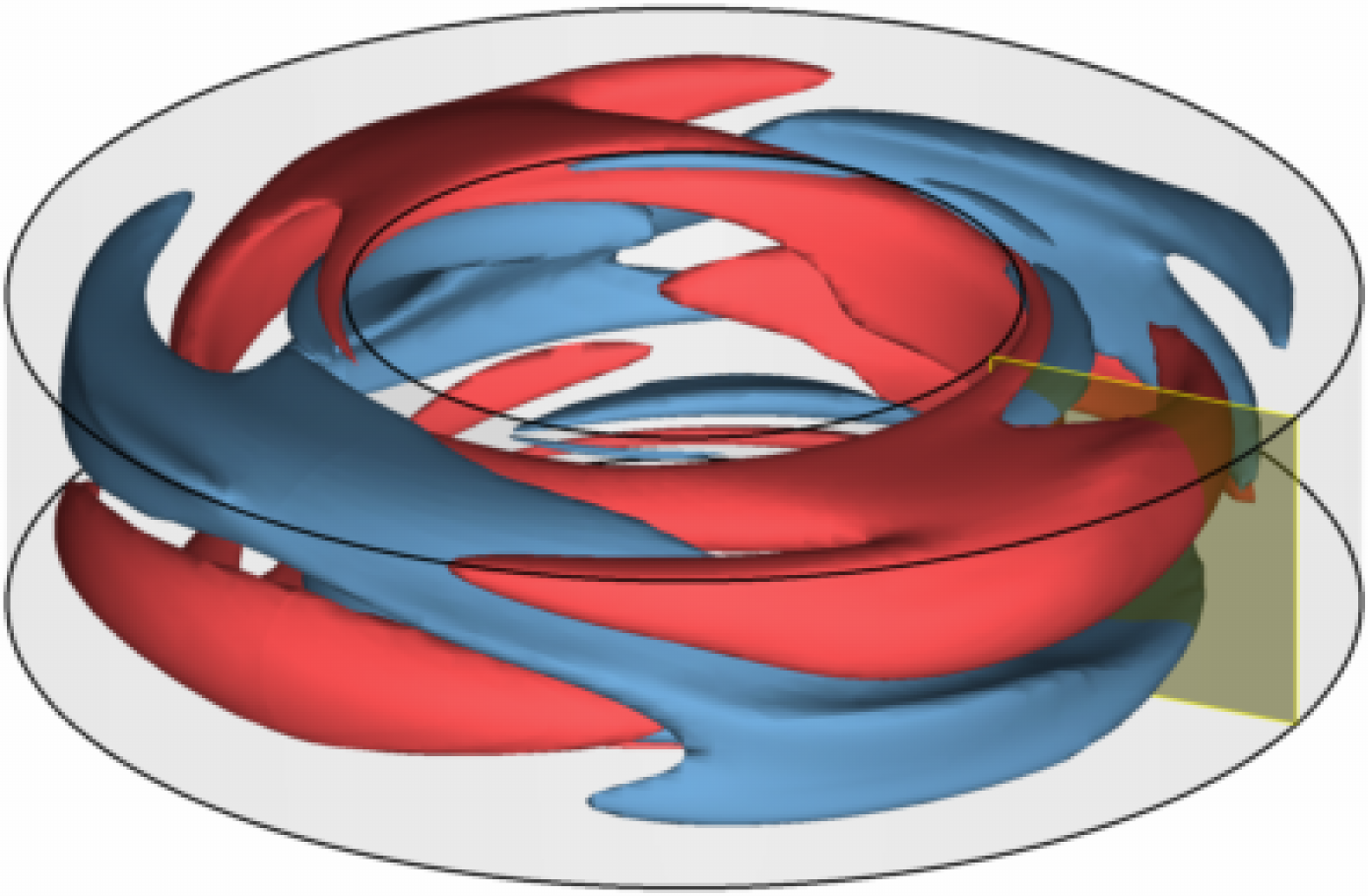}} \\
\subfloat[]{\includegraphics[width=0.4\textwidth]{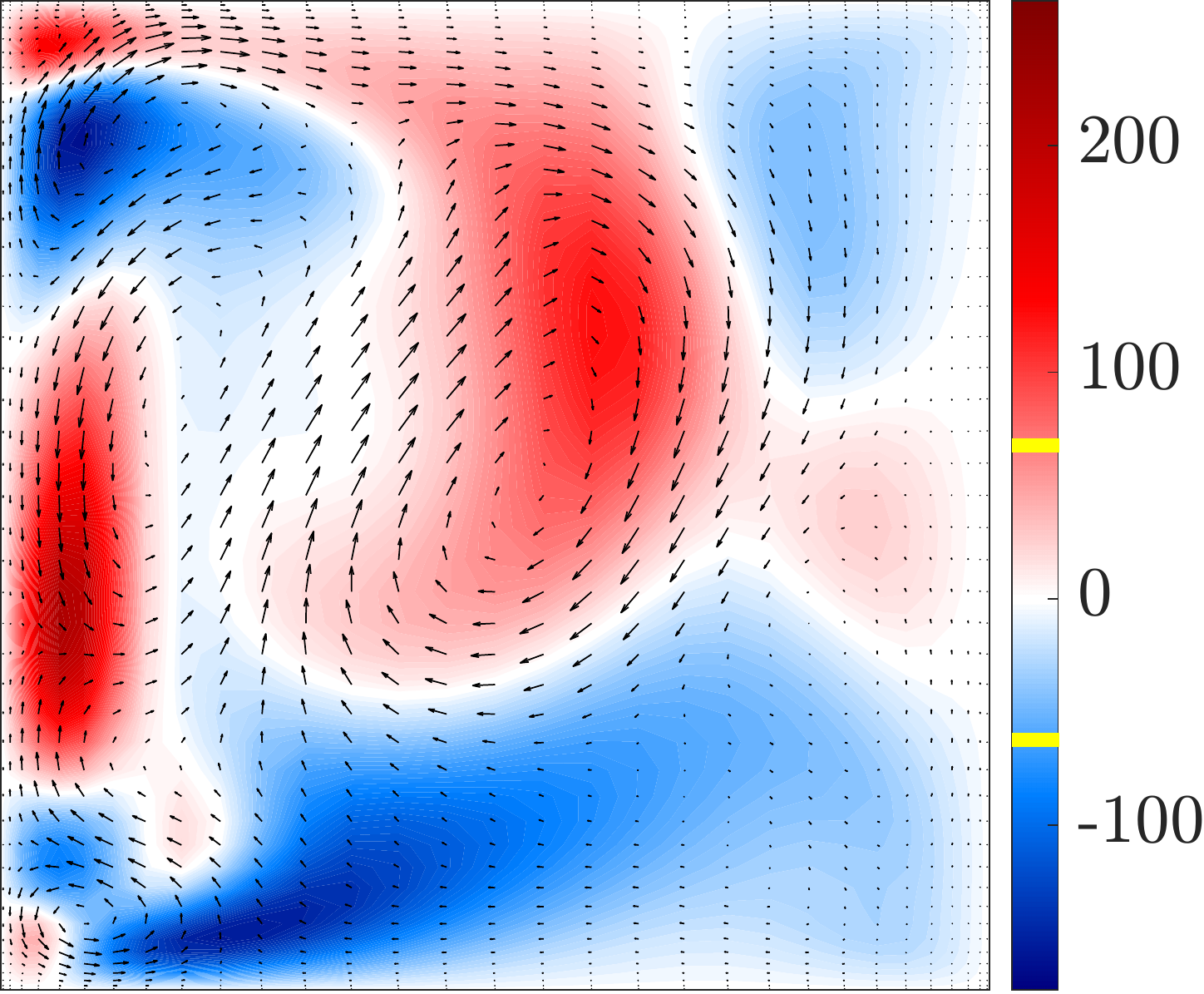}} \hspace{8mm}
\subfloat[]{\includegraphics[width=0.4\textwidth]{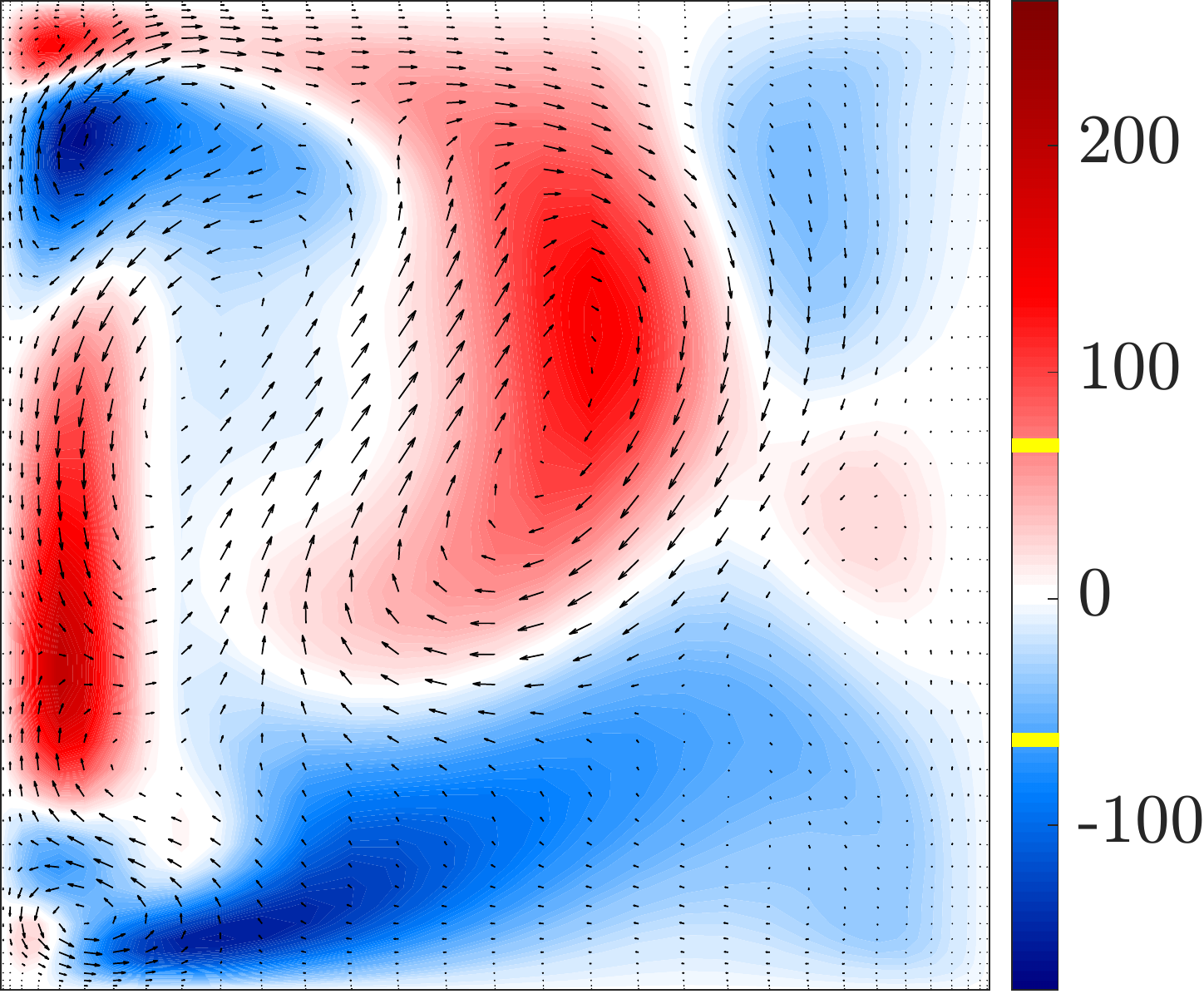}}
\caption{\label{fig:shadow_rpo05} \small 
Turbulent flow ${\bf u}^a(t)$ shadowing \rpo{05}. (a) Evolution of coordinates $\tau$, $\tilde{\phi}$, and $D$. Black (white) circles indicate instances when the corresponding shadowing criterion is satisfied (not satisfied). Red lines denote the temporal intervals during which a given criterion is satisfied. The gray bar shows the escape time, $\gamma_n^{-1}$. Snapshots of (b) the turbulent flow and (c) the ECS at the instant marked with the black vertical line in (a). Flow fields in the yellow cross-section for (d) the turbulent flow and (e) the ECS. A movie of this shadowing event is included as supplementary material.
}
\end{figure}

\begin{figure}
\vspace{5mm}
\center
\subfloat[]{\includegraphics[width=\textwidth]{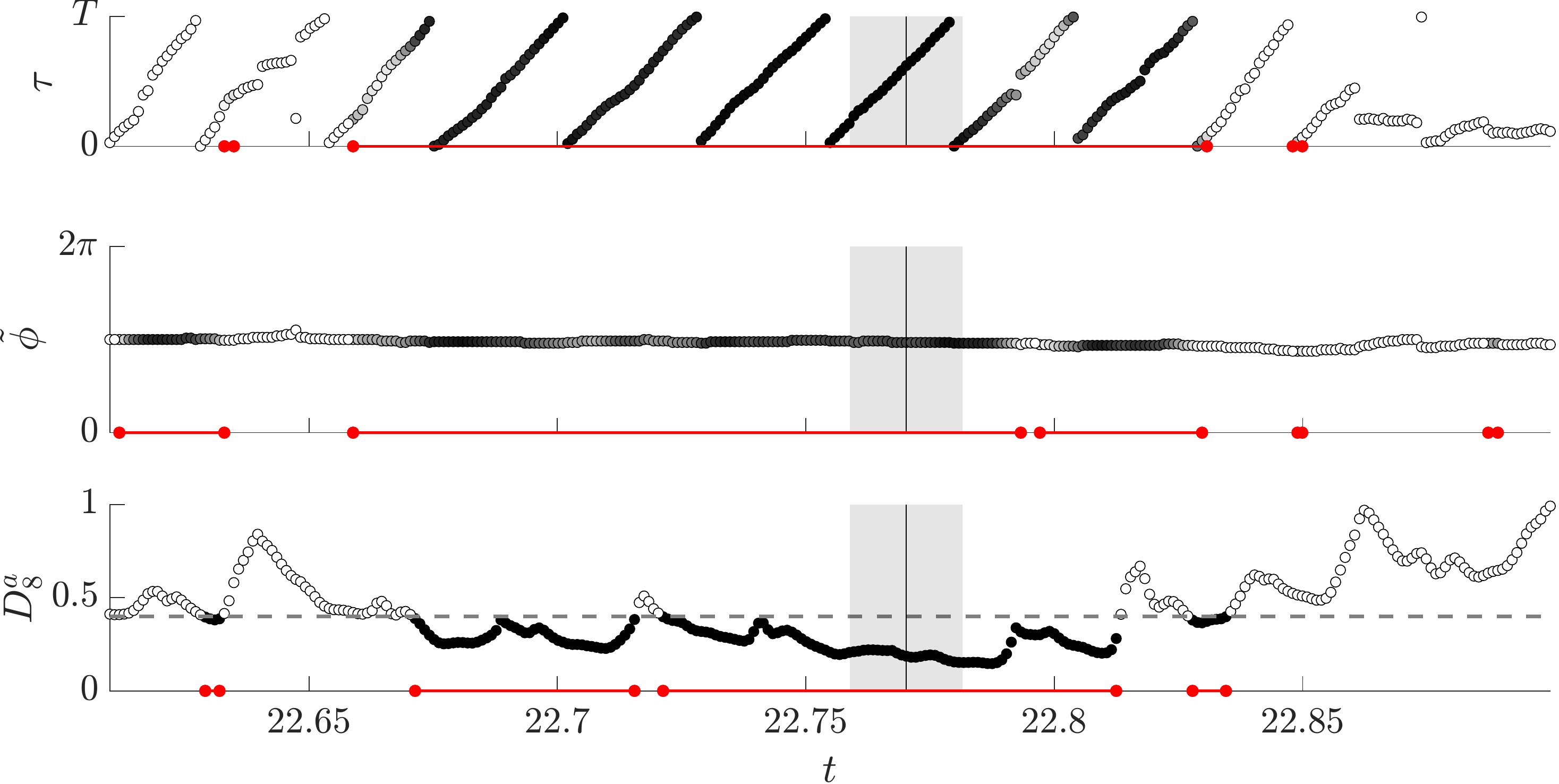}} \\
\subfloat[]{\includegraphics[width=0.45\textwidth]{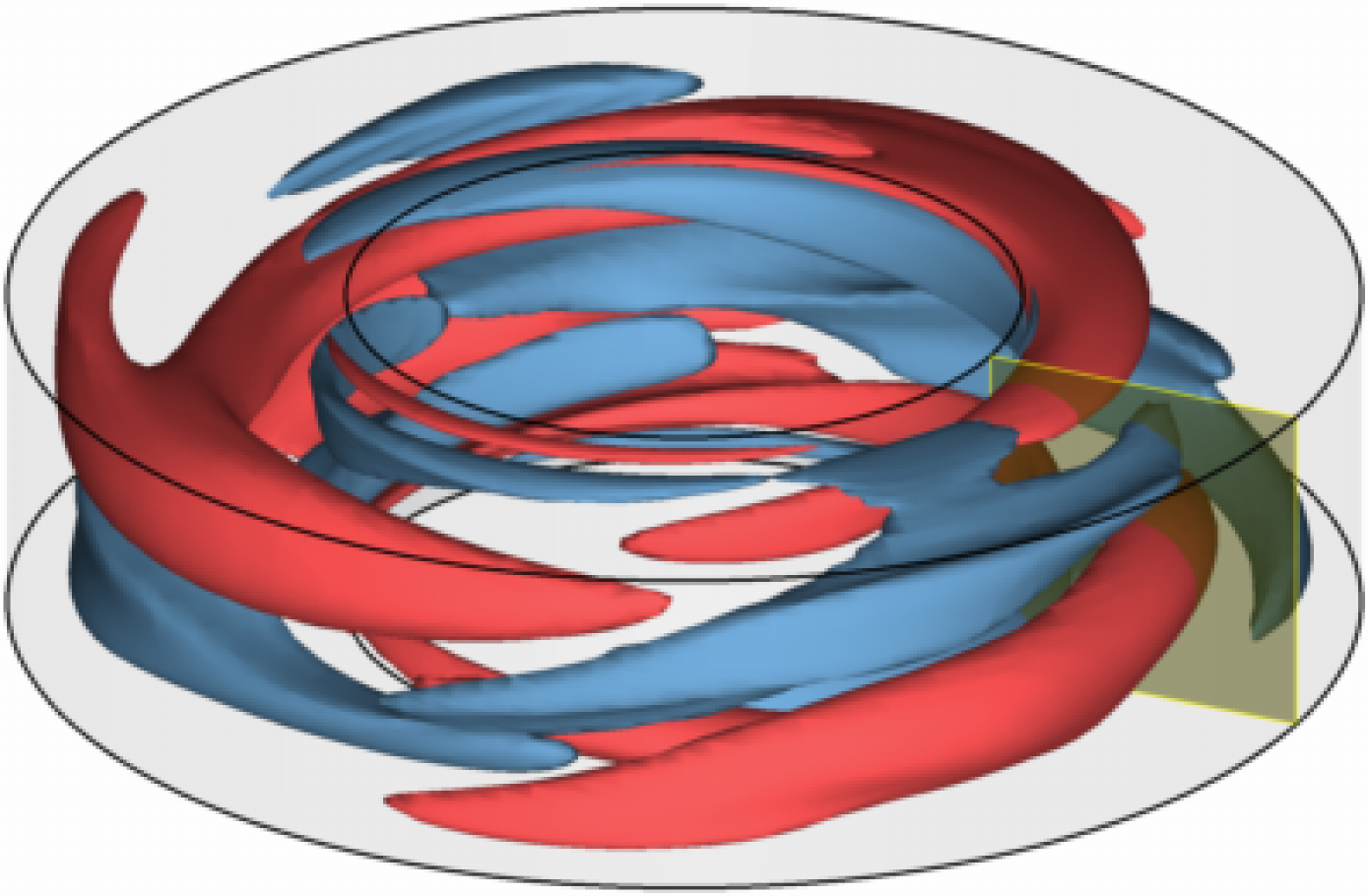}}  \hspace{3mm}
\subfloat[]{\includegraphics[width=0.45\textwidth]{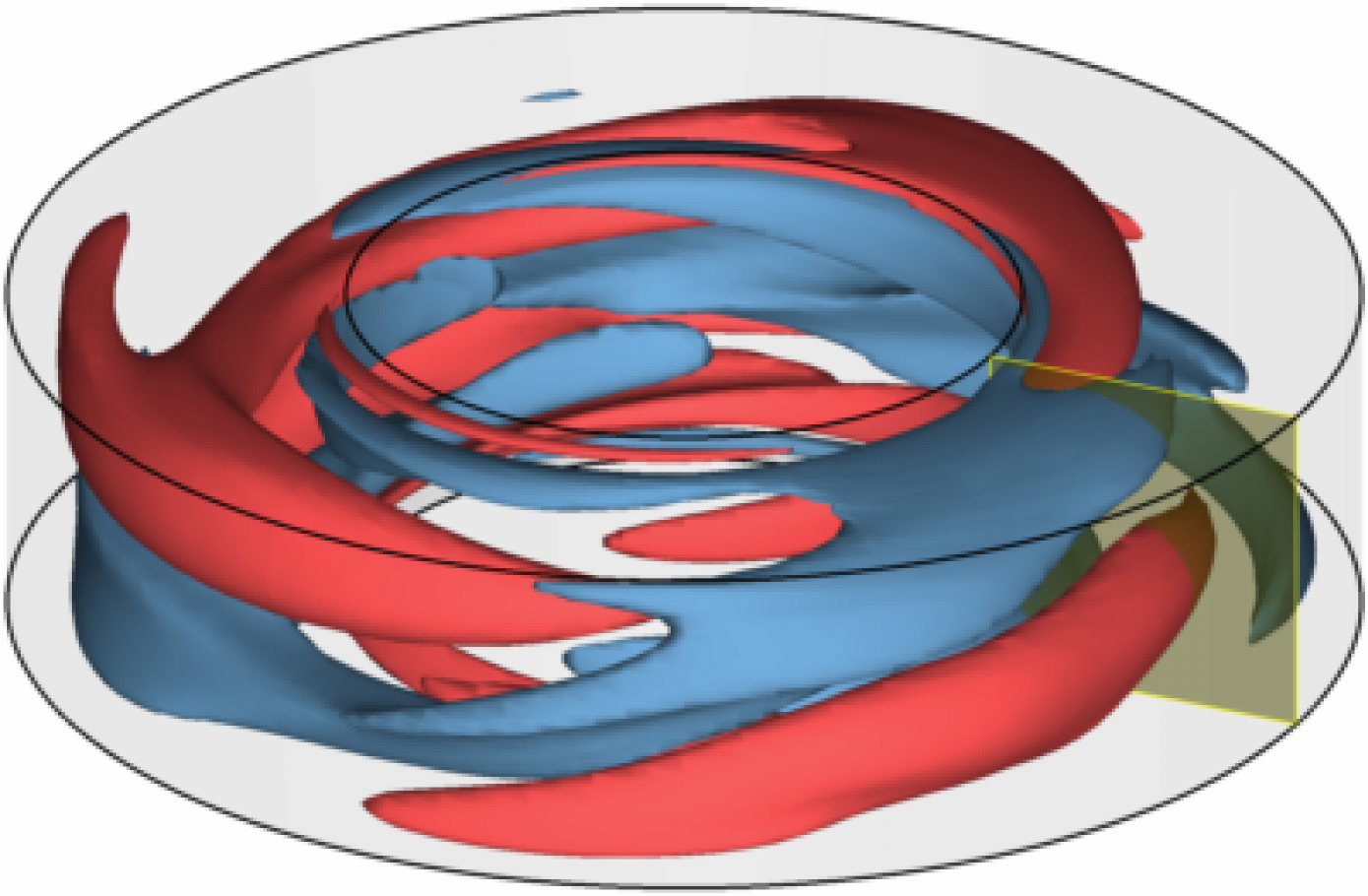}} \\
\subfloat[]{\includegraphics[width=0.4\textwidth]{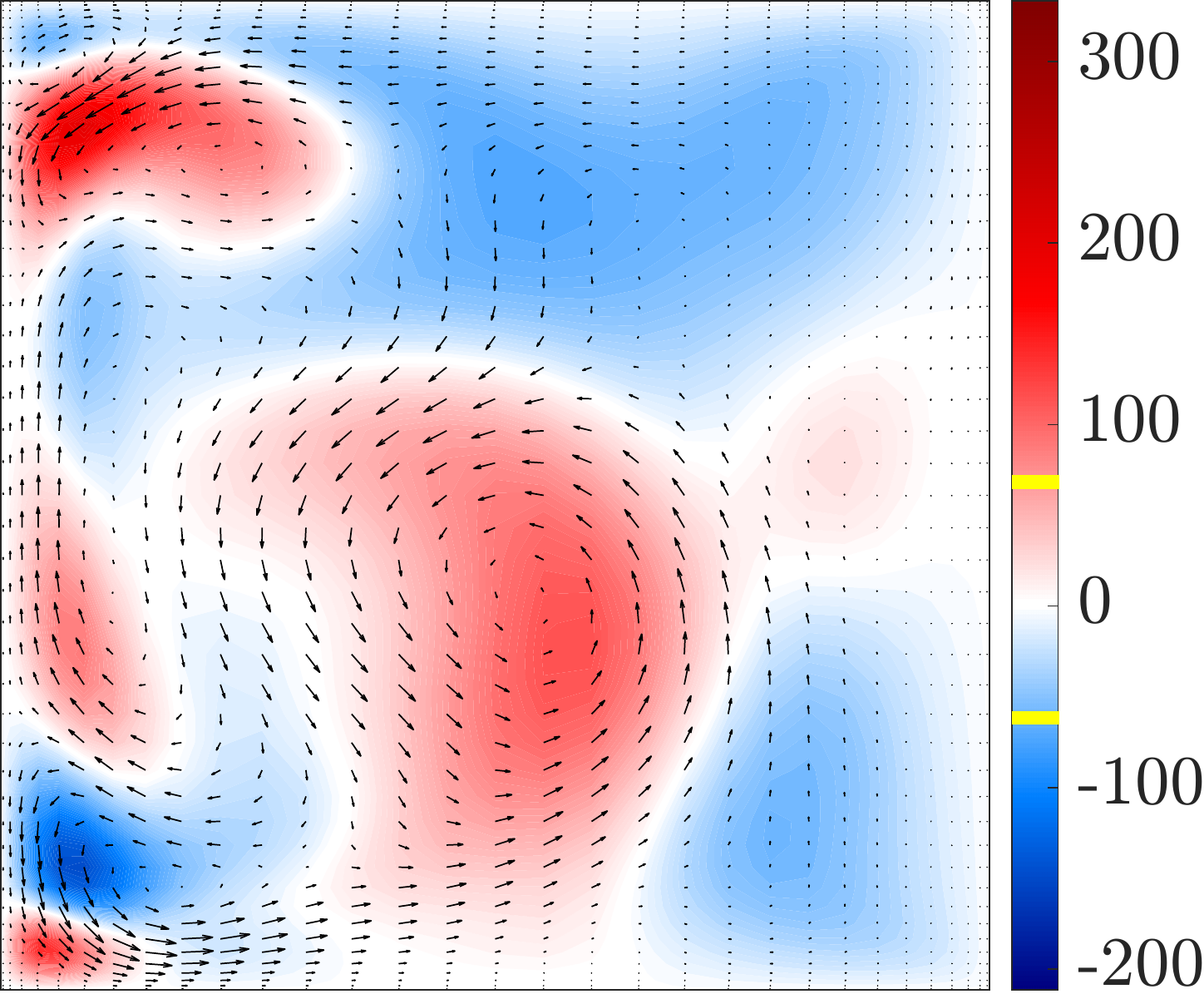}} \hspace{8mm}
\subfloat[]{\includegraphics[width=0.4\textwidth]{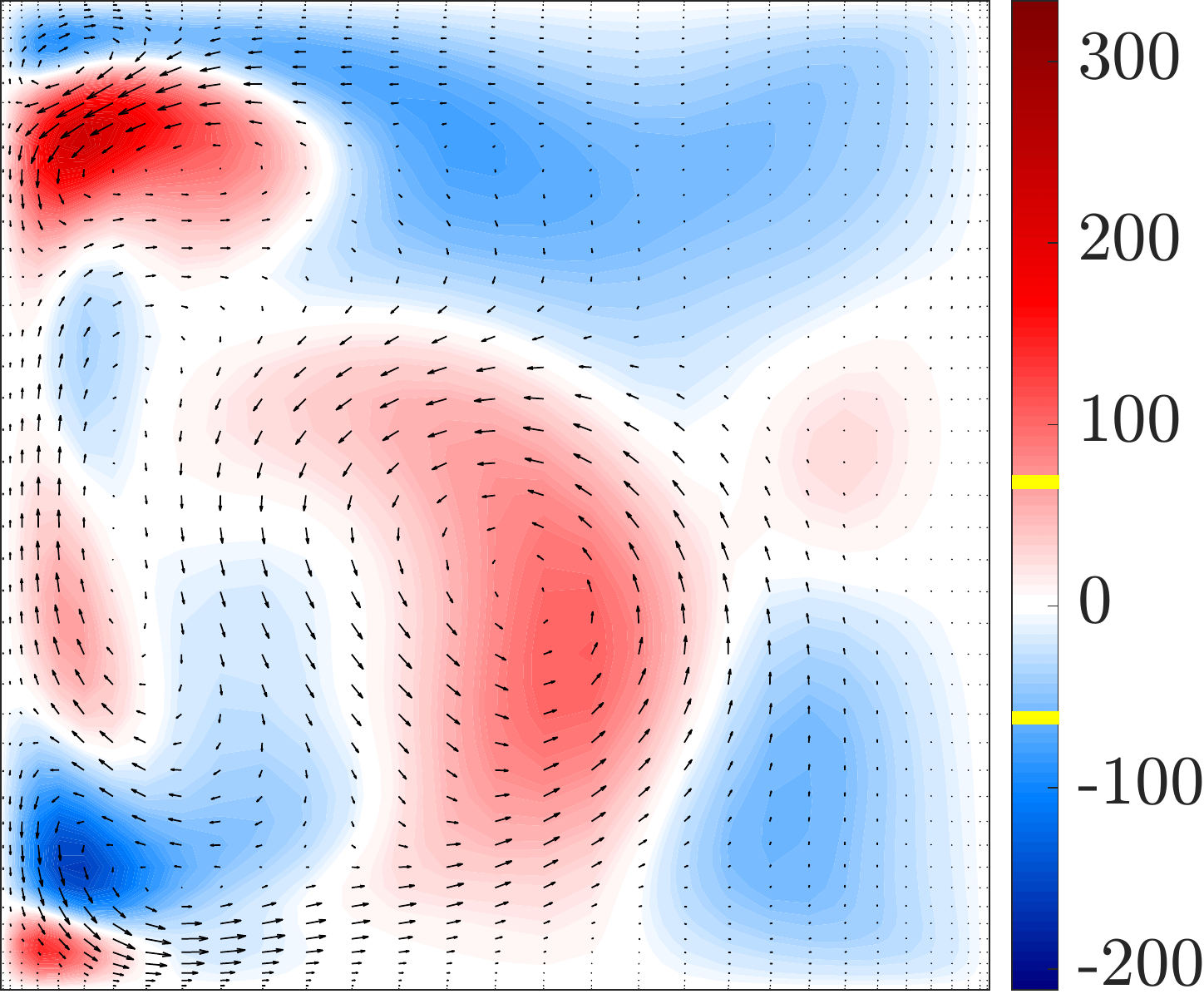}}
\caption{\label{fig:shadow_rpo05K} \small 
A shadowing event for the reflected copy of \rpo{05} in \lobe{1}. A corresponding movie is available as supplementary material.
}
\end{figure}

For instance, consider a shadowing event for \rpo{05} shown in \autoref{fig:shadow_rpo05} and its symmetry-related copy in \autoref{fig:shadow_rpo05K} (the corresponding movies are included as supplementary material). The same format is used here and below to illustrate shadowing events. Panel (a) shows the evolution of the three coordinates $\tau$, $\tilde{\phi}$, and $D$ with black (white) circles indicating instances when the corresponding shadowing criterion is satisfied (not satisfied). Red lines denote the temporal intervals when the criterion based on the corresponding coordinate is satisfied; for an interval to be considered an instance of shadowing, all three criteria must be met simultaneously. The gray bar shows the length of the interval $\gamma_n^{-1}$ for comparison with the temporal period of the ECS. Panels (b) and (c) compare, respectively, the spatial structure (contours of constant $u_\theta$ at the values denoted in yellow on the color bar in panels (d) and (e)) of the turbulent flow and the corresponding ECS in the entire domain at the instant denoted by the vertical black line in panel (a). Finally, panels (d) and (e) compare the velocity fields in the yellow cross-section shown in panels (b) and (c), respectively. In all panels, the turbulent mean flow is subtracted off. 

Note that, for both \rpo{05} and its symmetry-related copy, the respective \rpo{} is shadowed continuously for several periods (interval of around $6\gamma_n^{-1}$). During the shadowing intervals, all three criteria are satisfied to a high accuracy: the slope of $\tau(t)$ is near unity, the slope of $\tilde{\phi}(t)$ is near zero, and $D(t)$ stays below the threshold $\bar{D}$. As discussed previously, for $D(t)<\bar{D}$ the turbulent flow is visually almost indistinguishable from the corresponding ECS in the entire domain. \autoref{fig:shadow_rpo05}(b-c) illustrates this particularly convincingly for \rpo{05}. The similarity is not just qualitative, but quantitative, as \autoref{fig:shadow_rpo05}(d-e) shows. For small $D(t)$, it is expected that the turbulent flow evolves in the same manner as the nearby \rpo{}. However, when $D(t)$ briefly increases above threshold, the slopes of $\tau(t)$ and $\tilde{\phi}(t)$ remain virtually unchanged, suggesting that turbulence continues shadowing the same member of the \rpo{} family even when the corresponding flows are not very similar. Analogous statements apply to the reflected version of \rpo{05}, as \autoref{fig:shadow_rpo05K} illustrates.

\begin{figure}
\vspace{5mm}
\center
\subfloat[]{\includegraphics[width=\textwidth]{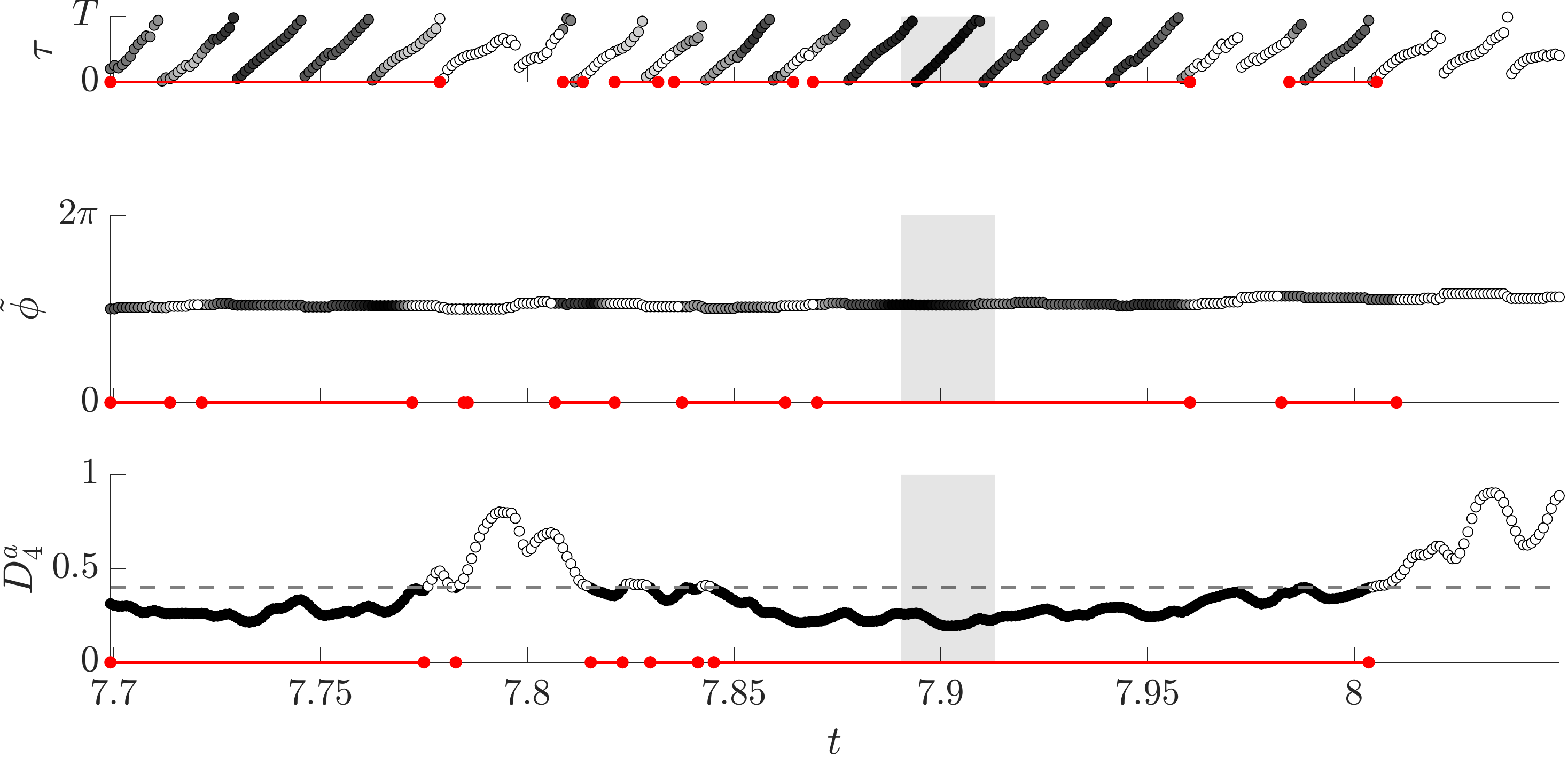}} \\
\subfloat[]{\includegraphics[width=0.45\textwidth]{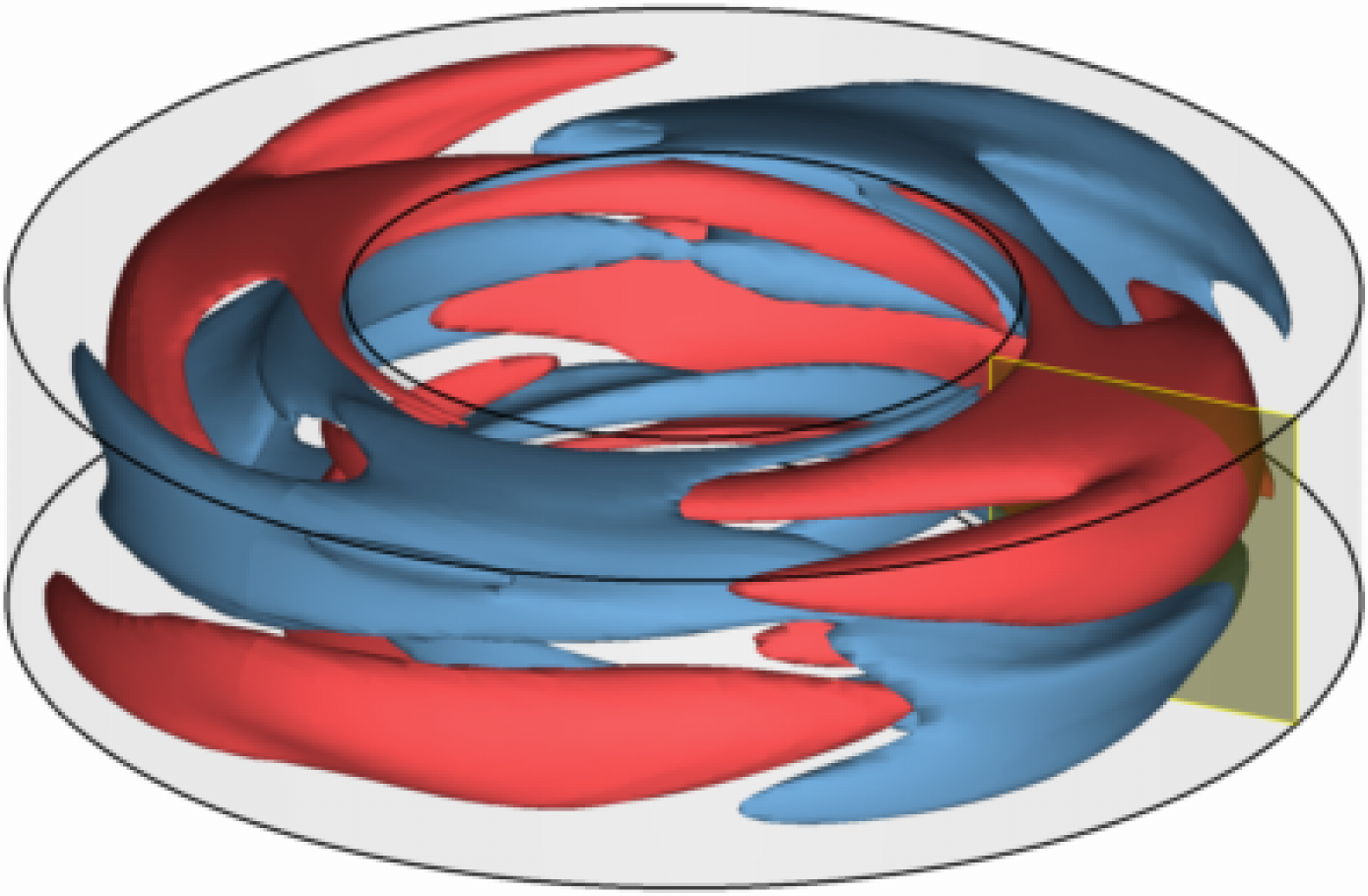}}  \hspace{3mm}
\subfloat[]{\includegraphics[width=0.45\textwidth]{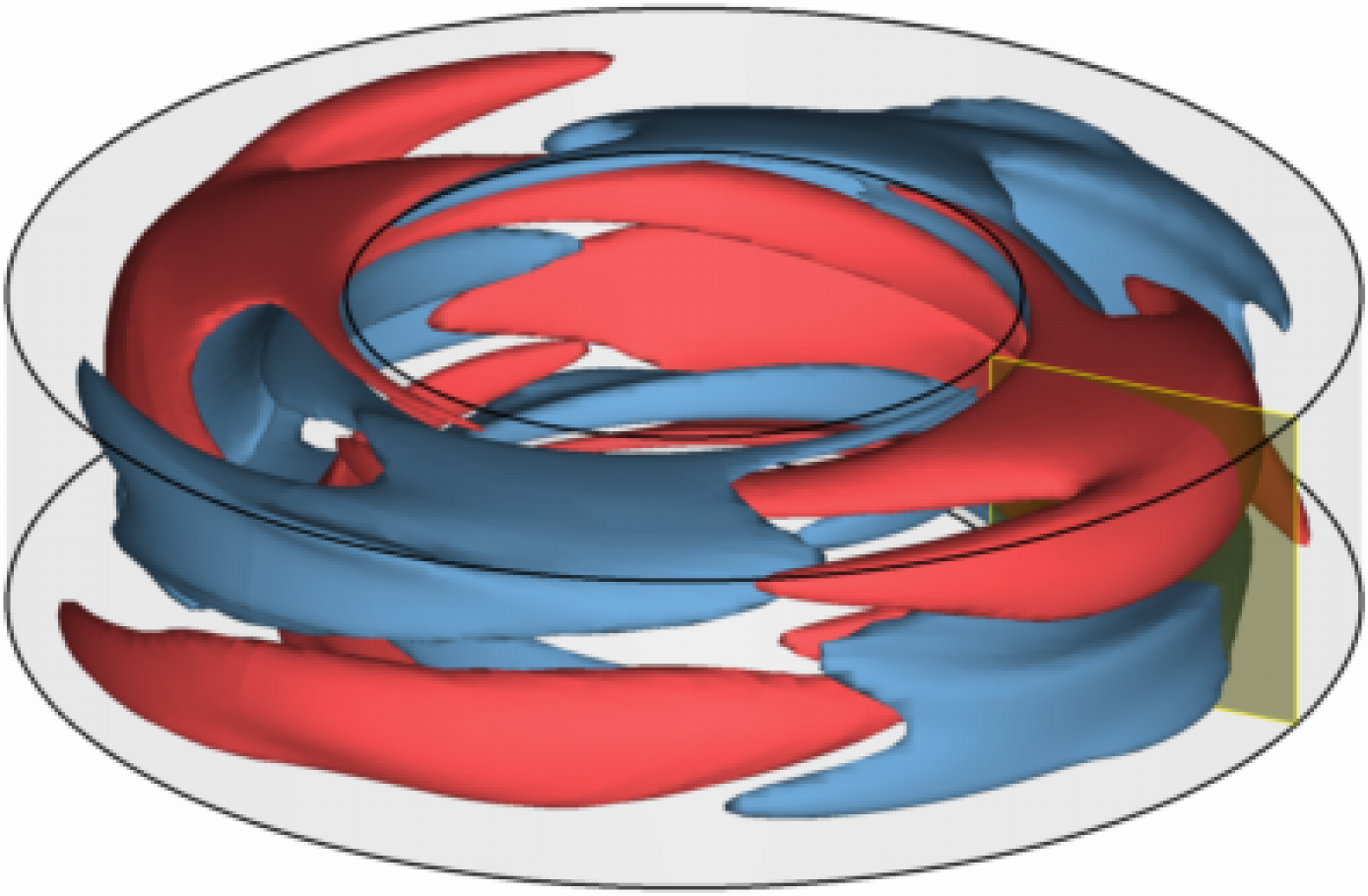}} \\
\subfloat[]{\includegraphics[width=0.4\textwidth]{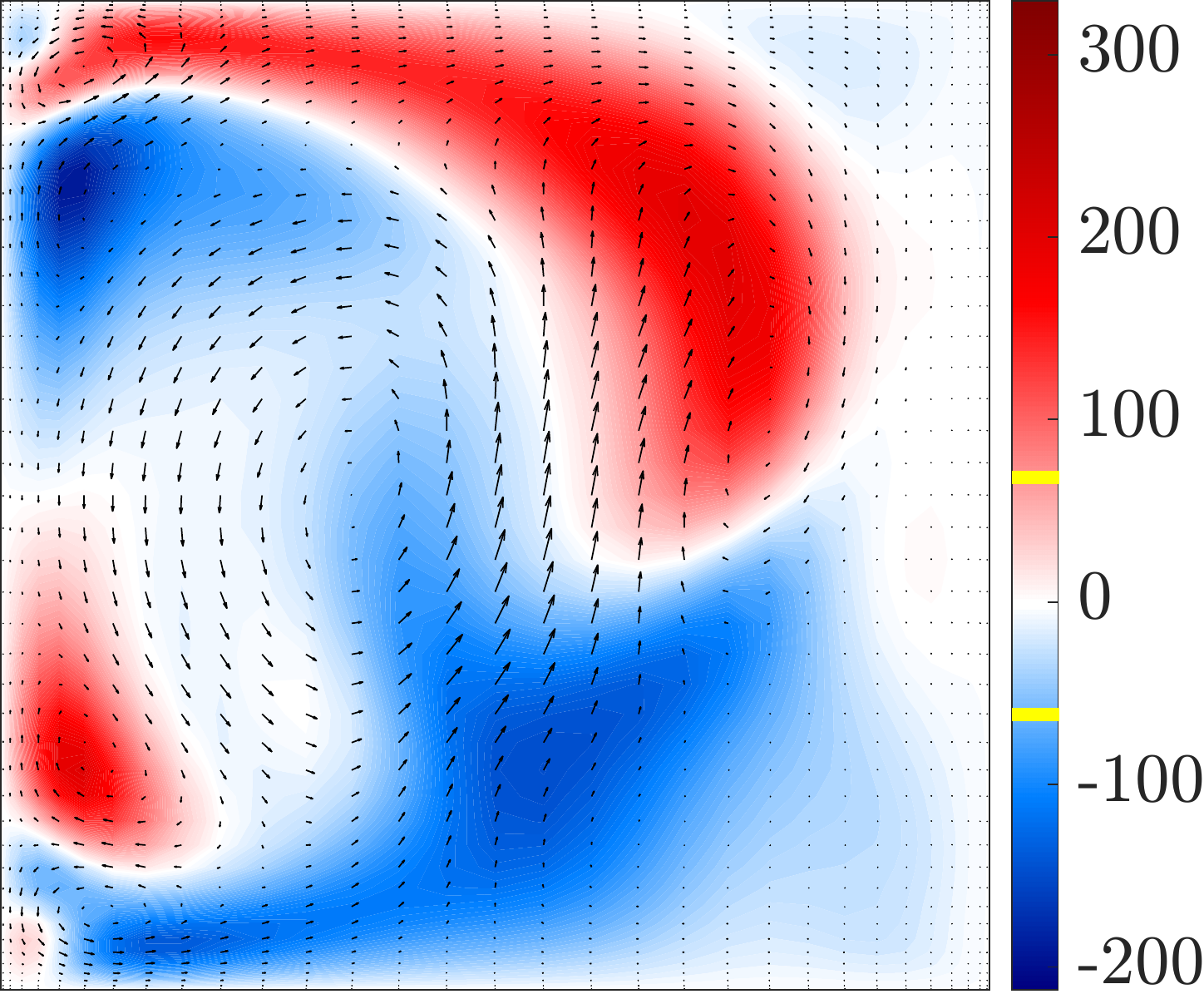}} \hspace{8mm}
\subfloat[]{\includegraphics[width=0.4\textwidth]{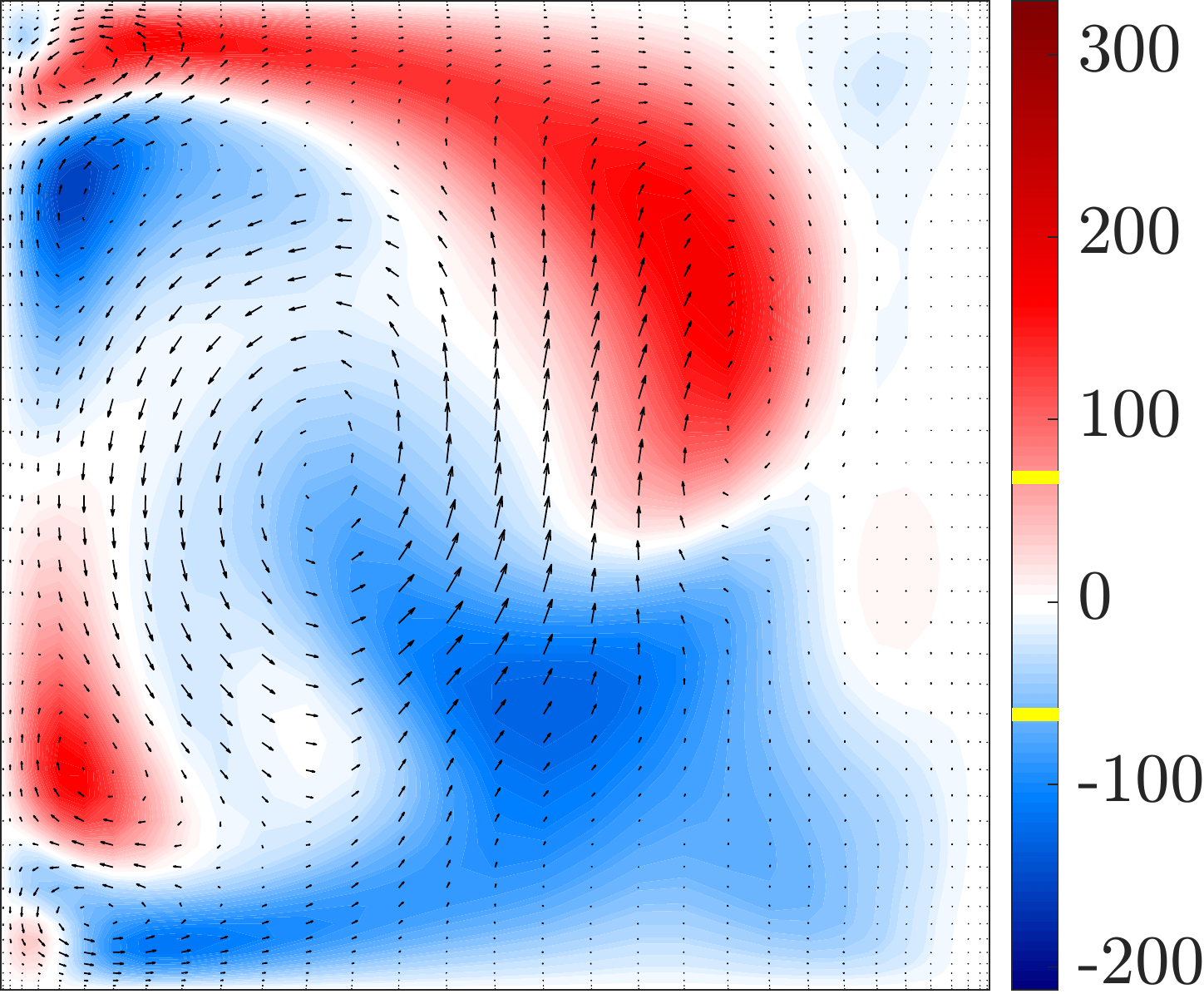}}
\caption{\label{fig:shadow_rpo01} \small 
A shadowing event for \rpo{01} in \lobe{1}. A corresponding movie is available as supplementary material.
}
\end{figure}

\autoref{fig:shadow_rpo01} presents a sequence of several shadowing events for \rpo{01}. These shadowing events are of duration comparable to those for \rpo{05} and its reflection, measured in units of $\gamma_n^{-1}$, but correspond to quite a few periods of \rpo{01}. Not only that, turbulence also visits the neighborhood of \rpo{01} much more frequently, as \autoref{fig:shadow_summary}(a) demonstrates. In fact, this solution is by far the most frequently visited \rpo{} in \lobe{1}. Note that \rpo{01} is symmetric with respect to $K_zR_{\pi/2}$, so reflection is equivalent to a rotation by $\pi/2$. Hence, both \rpo{01} and its reflection belong to the same solution family, and both rows in \autoref{fig:shadow_summary}(a) contain identical sequences of shadowing events. Also, just like in the case of \rpo{05}, turbulent flow is essentially indistinguishable from \rpo{01} during the shadowing episodes.

\begin{figure}
\vspace{5mm}
\center
\subfloat[]{\includegraphics[width=0.55\textwidth]{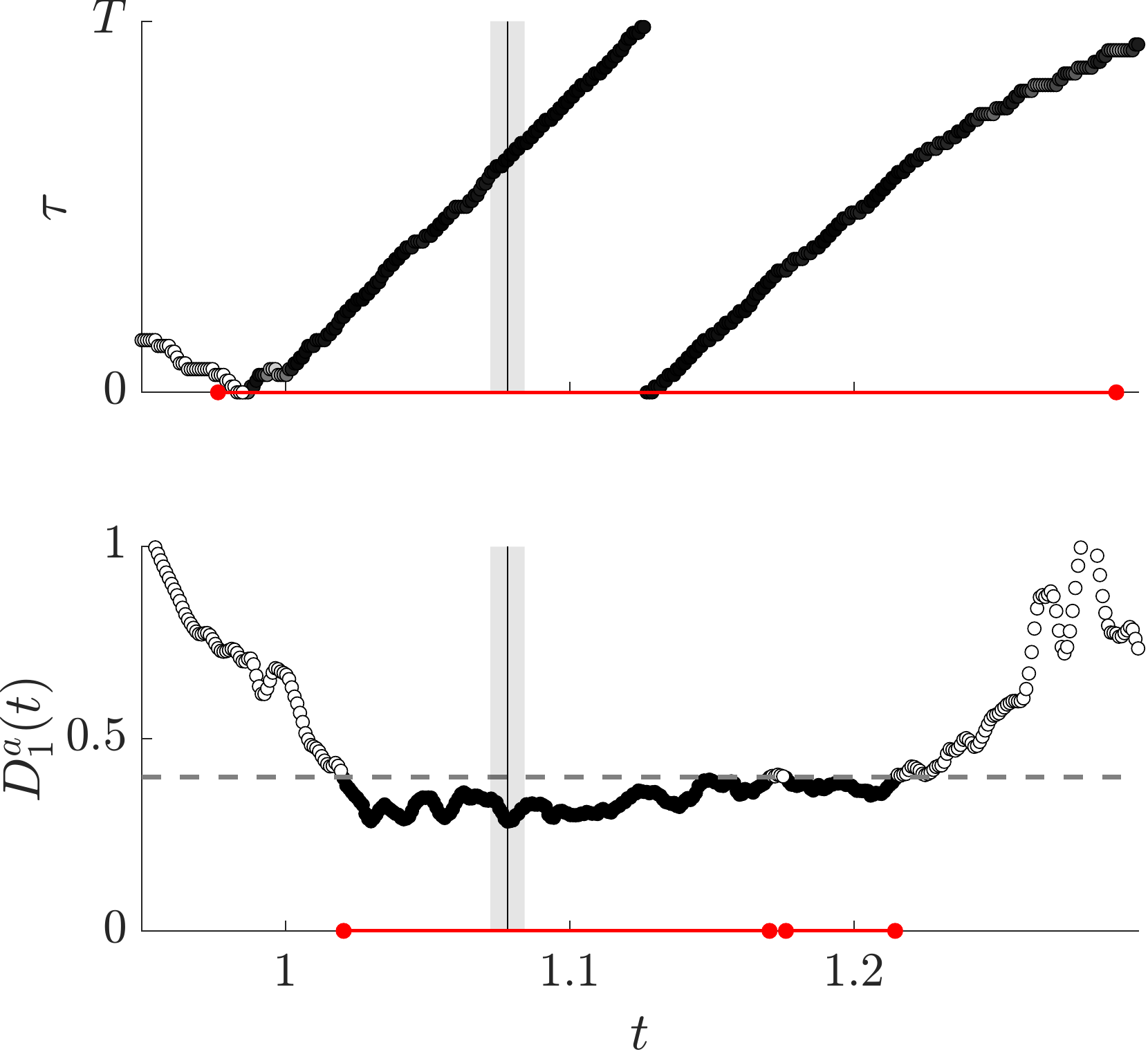}} \\
\subfloat[]{\includegraphics[width=0.45\textwidth]{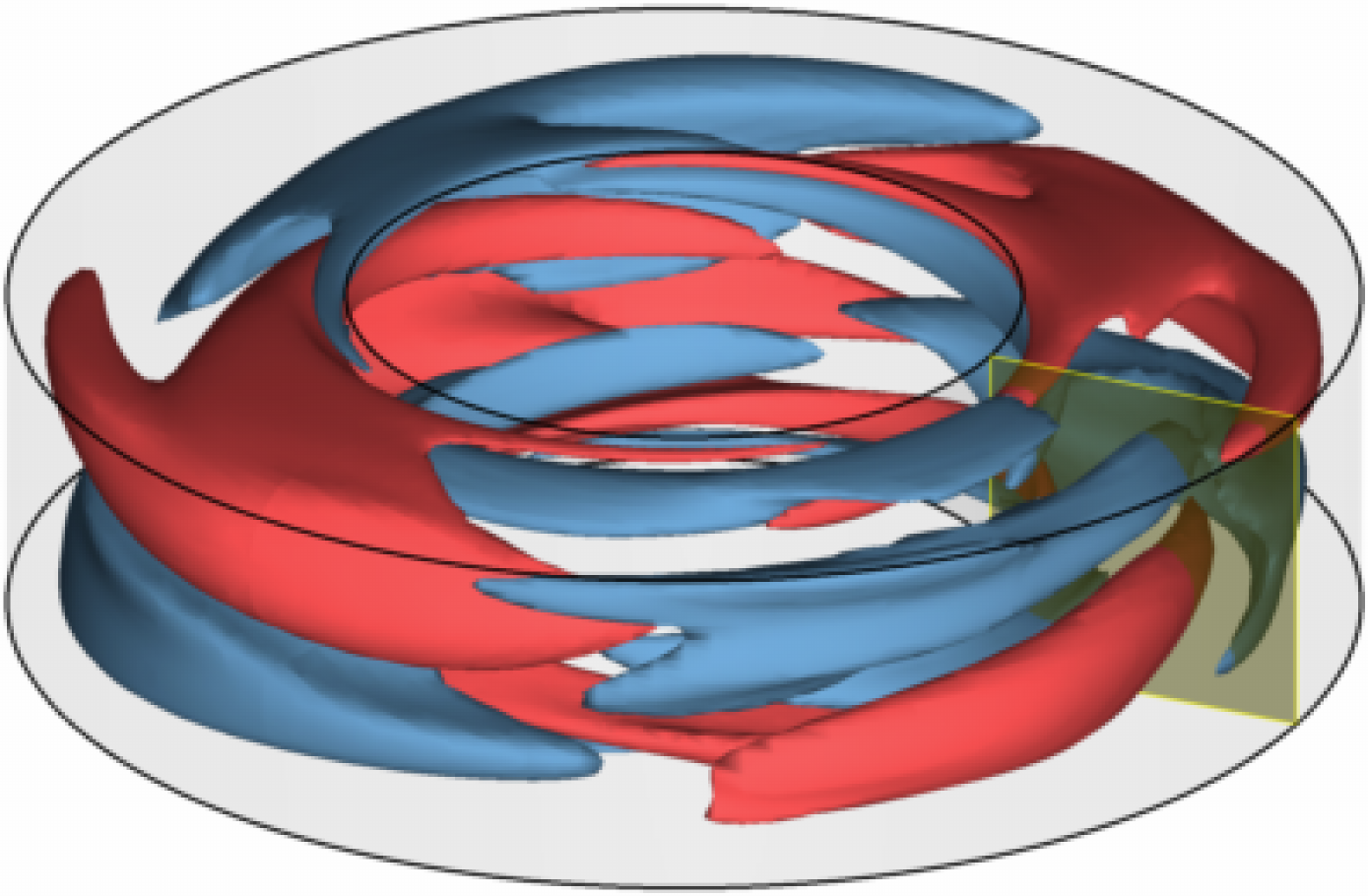}}  \hspace{3mm}
\subfloat[]{\includegraphics[width=0.45\textwidth]{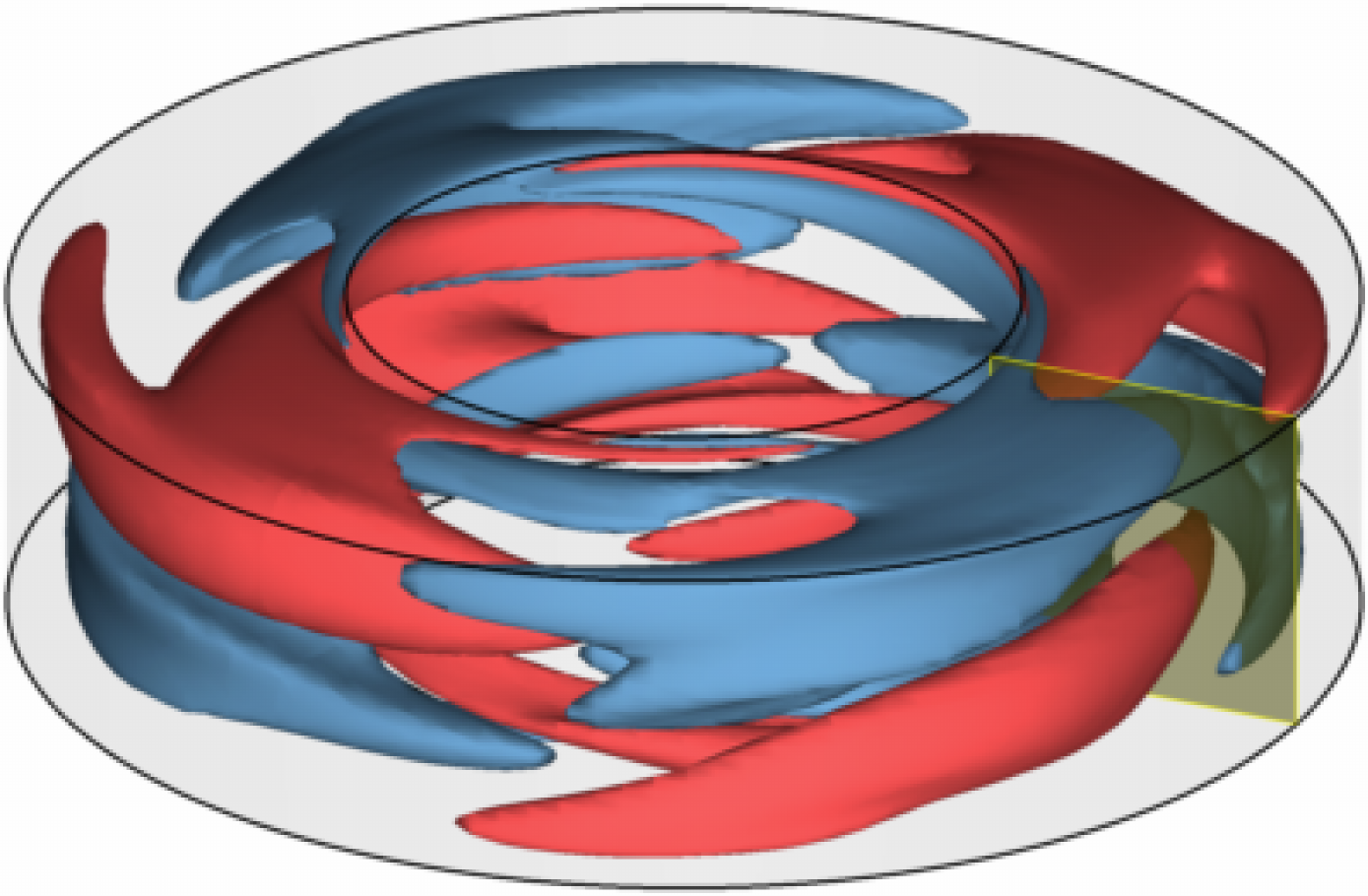}} \\
\subfloat[]{\includegraphics[width=0.4\textwidth]{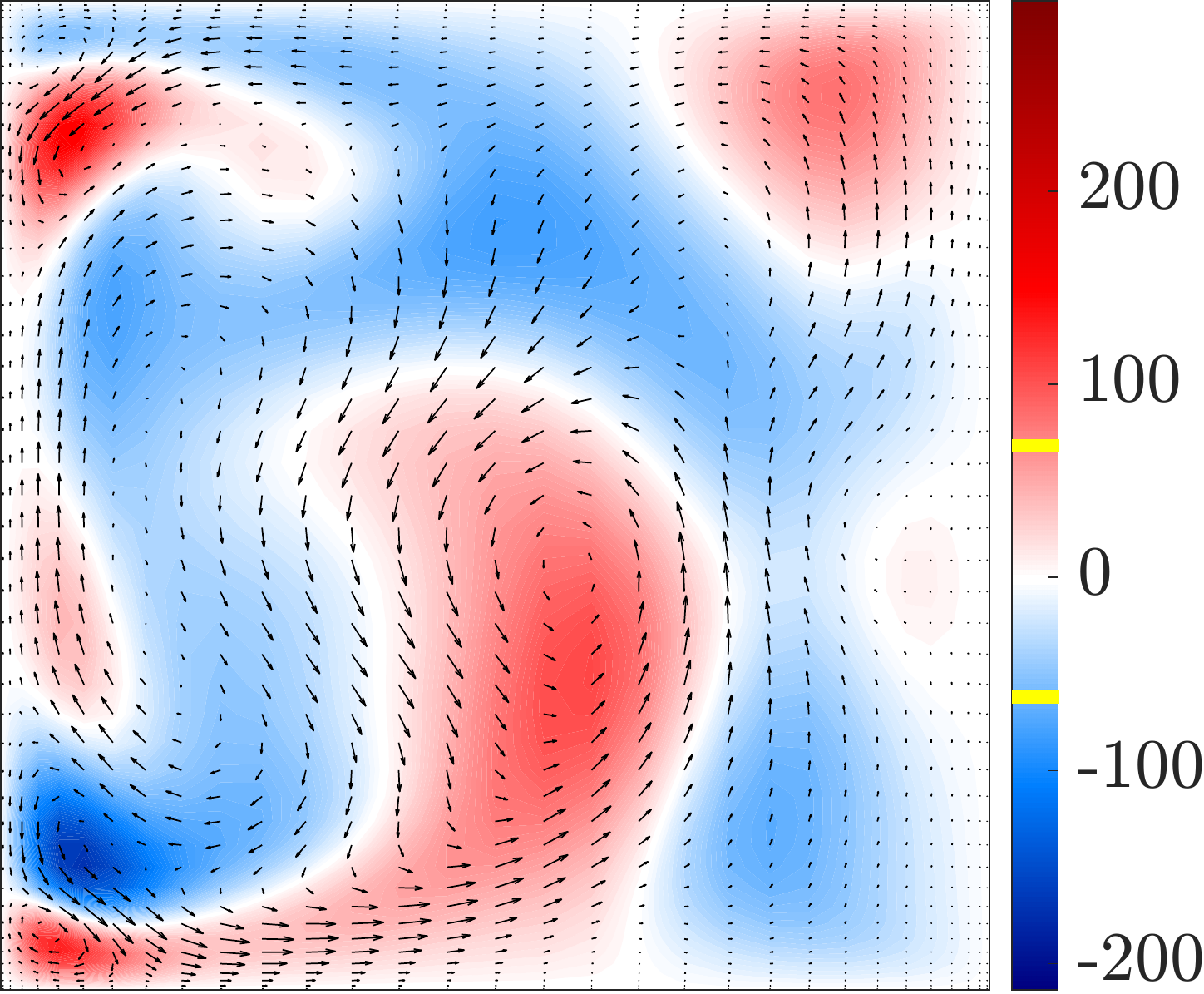}} \hspace{8mm}
\subfloat[]{\includegraphics[width=0.4\textwidth]{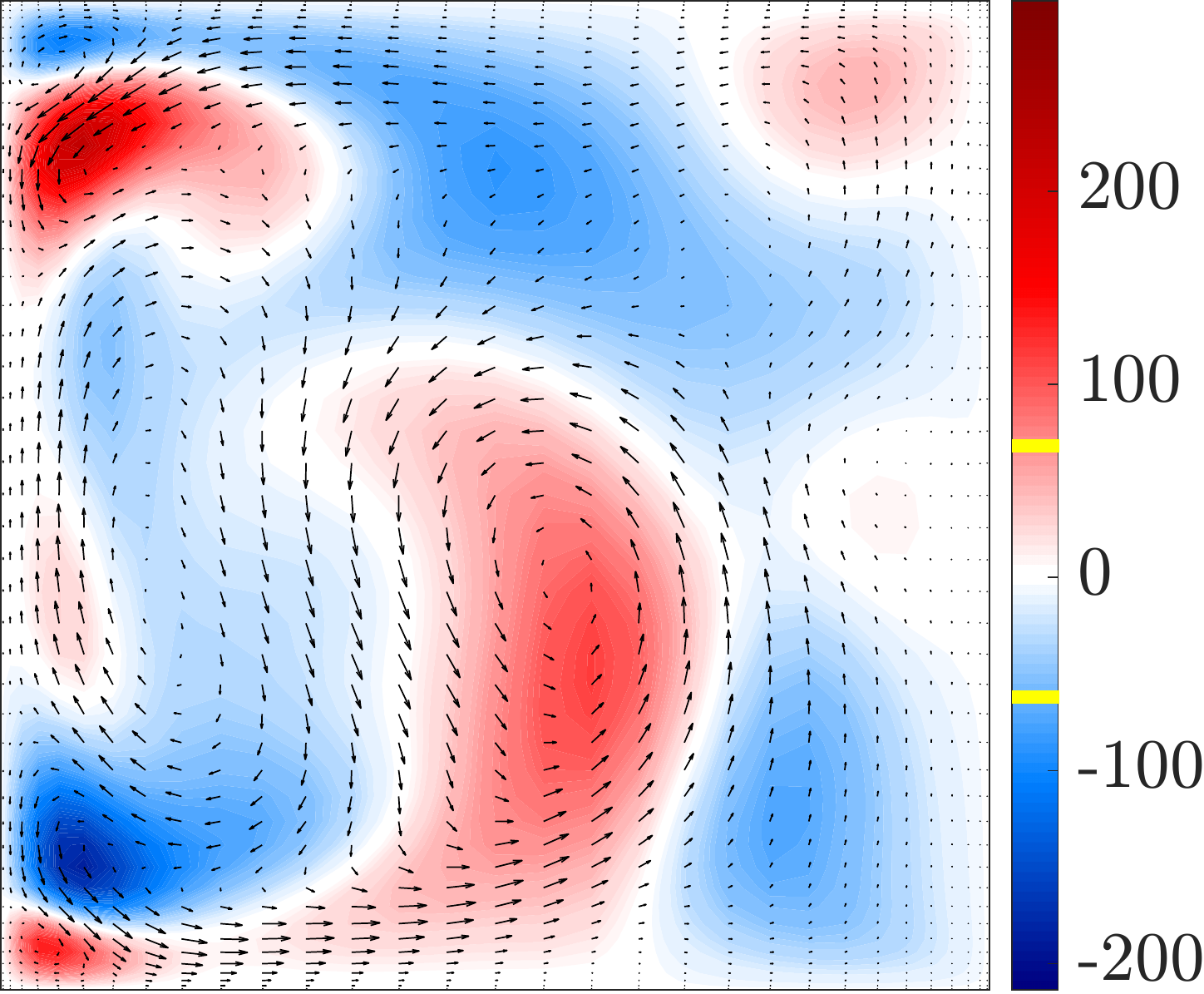}}
\caption{\label{fig:shadow_re01} \small 
A shadowing event for \re{01} in \lobe{1}. A corresponding movie is available as supplementary material.
}
\end{figure}

\autoref{fig:shadow_re01} illustrates shadowing of \re{01}. 
Rather unexpectedly, \re{01} is found to be shadowed even more frequently than \rpo{01}. In fact, when turbulent trajectory ${\bf u}^a(t)$ is inside \lobe{1}, it spends more time near \re{01} than any other ECS according to \autoref{fig:shadow_summary}(a).
Moreover, when \re{01} is shadowed, then typically so is \rpo{01}. This is not particularly surprising, as \rpo{01} is rather compact, and the two ECSs lie very close to each other, with the distance between them being roughly a half of $\bar{D}$ according to \autoref{fig:xcorr}(a). This is a consequence of \rpo{01} being born in a Hopf bifurcation of \re{01}, which happens to lie close in the parameter space. It should also be mentioned that the length of the shadowing intervals for \re{01} tends to be quite large (e.g., around $8\gamma_n^{-1}$ for the event shown in \autoref{fig:shadow_re01}) compared with typical shadowing intervals for the \rpo{}s embedded in \lobe{1}.

We have not found any shadowing events for \rpo{19} or its reflection, even though it appears to be embedded inside \lobe{1}. Indeed, the smallest value of $D(t)$ for this ECS is $0.42$, which is above our threshold $\bar{D}$. Every other \rpo{} (as well as \re{01}) embedded in \lobe{1} is being shadowed by turbulence. Recall that ${\bf u}^a(t)$ visits both \lobes{1}{2}. As expected, shadowing events are confined to the temporal intervals when turbulent flow is inside \lobe{1}. The shadowing criteria (a)-(c) are not satisfied when the turbulent trajectory is inside \lobe{2} as we have not identified ECSs in that region of the state space.

\begin{figure}
\vspace{5mm}
\center
\subfloat[]{\includegraphics[width=0.5\textwidth]{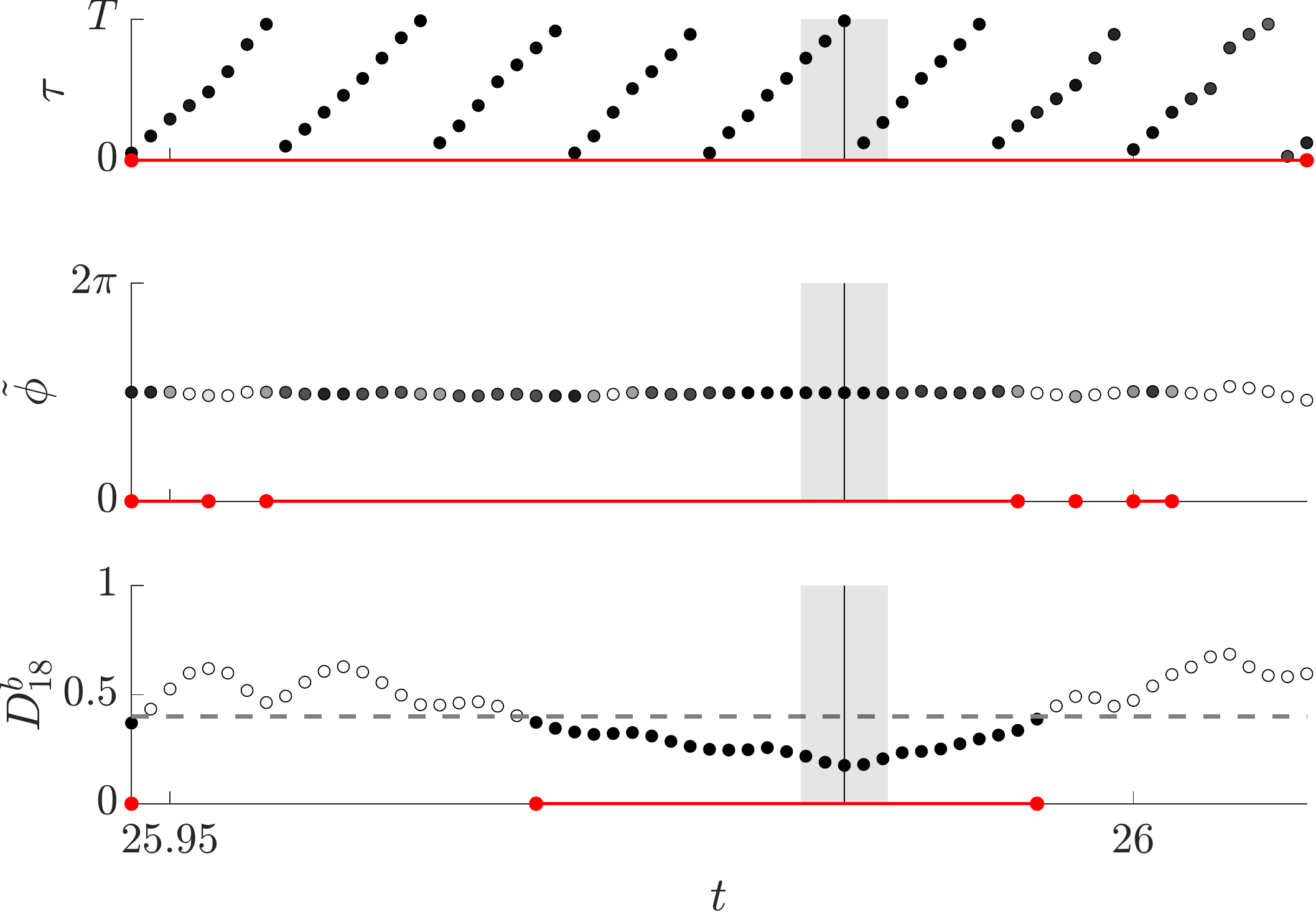}} \\
\subfloat[]{\includegraphics[width=0.45\textwidth]{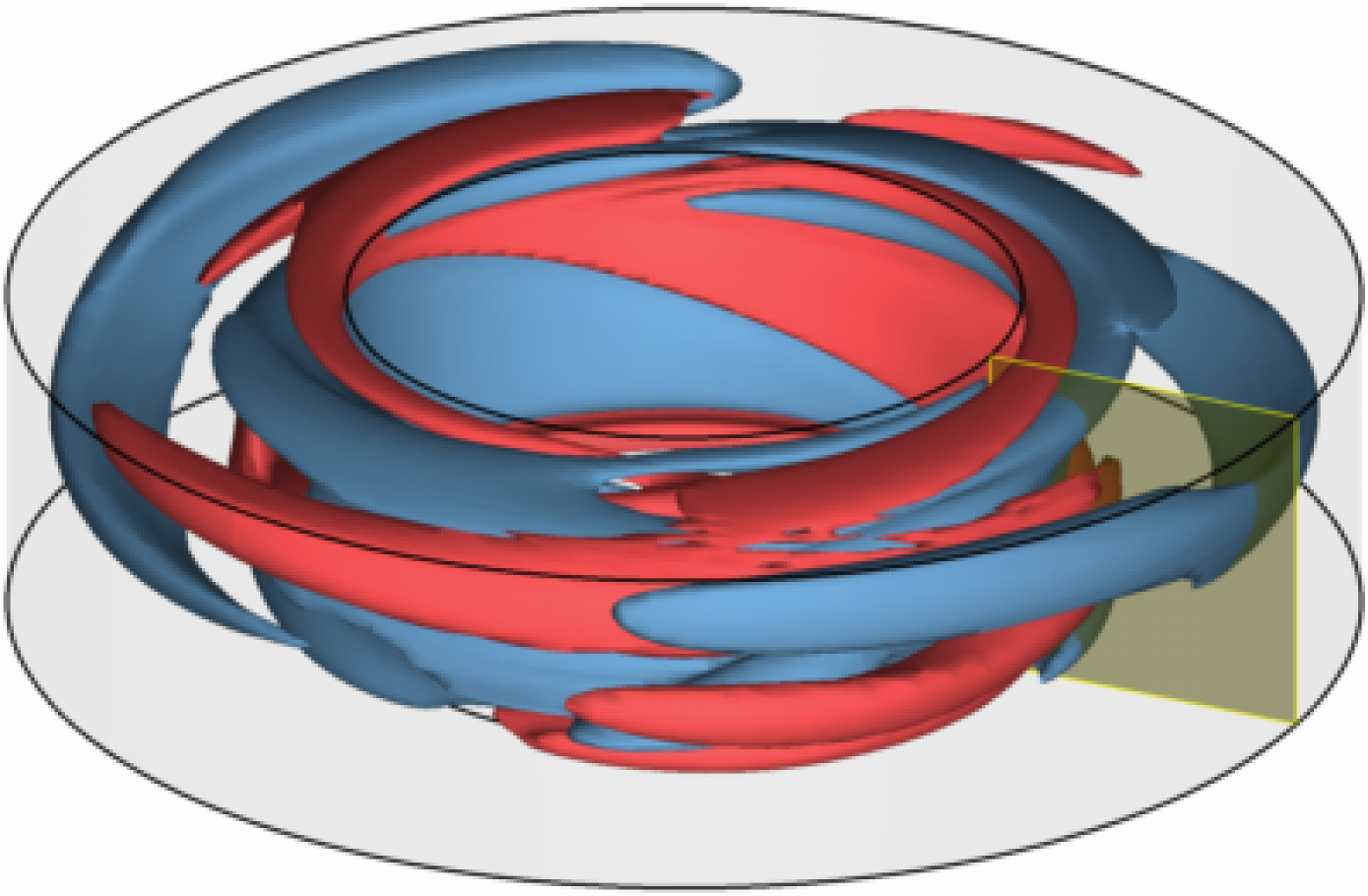}}  \hspace{3mm}
\subfloat[]{\includegraphics[width=0.45\textwidth]{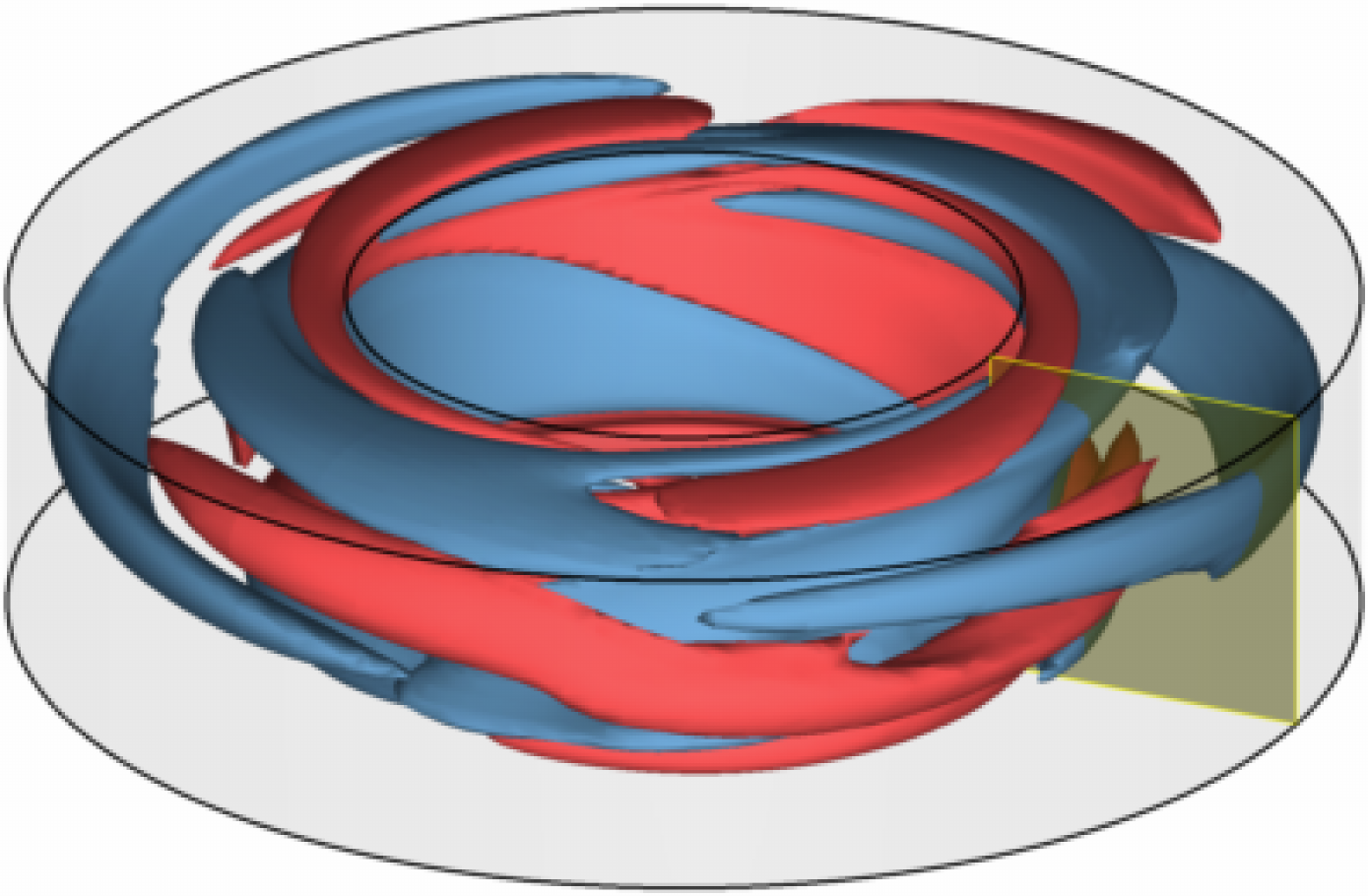}} \\
\subfloat[]{\includegraphics[width=0.4\textwidth]{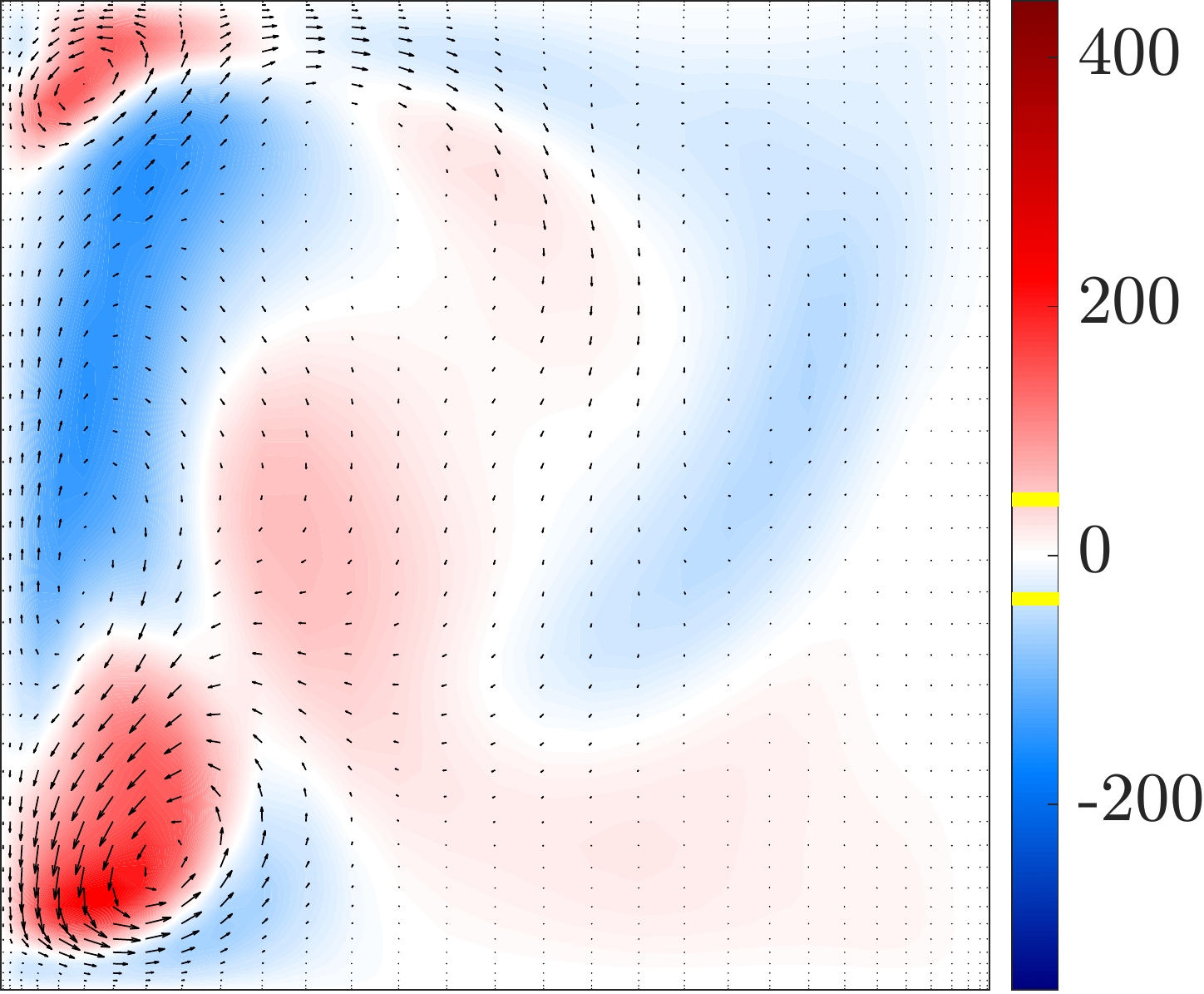}} \hspace{8mm}
\subfloat[]{\includegraphics[width=0.4\textwidth]{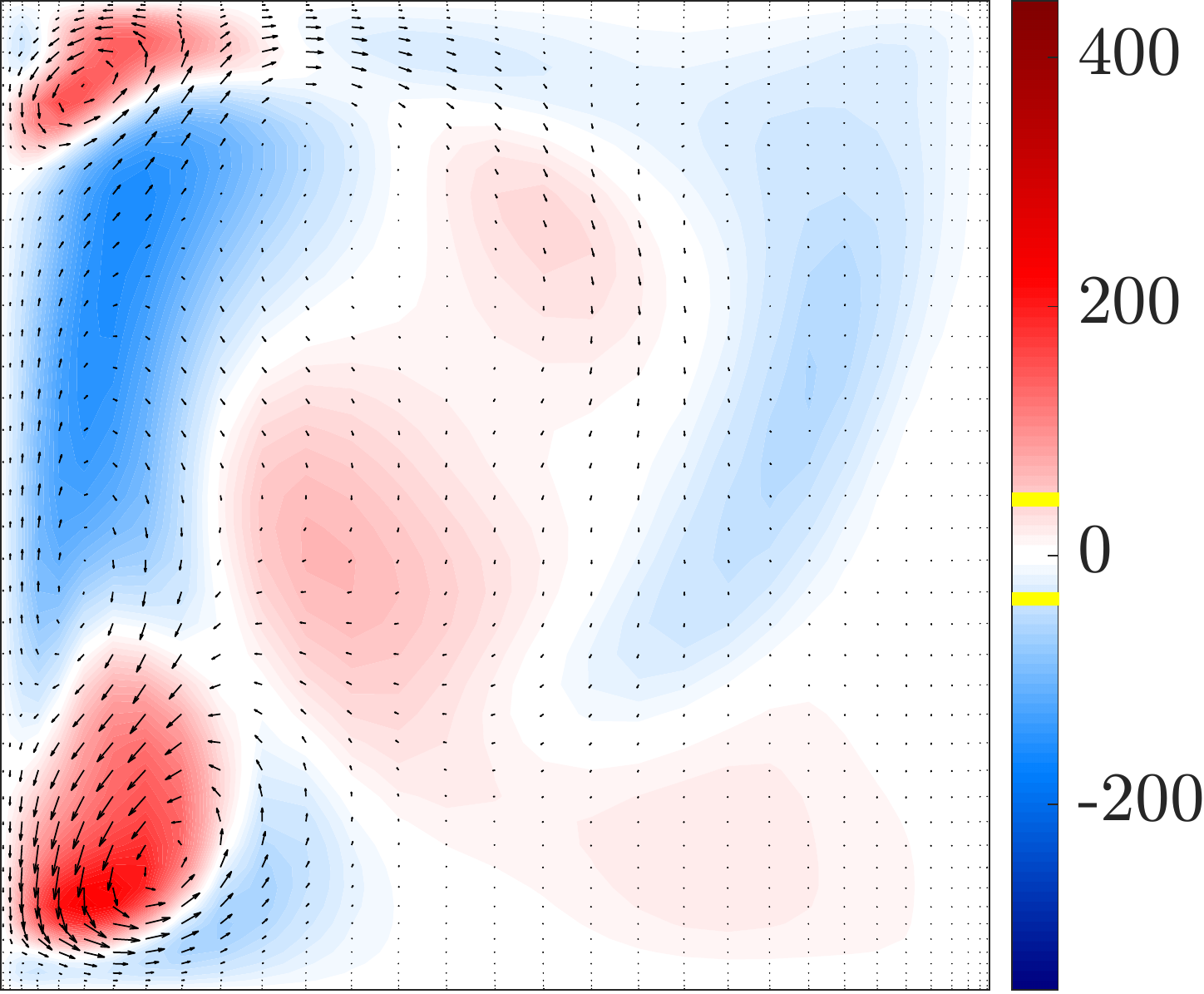}}
\caption{\label{fig:shadow_rpo15} \small 
A shadowing event for \rpo{15} in \lobe{3}.  A corresponding movie is available as supplementary material.
}
\end{figure}

For completeness, we have also performed a similar analysis for the turbulent trajectory ${\bf u}^b(t)$ which lies in \lobe{3}. It was found to shadow \rpo{13}-\rpo{16}, and \rpo{18}. Since this lobe breaks the reflection symmetry, none of the reflected copies of these \rpo{}s were shadowed. An example of shadowing for \rpo{15} is shown in \autoref{fig:shadow_rpo15}. It is worth emphasizing that ECSs in \lobe{3} are more unstable than those in \lobe{1}, as illustrated by both the larger number of unstable directions and the shorter escape time. As a result, there are fewer instances of shadowing and each shadowing event is nominally shorter. For instance, even though \rpo{15} is shadowed for almost three (very short) periods in \autoref{fig:shadow_rpo15}, this interval corresponds to only 0.022 in nondimensional units, compared with around $0.1$ for \rpo{01} and \rpo{05}. 

%
%

\section{Discussion}
\label{sec:disc}

We have considered a new regime of small-aspect-ratio Taylor-Couette flow where the two cylinders rotate in opposite directions, so both centrifugal and shear instabilities may induce and sustain turbulent flows. For the Reynolds numbers considered, we find that turbulent flows explore several distinct regions of the state space. These regions correspond to five lobes of the associated chaotic set that are dynamically connected to each other.

Discovery of \lobe{3} was prompted by the observation that a large number of ECSs were found in the same region of the state space outside of \lobes{1}{2}. While it is expected to find clusters of ECSs in regions of the state space inhabited by turbulence, to the best of our knowledge, this is the first example where a cluster of ECSs was found to predict the presence of a chaotic set supporting turbulent flow. ECSs inside \lobe{3} were found using both parameter continuation of ECSs inside \lobe{1} and using a Newton-Krylov solver with initial conditions confined to \lobe{1}. This suggests that computation of large sets of ECSs is a fairly robust way of identifying initial conditions that lead to turbulence, at least of transient variety.

The majority of the computed ECSs were found to be collocated with either \lobe{1} or \lobe{3}. Somewhat surprisingly, none of the ECSs were found to be collocated with \lobe{2}, although this may be due to the initial conditions for the Newton-Krylov solver lying mostly in \lobe{1}. Furthermore, some of the ECSs (e.g., \re{03} computed via continuation) were found to lie outside of all three lobes of the chaotic set, which is not entirely surprising. Of the ECSs collocated with one of the lobes, the majority are \rpo{}s, although \re{01} and \re{02} are also collocated with \lobes{1}{3}, respectively. More importantly, most of the ECSs collocated with one or the other lobe are found to be shadowed by turbulent trajectories. A similar result appears to also hold for turbulent pipe flow, although the respective study \citep{budanur2017} has only verified condition (a) for shadowing of a single \rpo{}.

Closeness and shadowing are often assumed to be synonymous in the literature, which has fostered confusion in the field. In fact, it is possible for the turbulent trajectory to be close to an ECS and not shadow it. In particular, closeness implies that $D$ is small regardless of whether the two solutions co-evolve. On the other hand, shadowing implies co-evolution, not just closeness, so that the ECS describes the dynamics of turbulent flow over some period of time. Hence, shadowing of an ECS implies that the ECS is dynamically relevant, while closeness to an ECS does not. Of course, for infinitesimally small $D$, closeness and shadowing do become equivalent. However, in practice turbulence never approaches any ECS infinitesimally closely.
Our results suggest that Euclidean distance $D$ between turbulent flow and an ECS family becoming small compared with its mean value is neither a necessary nor a sufficient condition for co-evolution. As Figures \ref{fig:shadow_rpo05}-\ref{fig:shadow_rpo01} illustrate, one routinely finds co-evolution in terms of variables $\tau$ and $\phi$ parameterising the group manifold, (i.e., conditions (b) and (c) being satisfied) for values of $D$ that are not particularly small. Similarly, small values of $D$ do not guarantee that $\tau$ and $\phi$ evolve as they should when turbulence shadows an \rpo{}. Nonetheless, there is a strong correlation between the three conditions (a), (b), and (c): when one is satisfied, more often than not so are the other two. 

More importantly, given that $D$ never becomes particularly small, we find many convincing examples of shadowing for both \rpo{}s and \re{}s on accessible time scales, despite our library of ECSs being relatively small. This is critical for the practical utility of an ECS-based framework for a dynamical description of turbulent flow. If the computed ECSs were rarely shadowed by turbulence, that would imply a serious problem with this framework, indicating that dynamically important solutions have not been found and possibly do not even correspond to either \rpo{}s or \re{}s. Admittedly, it is quite possible that (relative) quasi-periodic solutions may be more dynamically relevant than, say, the periodic ones. Of course, such a possibility cannot be excluded based on our results. Indeed, even for \lobe{1}, the neighborhoods of a dozen or so embedded \rpo{}s and \re{}s (plus their reflections), cover less than a half of this lobe, despite their relatively generous size, as illustrated in \autoref{fig:xcorr}(b). 

On the other hand, we found that multiple solutions may be shadowed simultaneously. As \autoref{fig:shadow_summary} illustrates, \re{01} and \rpo{01} are frequently shadowed at the same time. The same is true of \rpo{02} and \rpo{08} or \rpo{07} and \rpo{12}. This is not coincidental as the respective solutions are themselves close. For instance, \rpo{07} and \rpo{12} shadow each other for their entire duration, as illustrated by a movie provided in the Supplementary Material. Close solutions are all related via parameter continuation; groups of related solutions are indicated with brackets in \autoref{fig:xcorr}(a) and \autoref{fig:shadow_summary}(a).
For uniformly hyperbolic systems where periodic orbits are dense \citep{gaspard2005}, simultaneous shadowing of multiple unstable solutions is expected. The chaotic set underlying turbulence is not uniformly hyperbolic, as explained below. Despite this breakdown of uniform hyperbolicity, we find some RPOs to lie close enough to be shadowed simultaneously.

Unexpectedly, we found that the ECS that is shadowed the most by turbulent flow is a \re{}, not an \rpo{}. This result is potentially quite significant, as it suggests that an extension of periodic orbit theory to systems with continuous symmetries may have to include contributions from solutions other than \rpo{}s.
Since turbulence spends a large fraction of time in the neighborhood of \re{01}, excluding the contribution from this ECS to the average of any observable would greatly impact the corresponding temporal mean. 
Of course, it is possible that it is \rpo{01}, rather than \re{01}, that plays the important dynamical and statistic role, with \re{01} being shadowed because \rpo{01} is. Nonetheless, we chose a set of parameters that happens to be close to the bifurcation where \rpo{01} is born. As a result, \rpo{01} and \re{01} are not very distinct. A similar analysis to ours would have to be repeated further away from the bifurcation (e.g., at a higher \Rei) to determine which solution plays a more important role.  

Finally, we find that the shadowing property is robust to small changes in system parameters. As many of the solutions listed in \autoref{tab:ECSinTCF} were found through continuation in $Re_i$, there is a small deviation in this parameter from that describing turbulent flow. This variation emulates the discrepancies one might see when comparing numerically computed solutions to turbulent flows in experiment.

%
%

\section{Conclusions}
\label{sec:conc}

We set out to determine whether ECSs of one type or another are present and being shadowed by turbulent flow in a geometry with continuous symmetries and boundary conditions exactly matching a realistic experimental setup. Both questions were answered affirmatively for a small-aspect-ratio TCF with counter-rotating cylinders. We found two dozen \rpo{}s and \re{}s (not counting their symmetry-related copies) and determined that many of them are shadowed on accessible time scales, some rather frequently. The majority of ECSs we found are \rpo{}s, so it is not surprising that it is this type of ECSs that is shadowed the most frequently. What was unexpected is that the single most shadowed solution is a \re.

These results are quite significant, as they provide clear and unambiguous evidence supporting Hopf's picture of turbulence as a deterministic walk through neighborhoods of  various unstable solutions of the Navier-Stokes equation. The shadowing property implies that we can predict the evolution of turbulent flow over some interval of time, justifying the use of the ECS-based framework for a deterministic, dynamical description of turbulence. The length of this interval is not universal and is determined, as could be expected, by the degree of instability of the solution that is being shadowed.

Another key property of chaotic dynamics that has not been addressed fully is ergodicity. The number of ECSs we computed is insufficient for their neighborhoods to cover any of the lobes of the chaotic set, so we cannot make any conclusive statements regarding this property, except for one. For ECSs embedded into both \lobes{1}{3}, the number of unstable degrees is not constant. As shown by \citet{kostelich1997}, this implies that the dynamics are {\it not} uniformly hyperbolic in either lobe, so we should not expect the ergodic property to hold. This result is not exclusive to the flow we considered; the same observation applies to both pipe flow \citep{willis2013,budanur2017}, 2D Kolmogorov flow \citep{chandler2013,lucas2015} and its quasi-2D experimental realization \cite{suri2018,suri2020}. The implications of this for a dynamical theory of turbulence remain unclear.

This study focused almost exclusively on the dynamical significance of various ECSs. In fact, an ECS-based framework can also be used to connect a deterministic, dynamical description of fluid turbulence with a more traditional, statistical description. In particular, partition of the chaotic set into neighborhoods of various ECSs can be used to compute temporal means of any observable as a weighted sum over different ECSs. At present it remains unclear what types of ECSs should be included in the sum and how the weights should be computed. Initial studies \citep{kazantsev1998,chandler2013,lucas2015} aiming to address such issues had various limitations and were not conclusive, so further work in this area is needed. We will consider this issue in a subsequent publication.

\backsection[Acknowledgements]{The authors would like to thank Marc Avila for sharing his Taylor-Couette flow code and to gratefully acknowledge financial support by ARO under grant W911NF-15-1-0471 and by NSF under grant CMMI-1725587.}

\backsection[Declaration of interests]{The authors report no conflict of interest.}

\backsection[Supplementary data]{\label{SupMat}Movies illustrating various shadowing events are available at \newline \url{http://www.cns.gatech.edu/~roman/tcf/}.}

\backsection[Data availability statement]{Exact coherent structures and their key properties are available on GitHub at \url{https://github.com/cdggt/tcf/tree/main/eta0.50}.}

\bibliographystyle{jfm}
\bibliography{references}

\end{document}